\documentclass{aa}
\usepackage{txfonts}
\usepackage{graphicx}
\usepackage{lscape}
\usepackage{xcolor}

\usepackage{natbib}
\bibpunct{(}{)}{;}{a}{}{,}

\usepackage[normalem]{ulem}

\usepackage{pdflscape}
\usepackage{subfigure}
\usepackage{array}
\usepackage{multirow}
\usepackage{multicol}
\usepackage{longtable}
\usepackage{supertabular}

\usepackage{graphicx}
\usepackage{txfonts}
\usepackage{hyperref}
\begin{document} 

   \title{Searching for New Cataclysmic Variables in the Chandra Source Catalog}

   \author{Ilkham Galiullin\inst{1},
          Antonio C. Rodriguez\inst{2},
          Kareem El-Badry\inst{2},
          Paula Szkody\inst{3},
          Abhijeet Anand\inst{4},
          Jan van Roestel\inst{5},
          Askar Sibgatullin\inst{1},
          Vladislav Dodon\inst{1},
          Nikita Tyrin\inst{1},
          Ilaria Caiazzo\inst{2},
          Matthew J. Graham\inst{2},
          Russ R. Laher\inst{6},
          Shrinivas R. Kulkarni\inst{2},
          Thomas A. Prince\inst{7},
          Reed Riddle\inst{2},
          Zachary P. Vanderbosch\inst{2}
          \and Avery Wold \inst{2}
          }
   \institute{Kazan Federal University, Kremlevskaya Str.18, 420008, Kazan, Russia\\
     \email{IlhIGaliullin@kpfu.ru}
         \and
             Department of Astronomy, California Institute of Technology, 1200 E. California Blvd, Pasadena, CA, 91125, USA
        \and Department of Astronomy, University of Washington, 3910 15th Avenue NE, Seattle, WA 98195, USA
        \and Lawrence Berkeley National Laboratory, 1 Cyclotron Road, Berkeley, CA 94720, USA
        \and Anton Pannekoek Institute for Astronomy, University of Amsterdam, 1090 GE Amsterdam, The Netherlands
        \and IPAC, California Institute of Technology, 1200 E. California Blvd, Pasadena, CA 91125, USA
        \and Division of Physics, Mathematics, and Astronomy,
            California Institute of Technology, Pasadena, CA 91125, USA 
             }
             
    \authorrunning{Ilkham Galiullin, Antonio C. Rodriguez et al.}
    
   \date{Received XXX; accepted XXX}

% \abstract{}{}{}{}{} 
% 5 {} token are mandatory
 
  \abstract
   {}
     {Cataclysmic variables (CVs) are compact binary systems in which a white dwarf accretes matter from a Roche-lobe-filling companion star. In this study, we searched for new CVs in the Milky Way in the {\it Chandra} Source Catalog v2.0, cross-matched with {\it {\it Gaia}} Data Release 3 (DR3). }
     { We identified new CV candidates by combining X-ray and optical data in a color-color diagram called the ``X-ray Main Sequence".  We used two different cuts in this diagram to compile pure and optically variable samples of CV candidates. We undertook optical spectroscopic follow-up observations with the Keck and Palomar Observatories to confirm the nature of these sources.}   
     { We assembled a sample of 25,887 Galactic X-ray sources and found 14 new CV candidates. Seven objects show X-ray and/or optical variability. All sources show X-ray luminosity in the $\rm 10^{29}-10^{32}$ $\rm erg\ s^{-1}$ range, and their X-ray spectra can be approximated by a power-law model with photon indices in the $\rm \Gamma \sim 1-3$ range or an optically thin thermal emission model in the $\rm kT \sim 1-70$ keV range. We spectroscopically confirmed four CVs, discovering two new polars, one low accretion rate polar and a WZ~Sge-like low accretion rate CV. X-ray and optical properties of the other 9 objects suggest that they are also CVs (likely magnetic or dwarf novae), and one other object could be an eclipsing binary, but revealing their true nature requires further observations.}  
    {These results show that a joint X-ray and optical analysis can be a powerful tool for finding new CVs in large X-ray and optical catalogs. X-ray observations such as those by {\it Chandra} are particularly efficient at discovering magnetic and low accretion rate CVs, which could be missed by purely optical surveys. }

   \keywords{ novae, cataclysmic variables -- Stars: dwarf novae -- X-rays: binaries -- binaries: close -- white dwarfs -- binaries: eclipsing}

   \maketitle

%-------------------------------------------------------------------

\section{Introduction}

Cataclysmic variables (CVs) are compact semi-detached binaries in which a white dwarf (WD) accretes matter from a Roche-lobe filling companion star \citep{2003cvs..book.....W}. CVs have orbital periods, typically spanning 66 min -- 10 hr, except for ultra-compact AM Canum Venaticorum (AM CVn) stars, which possess $\lesssim 1$ hr periods \citep{2018A&A...620A.141R}. In magnetic CVs,  the magnetic field of the WD ($B\gtrsim 1$ MG) significantly influences the accretion onto the surface of the WD. Magnetic CVs are split into intermediate polars (IP) with truncated disks, and polars, where the strong field (B $\sim 10-250$ MG) prevents disk formation. Studies of CVs augment our understanding of various astrophysical phenomena, including their role as potential progenitors for Type Ia supernovae \citep{2014ARA&A..52..107M}. They also present optimal non-relativistic laboratories to study accretion processes \citep{2020AdSpR..66.1004H}. Furthermore, due to their short orbital periods, some CVs and many ultra-compact AM CVn systems are classified as prospective "verification binaries'' for future planned space-based gravitational wave observatories such as LISA \citep{2023arXiv230212719K}.

From the theoretical standpoint, CVs are predicted to form as a result of Common Envelope Evolution (CEE) \citep{1976IAUS...73...75P, 2013A&ARv..21...59I} that first brings the stars closer together. Angular momentum loss (AML) is the key factor in the long-timescale evolution of CVs. As AML occurs, the separation between the stars decreases, leading to a reduction in the orbital period. The evolutionary model considers two mechanisms for AML: Magnetic wind Braking (MB) \citep{1983ApJ...275..713R,1983A&A...124..267S} and Gravitational-wave Radiation (GR) \citep{1967AcA....17..287P, 1971ApJ...170L..99F}. Although the model initially aligned well with the observational data of that era, subsequent findings revealed numerous discrepancies between theory and observations \citep[e.g.,][]{1998PASP..110.1132P, 2009MNRAS.397.2170G, 2012MNRAS.419.1442P,2022MNRAS.517.4916E}. Consequently, empirical revisions were proposed to rectify these inconsistencies \citep[e.g.,][]{2011ApJS..194...28K, 2016MNRAS.455L..16S, 2021NatAs...5..648S}.

CVs exhibit a broad range of observational features, therefore detailed multiwavelength studies are necessary to draw accurate conclusions about their nature and population. X-ray emission analysis is a powerful tool for searching for and studying CVs. The first X-ray observations of CVs took place with HEAO-1 \citep{1981MNRAS.196....1C} and Einstein \citep{1984MNRAS.206..879C}. Early X-ray surveys led to the discovery of a unique, previously unidentified population of magnetic CVs \citep[e.g.,][]{1995A&A...297L..37H}. The all-sky X-ray survey conducted by the eROSITA telescope aboard the Spektrum-Roentgen-Gamma (SRG) mission \citep{2021A&A...656A.132S,2021A&A...647A...1P} is proving to be groundbreaking by cataloging millions of sources \citep{2024A&A...682A..34M} and has already been successful in discovering a collection of new and rare CVs due to its ability to reach much fainter fluxes \citep[e.g.,][]{2022schwope, 2022schwope_b, 2023ApJ...954...63R, 2024MNRAS.528..676G}.

Despite not being an all-sky X-ray survey, the {\it Chandra} observatory provides unique sub-arcsecond on-axis resolution which is comparable with optical surveys \citep{2000SPIE.4012....2W,2002PASP..114....1W}. The {\it Chandra Source Catalog v2.0} (hereafter, CSC2)  provides the uniformly calibrated, science-ready data products compiled from the archival Chandra data \citep{2010ApJS..189...37E}. Such data collection enables the study of properties of X-ray sources with spectral and timing analyses and classifying new objects in the catalog. In addition to new X-ray surveys and catalogs, all-sky optical photometric surveys provide a wealth of new data. The Zwicky Transient Facility (ZTF) is the deepest Northern sky time-domain photometric survey, conducted at Palomar Observatory on the Samuel Oschin 48-inch telescope. The 47 $\textrm{deg}^2$ field-of-view camera images the sky roughly every two days down to $\sim$20.5 mag in $g$, $r$, and $i$ filters \citep{bellm2019, graham2019, dekanyztf, masci_ztf}. 

In this paper, we aim to search for and study new CVs in the CSC2, cross-matched with the  third  {\it Gaia} Data Release (hereafter, {\it Gaia} DR3). In Section \ref{sec:gal_src}, we describe the construction steps of a sample of Galactic X-ray sources based on the CSC2--{\it Gaia} DR3 catalog. In Section \ref{sec:cv_selection}, we describe the selection of new CV candidates in the CSC2--{\it Gaia} DR3 catalog using two different cuts in the X-ray and optical color-color diagram, known as the ``X-ray Main Sequence". We compile pure and optically variable samples of CV candidates. In Section \ref{sec:dataandanalysis}, we present the follow-up observations conducted with the 5m Hale Telescope at Palomar Observatory and the 10m Keck I Telescope on Mauna Kea. We also present detailed X-ray timing and spectroscopic analyses. In Section \ref{sec:results}, we present the sample of 14 CV candidates found in the CSC2--{\it Gaia} DR3 catalog. We separately discuss the nature of four spectroscopically confirmed CVs and four variable CV candidates. Our results are summarized in Section \ref{sec:summary}.

%--------------------------------------------------------------------
\section{Sample of Galactic X-ray sources}
\label{sec:gal_src}

We began by constructing a sample of Galactic X-ray sources from the publicly available CSC2 \citep{2010ApJS..189...37E}. We cross-matched CSC2 with the {\it Gaia} DR3  catalog \citep{2023A&A...674A...1G}, keeping only objects with significant parallaxes and proper motions. We identified the known sources in our sample by cross-matching with the Simbad database and other publicly available CV catalogs. All sample creation steps are presented in detail below.

\subsection{{\it Chandra} Source Catalog}
As a first step, we used the master source table from the CSC2 with a total of about $\sim 317,000$ unique source detections. We set the following criteria\footnote{For detailed flag descriptions see \citet{2010ApJS..189...37E} and CSC2 web-page: \url{https://cxc.cfa.harvard.edu/csc/columns/flags.html}} to select the point-like X-ray sources with measured parameters:

\begin{enumerate}
    \item $\rm extent\_flag=False$; only point-like sources.
    \item $\rm conf\_flag=False$; source regions do NOT overlap.
    \item $\rm pileup\_flag = False$; ACIS pile-up fraction does NOT exceed $\sim 10\%$ in all observations.
    \item $\rm sat\_src\_flag=False$; the source is NOT saturated in all observations.
    \item $\rm streak\_src\_flag=False$; the source is NOT located on an ACIS readout streak in all observations.
    \item $\rm dither\_warning\_flag = False$; highest statistically significant peak in the power spectrum of the source region count rate does NOT occur at the dither frequency of the observation.
    \item $\rm man\_add\_flag=False$ and $\rm man\_inc\_flag=False$; the source was NOT manually added nor included in the catalog via human review.
    \item $\rm man\_match\_flag=False$; source detections were NOT manually matched between overlapping stacked observations via human review.
    \item $\rm man\_pos\_flag=False$; best-fit source position was NOT manually modified via human review.
    \item $\rm man\_reg\_flag=False$; source region parameters were NOT manually modified via human review.
    
\end{enumerate}

Using these criteria, we selected 271,438 unique objects from CSC2.

\subsection{Cross-match with {\it Gaia}}

For the next step, we cross-matched 271,438 objects with the {\it Gaia} DR3 catalog \citep{2023A&A...674A...1G} to select only Galactic sources. As a search radius for each object, we used the individual X-ray positional error (with a search radius equal to $\rm err\_ellipse\_r0$, the semi-major axis of the $\rm 95\%$ confidence ellipse). We only kept sources meeting the following criteria:
\begin{enumerate}
    \item $\rm RUWE < 1.4$ \citep[as recommended in][]{2023A&A...674A...1G}; good fit to the astrometric observations.
    \item Only ONE {\it Gaia} object must be within the search radius.
    \item Parallax and proper motion are measured with a signal-to-noise ratio $\rm S/N \ge 3$.
\end{enumerate}

These selection steps reduced the number of sources to 26,079. To remove any potential  high proper motion stars, we applied an additional filtering criterion:

\begin{enumerate}
    \item $\rm pm\times \Delta t \le err\_ellipse\_r0$;  The angular motion of the {\it Gaia} object (caused by a proper motion (pm)) during the time interval $\Delta t$ does not exceed the search radius ($\rm err\_ellipse\_r0$).   The time interval $\Delta t$ is an epoch difference between {\it Gaia} DR3 (Epoch 2016) and CSC2 (Epoch 2000) catalogs.
\end{enumerate}

This filter finally resulted in 25,887 objects in the CSC2--{\it Gaia} DR3 catalog.

\subsection{Identification of Known Sources}
\label{sec:simbad}

To identify known sources in the CSC2--{\it Gaia} DR3 catalog, we cross-matched our objects with the Simbad database \citep{2000A&AS..143....9W} using {\it Gaia} coordinates and a 3$\arcsec$ search radius. We grouped our objects based on the Simbad hierarchical structure\footnote{\url{http://simbad.cds.unistra.fr/guide/otypes.htx}}, making seven general types: STARS (single and binaries), ACCRETING BINARIES (X-ray binaries, CVs, Nova and Symbiotic Stars), GALAXY, ISM, PROPERTIES, BLENDS (not well-defined objects) and UNKNOWN (object is not presented in Simbad database). In Appendix \ref{appendix:PandC}, we present a detailed description of our general object types compiled from the Simbad hierarchical structure. We found Simbad classifications for 12,181\footnote{While creating the catalog of known systems, we had multiple matches (more than one) within the search radius for 810 of the 12,181 objects, giving about 7\% of possible spurious associations in the catalog. However, for most objects with multiple matches a Simbad database provided a similar main object type (as defined in Table \ref{tab:types}). We adopted the Simbad object type from the nearest source for those objects.}  of the 25,887 objects. 32 of these 12,181 were known CVs or CV candidates. 

To check that we did not miss any known CVs, we independently cross-matched our 25,887 objects with the following catalogs using a 2$\arcsec$ search radius: i) {\it Ritter and Kolb catalog} \citep{2003ritterkolb} (Final version; December 31, 2015); ii) {\it Cataclysmic Variables Catalog} (2006 edition; \cite{2001PASP..113..764D}); iii) {\it The Open Cataclysmic Variable catalog} \citep{2020RNAAS...4..219J} (only confirmed CVs are kept in the list). We combined the information from the Simbad database and these three CV catalogs to make the final list of known CVs. Out of 25,887 sources, we finally compiled a sample of 40 known CVs. This sample of 40 consists of the CVs from Simbad plus those from the three catalogs mentioned above.

We note that 52 objects (out of 25,887) were classified as GALAXY. These objects are primarily quasars and galaxies hosting active galactic nuclei that show significant parallax and proper motion. The high proper motions and parallaxes of these extragalactic objects could be due to variability-induced motion \citep[e.g.,][]{2022A&A...660A..16S, 2022ApJ...933...28M, 2023AstL...49..271K}. Detailed investigation of these objects is beyond the scope of this paper.

\section{CV Selection with the X-ray Main Sequence}
\label{sec:cv_selection}

\begin{figure*}
    \centering
    \includegraphics[width=1\textwidth]{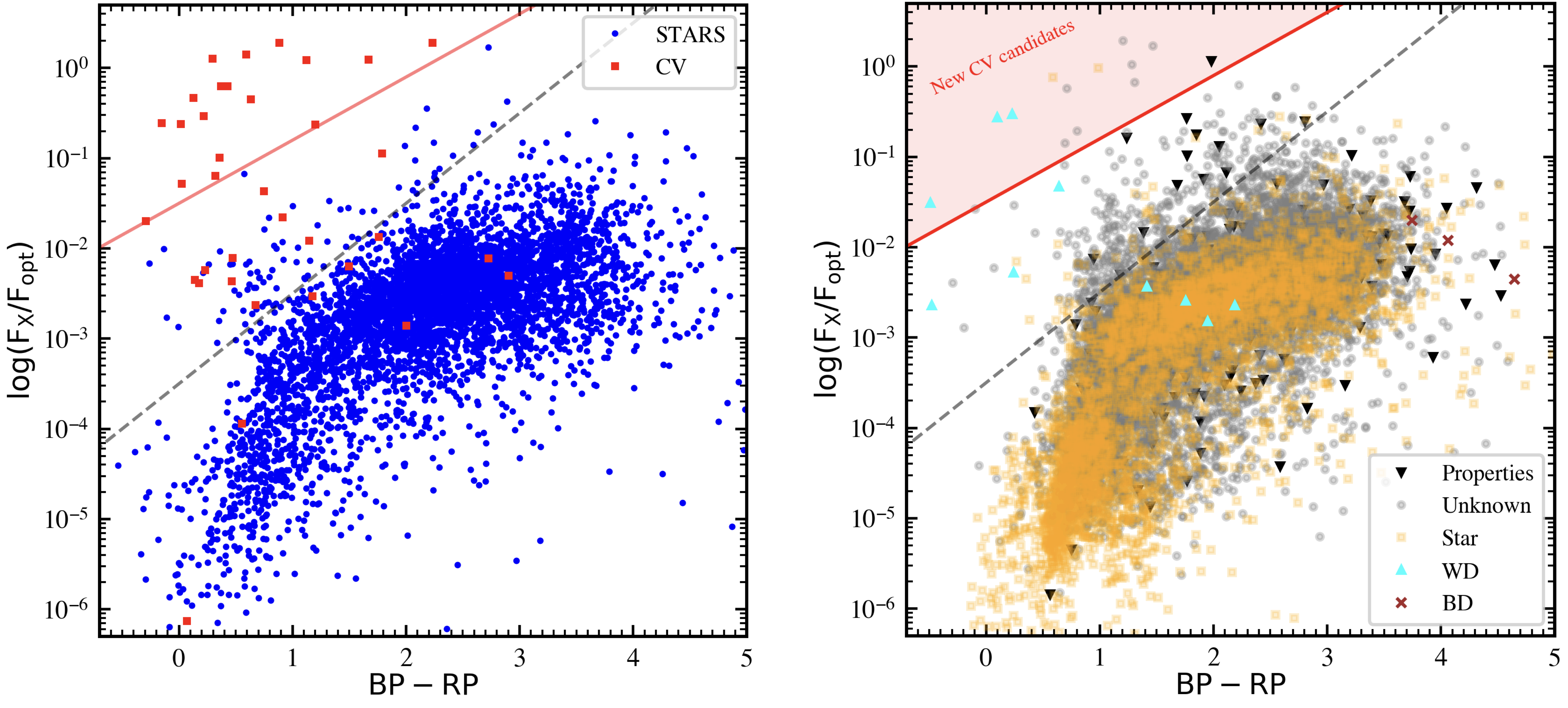}
    \caption{ X-ray-to-optical flux ratio vs Gaia (BP--RP) optical color  \citep[X-ray Main Sequence;][]{2024PASP..136e4201R} for objects in the CSC2--{\it Gaia} DR3 catalog. {\it Left panel:} Known CVs (red) and stellar objects (STARS, blue) in the CSC2--{\it Gaia} DR3 catalog. {\it The dashed black line} shows an empirical cut from \cite{2024PASP..136e4201R}. {\it The solid red line} shows a cut to select a pure sample of CV candidates. {\it Right panel:} Objects in the CSC2--{\it Gaia} DR3 catalog with the Simbad classifications (which may be imprecise). {\it Color symbols} correspond to the different object types based on the Simbad hierarchical structure (see Table \ref{tab:types} in Appendix \ref{appendix:PandC}). Objects above {\it the solid red line} and variable objects above {\it the dashed black line} are classified as new CV candidates (see Table \ref{tab:new_cvs} and Table \ref{tab:app_new_cvs}). For more details, see Section \ref{sec:cv_selection}. 
    }
    \label{fig:fxvsfopt}
\end{figure*}

We combined X-ray and optical data to search for new CV candidates in the CSC2--{\it Gaia} DR3 catalog. We used the ``X-ray Main Sequence", a phase space of  X-ray-to-optical flux ratio ($\tt F_{X}/F_{opt}$) and a {\it Gaia} optical BP--RP color \citep{2024PASP..136e4201R}, to select CV candidates and create a pure sample of these objects. The empirical cut in the ``X-ray Main Sequence" is given by the following functional form,
\begin{gather}
  \tt log(F_{X}/F_{opt})_{cut}= A + B \times (BP-RP),
    \label{eq:fx_fopt}
\end{gather}
where ${\tt A}= -3.5$ and ${\tt B}=1$ are linear parameters from \cite{2024PASP..136e4201R}. We computed the $\tt F_{X}/F_{opt}$ ratio based on X-ray and {\it Gaia} optical fluxes. The X-ray fluxes were calculated in the 0.5--7 keV energy band from CSC2. The aperture-corrected ($90\%$ enclosed counts fraction) energy flux is calculated using the power-law model with a photon-index of $\Gamma=2$ and the individual Galactic hydrogen column density $\rm N_{H(Gal)}$  in the direction of the source. {\it Gaia} G band magnitudes were converted to optical fluxes using the AB magnitude system and a central wavelength of 6,000~\AA. Out of 25,887 sources in the CSC2–{\it Gaia} DR3 catalog, we selected only sources with measured X-ray fluxes and optical magnitudes:
\begin{enumerate}
\item $\rm flux\_powlaw\_aper90\_b/error \ge 2$; Source X-ray flux in the 0.5--7 keV energy band is measured with signal-to-noise ratio more than $\rm S/N \ge 2$.
\item {\it Gaia} G band, $\rm BP$ and $\rm RP$ magnitudes are measured having signal-to-noise ratio more than $\rm S/N \ge 2$.
\end{enumerate}
These selection steps finally resulted in 19,665 sources (with 36 known CVs).

\subsection{First Selection: A pure sample of CV candidates }
\label{sec:first_cut}

To compile a pure sample of CV candidates and avoid contamination from stellar objects, we searched for optimal values for linear parameters ${\tt A}$ and ${\tt B}$ in Eq.~\ref{eq:fx_fopt}. We created two samples of known sources classified based on the Simbad hierarchical structure: one consisting of CVs (36 sources) and the other of stellar objects (STARS, 4781 sources). For the STARS sample, we included only objects with sub-types classified as "YSO", "Binaries", "Massive Stars", "Main-sequence Stars", or "Evolved Stars" (see Table \ref{tab:types}). The left panel of Figure \ref{fig:fxvsfopt} shows an $\tt F_{X}/F_{opt}$ and {\it Gaia} BP--RP color diagram for known CVs (red squares) and STARS (blue dots). It's evident that known CVs are primarily located in the upper left corner of the diagram and could be visually separated from STARS. We computed the purity and completeness using a sample of these known CVs and STARS (see Appendix \ref{appendix:PandC} for more details). Based on our completeness/purity analysis we selected ${\tt A} = -1.5$, and ${\tt B}=0.7$ that gives the highest purity ($\approx$100 \%) and $\approx$52\% completeness.
  
We only searched for new CV candidates among objects without precise classification from the Simbad database. Out of 19,665 objects, we selected 13,956 sources classified as one of the following object types: i) UNKNOWN (9,868 sources); ii) PROPERTIES (355 sources); iii) STARS, sub-type "Star" (3,720 sources); iv) STARS, sub-type "WD" (10 sources); v) STARS, sub-type "BD" (3 sources). The right panel of Figure \ref{fig:fxvsfopt} shows the $\tt F_{X}/F_{opt}$ and {\it Gaia} BP--RP color diagram for these objects. We found 11 new CV candidates (out of 13,956 objects) with our selection based on our most optimal values for  ${\tt A} = -1.5$ and ${\tt B}=0.7$ (see objects above the solid red line in the right panel of Figure \ref{fig:fxvsfopt}).

We note that we independently found three objects (recently confirmed as CVs) as CV candidates using the $\tt F_{X}/F_{opt}$ and Gaia (BP–RP) color diagram. During the writing of this manuscript, we learned that these systems had been previously discovered. However, these objects have only been mentioned in one or two papers, so we included those objects in our analysis and discussed them individually in Appendix \ref{app:knownCVs}.

\subsection{Second Selection: An optically variable sample of CV candidates with ZTF photometry}
\label{sec:second_cut}

Many CVs appear as periodic sources in optical light curves. Furthermore, accretion can lead to outbursts, which can also be observed in their optical light curves, particularly in non-magnetic CVs. Therefore, even after compiling a sample of 11 CV candidates based purely on an $\tt F_{X}/F_{opt}$ and {\it Gaia} BP--RP color diagram, we also searched for possible variability in a larger sample of sources. We used the empirical cut on the $\tt F_{X}/F_{opt}$ and {\it Gaia} BP--RP color diagram from \cite{2024PASP..136e4201R} (${\tt A} = -3.5$ and ${\tt B}=1$) and searched for possible variability in 375 sources (see objects above the dashed black line in the right panel of Figure \ref{fig:fxvsfopt}).

We searched for periodic and outbursting-type variability in sources based on ZTF optical photometry for both  $g$ and $r$ light curves, considering only data with a signal-to-noise ratio greater than 5, $\rm SNR_{ZTF}$ > 5. We retrieved ZTF light curves for our objects and used the barycentric corrected modified Julian date (MJD) in our analysis. Light curves have a photometric precision of $\sim0.015$ mag at the bright limit of 14 mag, decreasing to a precision of $\sim0.1$ mag at the faint limit of 20.5 mag. We searched for periodic variability using the Lomb-Scargle periodogram, as implemented in the \texttt{gatspy} tool to search for periods ranging from 5 min to 10 days. Once the period search was performed, we excluded all sources with orbital periods that are multiples or fractions of the sidereal day. To first order, the periods are reported up to a precision of 30 sec, which is determined by the exposure time. To calculate the significance of a period, we first computed the median absolute deviation (MAD) of the periodogram. Then, we took the periodogram value at the highest peak and divided it by the MAD. This ratio served as our significance metric. We plotted the distribution of significance metrics and kept only those sources in the upper $80^\textrm{th}$ percentile of the distribution. To search for outbursting sources, we looked for ZTF light curves with more than 5 outburst points $\Delta mag\ge 1$, with each point more than 3 standard deviations from the median.

\begin{figure*}
    \centering
    \includegraphics[width=0.75\textwidth]{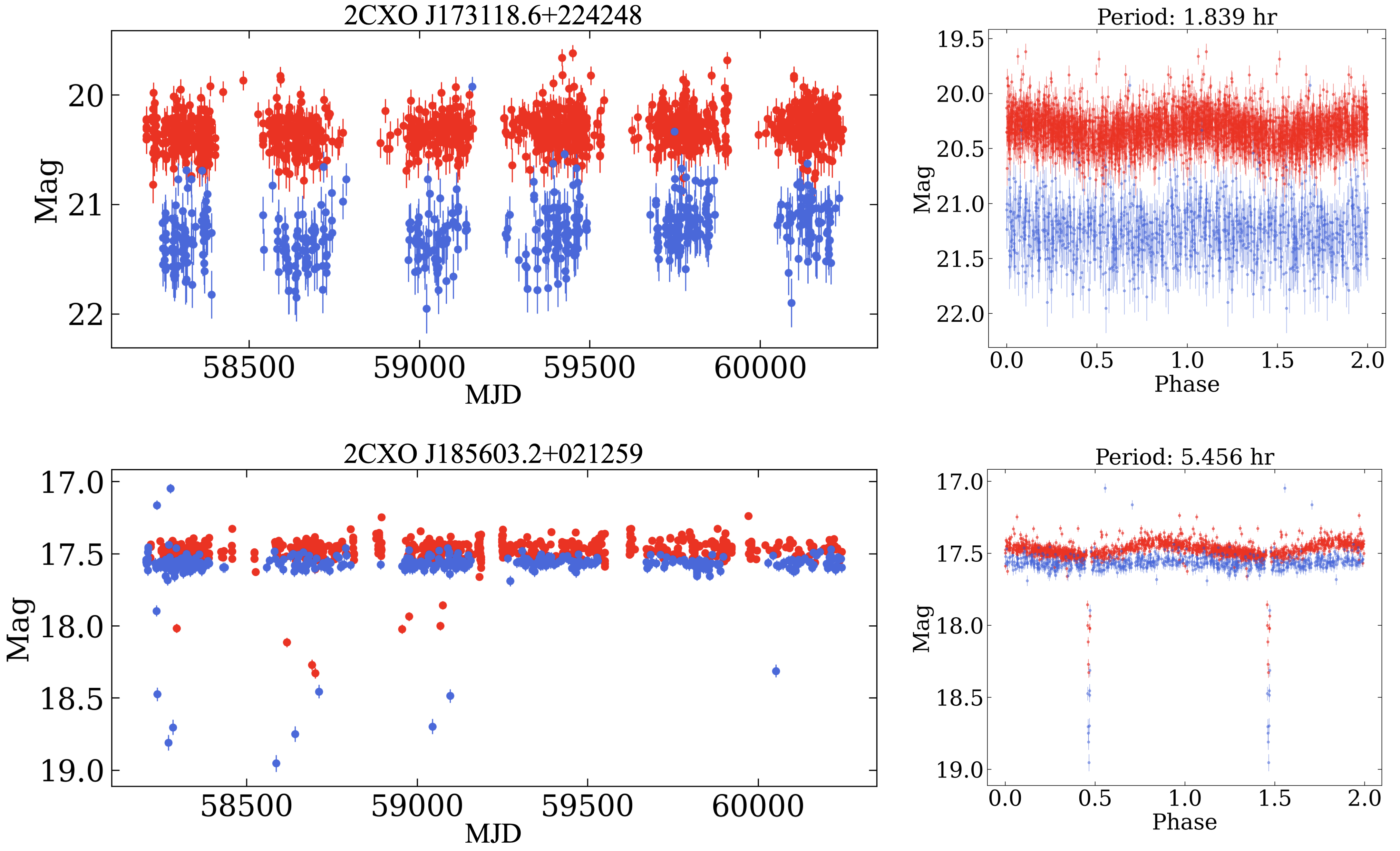}
    \caption{ZTF long-term (left) and phase-folded (right) light curves for 2CXO J173118.6+224248 (top panel) and 2CXO J185603.2+021259 (bottom panel) in $g$ (blue) and $r$ (red) filters.
    }
    \label{fig:ZTFLC}
\end{figure*}

In total, we had 178 (out of 375) sources with ZTF photometry. Using our selection criterion described above, we found five periodic sources (see Figures \ref{fig:ZTFLC}, \ref{fig:j0440_ztfspec}, \ref{fig:j1237_ztfspec} and \ref{fig:J1821_ztfspec}). With ZTF photometry, only one outbursting source (2CXO J024131.0+593630) was identified, which we discuss in Appendix \ref{app:knownCVs}. Based on the analysis of optical ZTF light curves, we extended our final sample of CV candidates to include a total of 14 candidates\footnote{Two periodic sources were already presented in the sample of 11 new CV candidates (selected by $\tt F_{X}/F_{opt}$ and {\it Gaia} BP--RP color). Therefore, we added only three variable sources (out of five) in our final sample to include 14 CV candidates in total.}.  

\begin{figure*}
    \centering
    \includegraphics[width=0.8\textwidth]{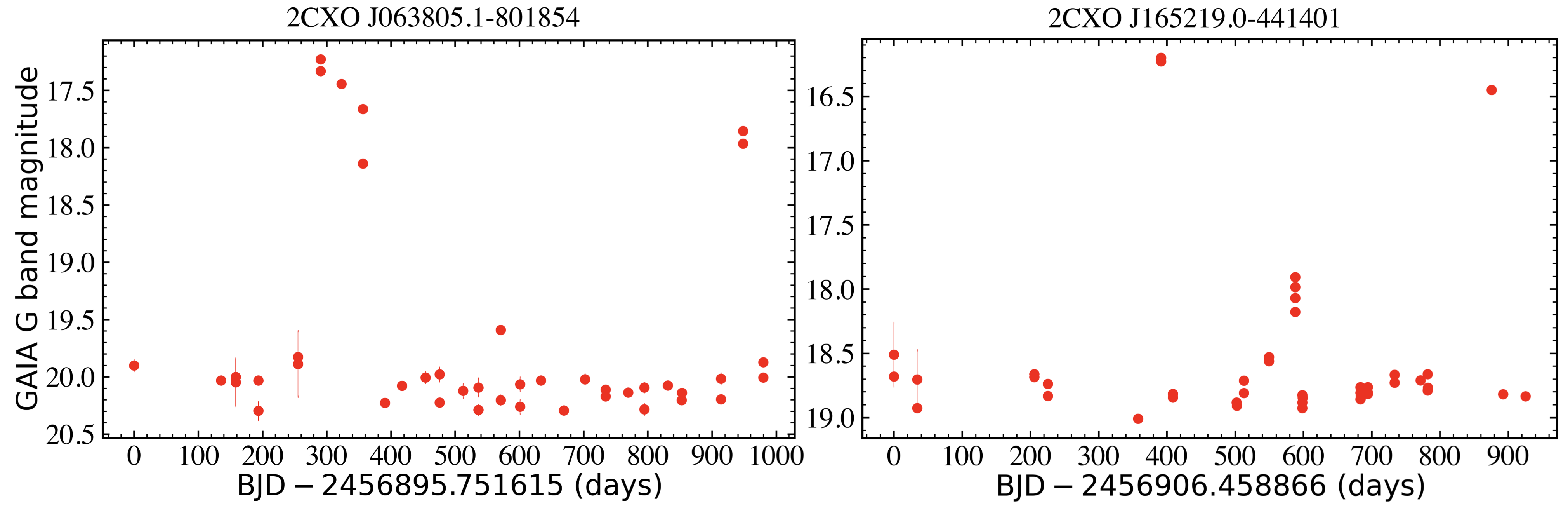}
    \caption{{\it Gaia} long-term light curves in G band for two sources in our sample. {\it Panels:}  2CXO J063805.1--801854 (left) and 2CXO J165219.0--441401 (right).
    }
    \label{fig:gaiaLC}
\end{figure*}

Out of 14 sources, 2 objects were marked as variables with a {\it Gaia} flag ($\tt phot\_variable\_flag=VARIABLE$). These objects show outbursting-type variability. Figure \ref{fig:gaiaLC} shows optical light curves based on {\it Gaia} photometry ($\rm SNR_{{\it Gaia}}$ > 3). Table \ref{tab:new_cvs} (column 9) summarizes the variability properties of the sources in the final sample. We note that the number of spurious matches in our final sample is effectively zero due to the $\sim$1--2" localization of X-ray sources made possible by {\it Chandra}. Furthermore, our candidates that show optical variability are even less likely to be false matches, since optical variable sources, particularly accreting compact objects, are typically associated with X-ray emission (e.g., see Figure 2 of \cite{2024PASP..136e4201R}).

\begin{table*}
\tiny
\caption{List of new CVs and their candidates found in the CSC2--{\it Gaia} DR3 catalog.}
\label{tab:new_cvs}
\renewcommand\arraystretch{1.5}
\setlength\tabcolsep{3.pt}
\centering
\begin{tabular}{lccccccccllcc}
\hline
Object ID & R.A. & DEC. & $\rm R_{95}$ & {\it Gaia} DR3 & G & BP--RP & $\tt F_{X}/F_{opt}$ 
& $\tt Var.$ 
& d 
& $\rm L_X$ 
& Obj. 
& Opt.  
\\

(2CXO) &  &  & ($\arcsec$) & source\_id & (mag) & (mag) & 
& 
& (pc) 
& ($10^{30}$ $\rm erg\ s^{-1}$) 
& Type 
& Spec.  
\\

(1) & (2) & (3) & (4) & (5) & (6) & (7) & (8)& (9)& (10)& (11)  
& (12) 
& (13)  
\\

\hline
\hline
\multicolumn{13}{l}{\textit{CV candidates selected via $\tt F_{X}/F_{opt}$ ratio and {\it Gaia} (BP-RP) color (First cut) }}\\

J044147.9--015145 & 04 41 47.9 & --01 51 45.7 & 0.92 & 3229135015053480064 & 20.27 & 0.10 & 0.27 
& -- 
& $249\pm42$ 
& $0.31\pm{0.12}$ 
& 4 
& \checkmark  
\\

J063805.1--801854 & 06 38 05.2 & --80 18 54.1 & 0.75 & 5207809444546340096 & 20.09 & 0.59 & 0.76 
& $\rm Xt$/Ga 
& $858\pm181$ 
& $23.73\pm{10.10}$ 
& 3 
& 
\\

J094607.7--311550 & 09 46 07.8 & --31 15 50.9 & 4.61 & 5440027429908157440 & 19.62 & --0.49 & 0.03 
& -- 
& $498\pm91$ 
& $0.39\pm{0.18}$ 
& 3 
& 
\\

J104435.5--594045 & 10 44 35.6 & --59 40 45.7 & 1.03 & 5350359031173784704 & 19.03 & --1.91$^*$ & 0.01  
& -- 
& $755\pm161$ 
& $0.22\pm{0.11}$ 
& 3 
& 
\\

J123727.5+655211 & 12 37 27.5 & +65 52 11.9 & 0.87 & 1680514565693517696 & 19.18 & 1.28 & 1.05 
& 2.12$^h$/X 
& $593\pm57$ 
& $43.64\pm{8.48}$ 
& 1 
& \checkmark  
\\

J155926.5--754311 & 15 59 26.6 & --75 43 11.5 & 6.97 & 5781613222711313152 & 20.52 & 0.80 & 0.20 
& -- 
& $409\pm108$ 
& $0.49\pm{0.31}$ 
& 1 
& 
\\

J165219.0--441401 & 16 52 19.0 & --44 14 01.8 & 0.72 & 5964348898459271040 & 18.79 & 1.98 & 1.16 
& Ga 
& $1839\pm572$ 
& $303.87\pm{195.30}$ 
& 2 
& 
\\

J173332.5--181735 & 17 33 32.6 & --18 17 35.8 & 0.71 & 4123404591191472384 & 18.61 & 1.20 & 1.93  
& -- 
& $855\pm153$ 
& $171.65\pm{62.06}$ 
& 1 
& 
\\

J173555.3--292530 & 17 35 55.3 & --29 25 30.2 & 2.65 & 4060073129401992832 & 20.47 & 0.71 & 0.57  
& -- 
& $127\pm26$ 
& $0.13\pm{0.07}$ 
& 1 
& 
\\

J182117.2--131405 & 18 21 17.2 & --13 14 05.8 & 1.49 & 4152650221755630464 & 20.22 & 0.69 & 0.15 
& 2.01$^h$ 
& $189\pm30$ 
& $0.06\pm{0.03}$ 
& 1 
& \checkmark 
\\

J190823.2+343328 & 19 08 23.2 & +34 33 28.1 & 2.90 & 2044505592269020416 & 20.47 & 1.30 & 0.67  
& -- 
& $373\pm103$ 
& $1.84\pm{1.05}$ 
& 1 
& 
\\

\hline
\hline
\multicolumn{13}{l}{\textit{CV candidates selected via period search (Second cut)}}\\

J044048.3+292434 & 04 40 48.3 & +29 24 34.4 & 0.72 & 158134751604204416 & 18.89 & 2.22 & 0.62 
& 1.30$^{h**}$ 
& $296\pm21$ 
& $7.05\pm{1.41}$ 
& 1
& \checkmark  
\\

J173118.6+224248 & 17 31 18.7 & +22 42 48.8 & 1.63 & 4557025073462888960 & 19.70 & 2.51 & 0.29  
& 1.84$^{h**}$ 
& $488\pm75$ 
& $2.40\pm{0.84}$  
& 1 
& 
\\

J185603.2+021259 & 18 56 03.3 & +02 12 59.8 & 2.18 & 4267156871660419712 & 17.06 & 1.33 & 0.05  & 5.46$^h$ & $211\pm4$ & $0.25\pm{0.09}$ & 1 & \\

\hline
\hline

\end{tabular}
\flushleft
\tablefoot{ {\it Data from CSC2}: (1) Source name, (2) and (3) Source position: right ascension and declination, (4) Major radius of the 95\% confidence level position error ellipse; {\it Data from {\it Gaia} DR3}: (5) Source ID, (6) and (7) G band and (BP--RP) magnitudes; (8) X-ray-to-optical flux ratio; (9) Variability flag. Orbital periods (in hr) are computed based on ZTF photometry in this study (see Section \ref{sec:second_cut}). If object is marked as a variable in: "Ga" --  {\it Gaia} DR3 catalog; "X" -- CSC2 catalog; "$Xt$"-- X-ray timing analysis revealed a periodic signal (see Section \ref{sec:Xvariability}); (10) Distance and its $1\sigma$ uncertainty computed from the {\it Gaia} parallax; (11) Absorption-corrected X-ray luminosity in the 0.5--7 keV energy band and its $1\sigma$ uncertainty; (12) Main object type based on Simbad database (see Section \ref{sec:simbad} and Table \ref{tab:types}): 1 -- UNKNOWN; 2 -- PROPERTIES;  3 -- STARS, sub-type "Star"; 4 -- STARS, sub-type "WD"; (13) Source has an optical spectrum (see Section \ref{sec:opt_spectra}). ($^*$) Indicates likely unreliable value due to large measurement errors; therefore, CV candidate 2CXO J104435.5--594045 ($\rm M_{G} = 8.9$) is not shown in Figures \ref{fig:fxvsfopt} and \ref{fig:HR} for illustrative purposes. (**) With our current data, we cannot distinguish between the two orbital periods of the system. The orbital period can be either the current value or two times it.
 }
\end{table*}

\begin{table*}
\tiny

\renewcommand\arraystretch{1.3}
	\centering
	\caption{Data acquired  for CV candidates.}
    \label{tab:data}
    \setlength\tabcolsep{3.2pt}
	\begin{tabular}{lcccc}
 \hline
Object ID (2CXO) & Data Type & Date & Instrument & Finding \\

		\hline
        \hline

J044048.3+292434 & Identification Spectrum & 22 Nov. 2022 & P200/DBSP & Hydrogen Balmer and Helium, and high-excitation HeII 4686$\AA$ emission lines.\\

&  & 22 Nov. 2022 & P200/DBSP & {\parbox{8cm}{\vspace{7pt}\centering Two spectra were taken 1.4 hr apart and show no significant radial velocity shift. } }\\

J044147.9--015145 & Identification Spectrum & 16 Nov. 2022 & Keck I/LRIS & {\parbox{9cm}{\vspace{7pt}\centering  Hydrogen and Helium emission, along with WD Balmer absorption lines. No detectable HeII 4686$\AA$ emission line is seen.}}\\

J123727.5+655211 & Identification Spectrum & 22 Nov. 2022 & P200/DBSP & {\parbox{9cm}{\vspace{7pt}\centering Hydrogen, Helium, and high excitation HeII 4686$\AA$ emission lines.}}\\

J182117.2--131405 & Identification Spectrum & 22 May 2023 & Keck I/LRIS & {\parbox{9cm}{\vspace{7pt}\centering The spectrum is dominated by broad humps which correspond to cyclotron harmonics, and no significant  $H\beta$ and He II 4686$\AA$ emission lines are detected.}}\\

\hline
\hline
	\end{tabular}
\end{table*}

\section{Data and Analysis}
\label{sec:dataandanalysis}

\begin{figure*}
\centering
\setlength{\tabcolsep}{1pt}
\begin{tabular}{cc}

    \includegraphics[width=0.5\linewidth]{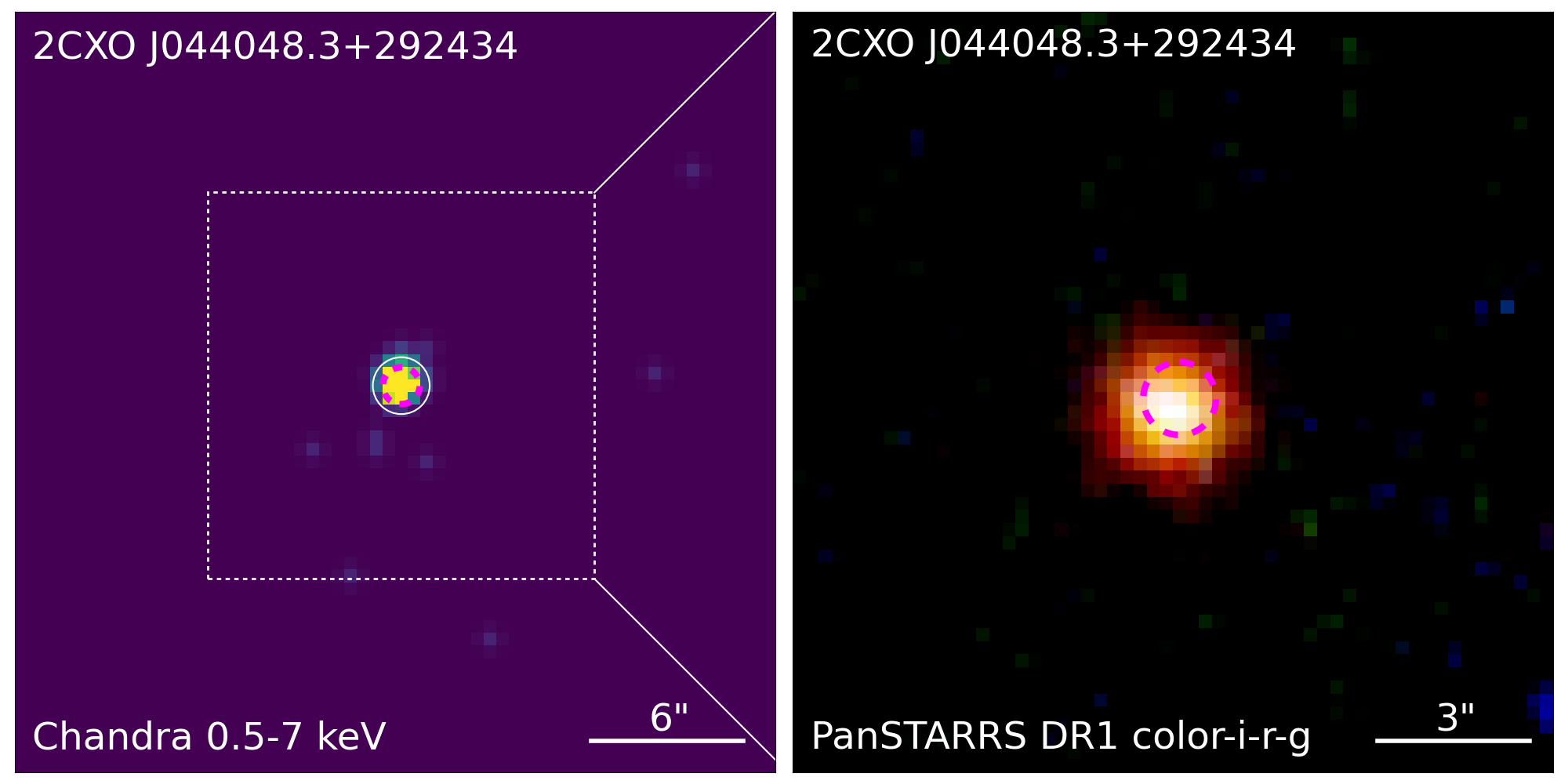} &
    
    \includegraphics[width=0.5\linewidth]{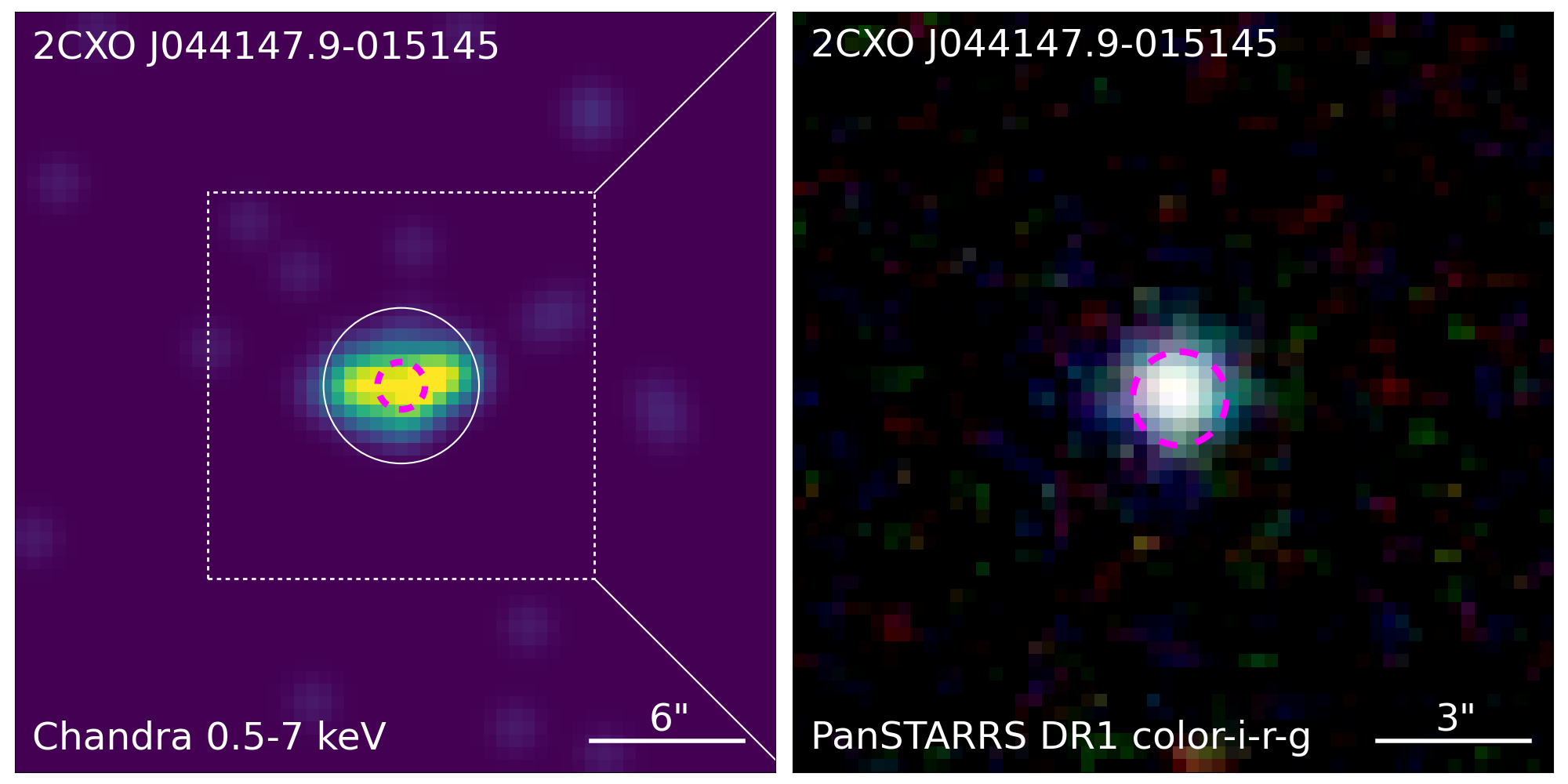} \\[0.01\tabcolsep]
    
    \includegraphics[width=0.5\linewidth]{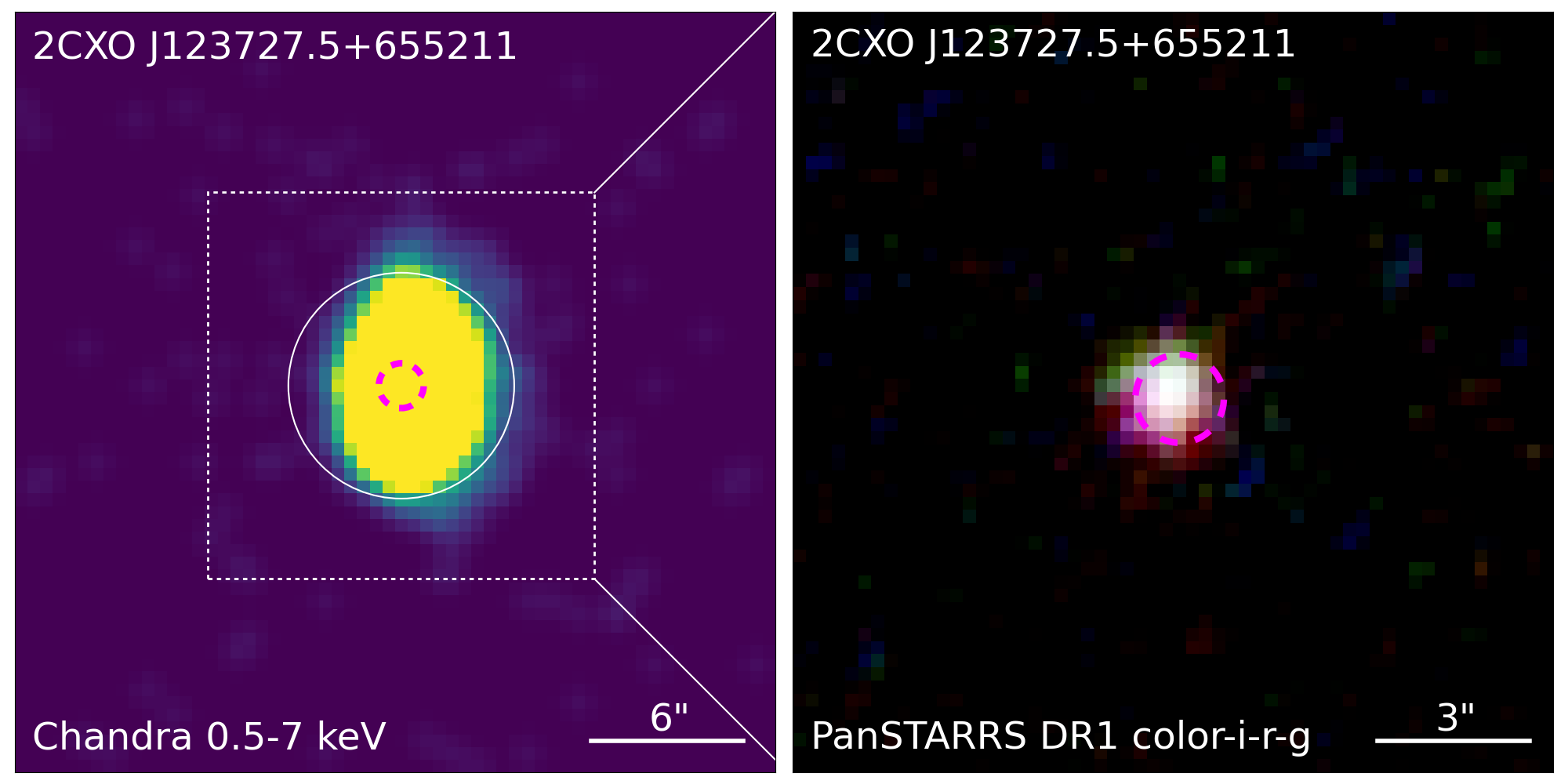} &

    \includegraphics[width=0.5\linewidth]{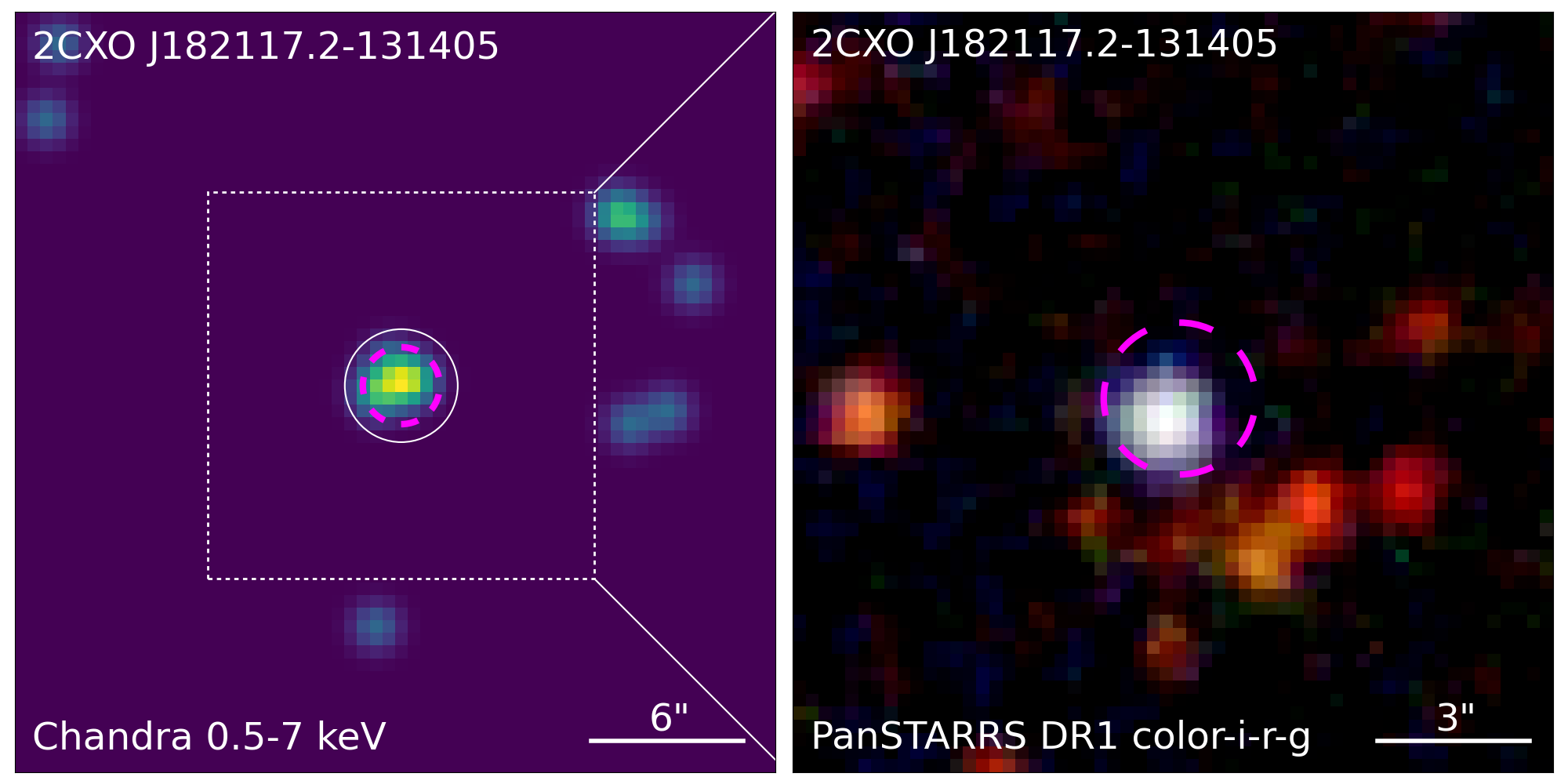} \\[0.01\tabcolsep]

\end{tabular}
\caption{X-ray and optical images for four new spectroscopically confirmed CVs from CSC2--{\it Gaia} DR3 catalog. {\it Left:} False-color {\it Chandra} X-ray images in the 0.5--7 keV energy band. The images were smoothed using a Gaussian kernel with different widths equal to $\rm 1-3$ times the PSF radius, depending on the source counts. {\it Right:} Composite optical images based on Pan-STARRS color i-r-g data. {\it The white boxes} show the field of view of the optical images on the right. {\it Circles}: a PSF radius equal to the 90\% of
encircled counts fraction (solid, white color), and a radius equal to the major semi-axis of the error ellipse (95\% localization error) centered on the source X-ray position (dashed, magenta color).} 
\label{fig:fov}
\end{figure*}

\begin{figure*}
    \centering
    \includegraphics[width=0.85\textwidth]{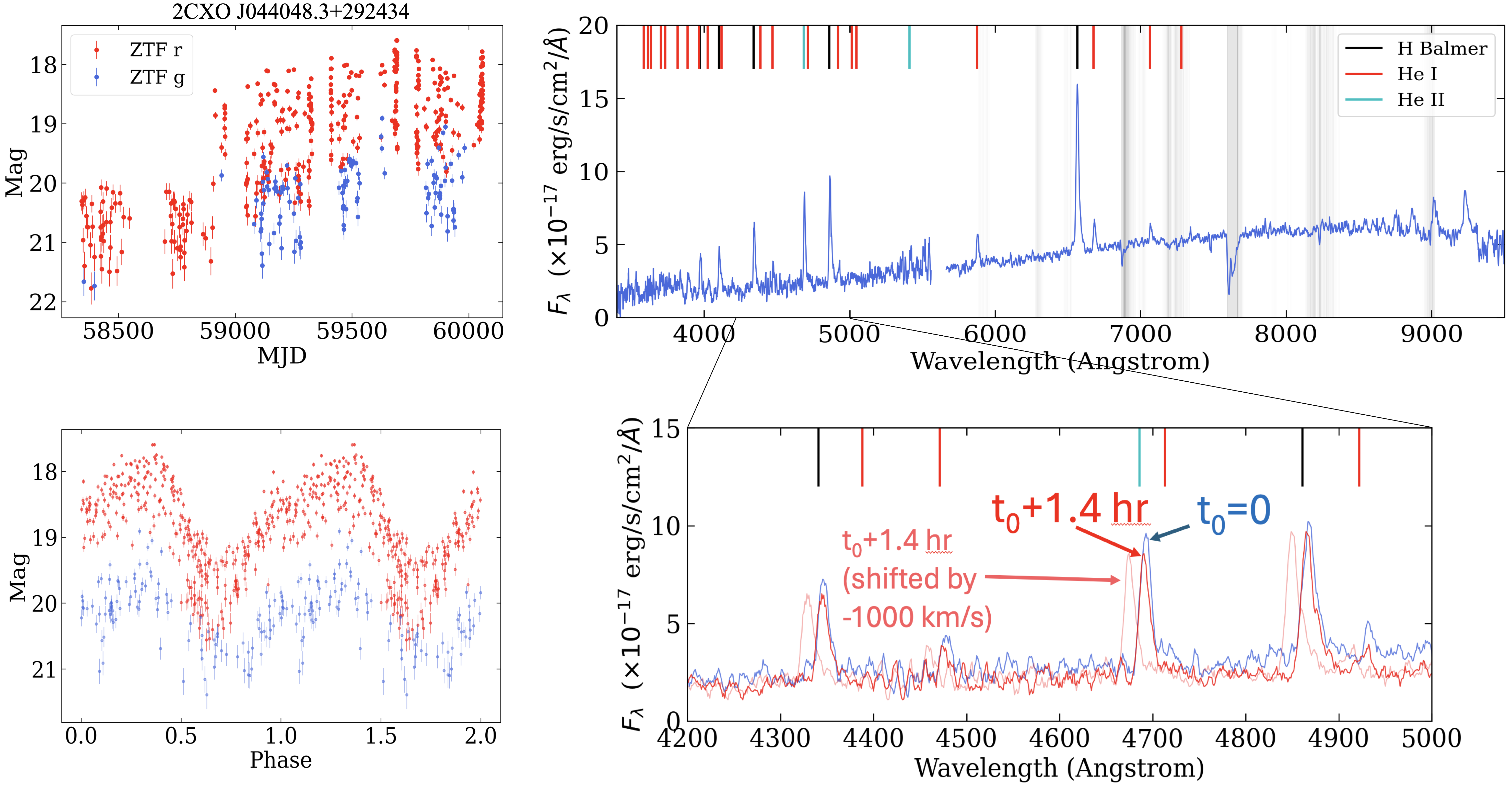}
    \caption{ZTF long-term (top left) and phase-folded (bottom left) light curves in $g$ and $r$ filters and the DBSP optical spectrum of 2CXO J044048.3+292434 (right). {\it Grey lines} represent the Keck Telluric Line List. The optical spectrum (top right) shows Hydrogen Balmer and Helium emission lines and high-excitation He II 4686$\AA$ emission line, giving a high line ratio  $\rm He II\ / H\beta\approx 0.64$ (see Table \ref{tab:EW_lines}).  Optical and X-ray properties indicate that the object is a polar. Bottom right panel: Since two spectra taken 1.4 hr apart do not show a significant radial velocity shift (a $\approx$1000 km/s shift would be expected if this were half of the orbital period and the system were viewed edge-on), the 1.3 hr period is preferable. However, time-resolved spectroscopy for 2.6 hr would be the only way to confirm the orbital period of this system, leading us to currently label the orbital period as either 1.3 or 2.6 hr (see Section \ref{sec:J0440} for more details).}
    \label{fig:j0440_ztfspec}
\end{figure*}

\begin{figure*}
    \centering
    \includegraphics[width=1.0\textwidth]{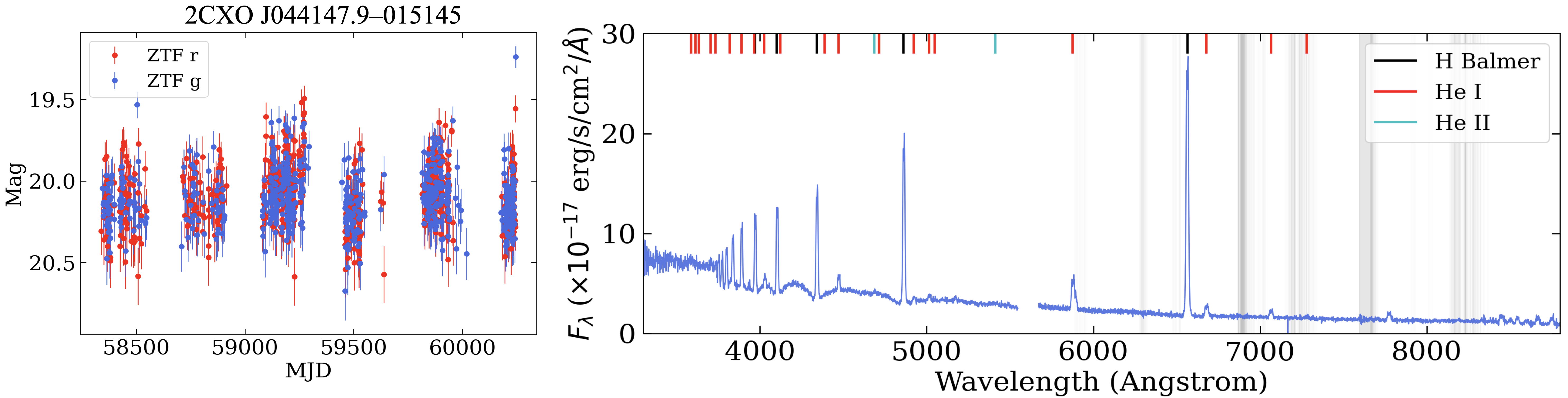}
    \caption{ ZTF long-term light curves in $g$ and $r$ filters (left) and DBSP optical spectrum of 2CXO J044147.9--015145 (right). {\it Grey lines} represent the Keck Telluric Line List. The optical spectrum shows WD Balmer absorption lines with no sign of the donor signature, weakly double-peaked Hydrogen and Helium emission lines, and no detectable He II 4686$\AA$ emission line, giving a $\rm 3\sigma$ upper limit for the line ratio $\rm He II\ / H\beta < 0.05$ (see Table \ref{tab:EW_lines}). Optical and X-ray properties suggest that the object is a CV (WZ~Sge type) (see Section \ref{sec:J04414} for more details).}
    \label{fig:J0441_ztfspec}
\end{figure*}

\begin{figure*}
    \centering
    \includegraphics[width=0.75\textwidth]{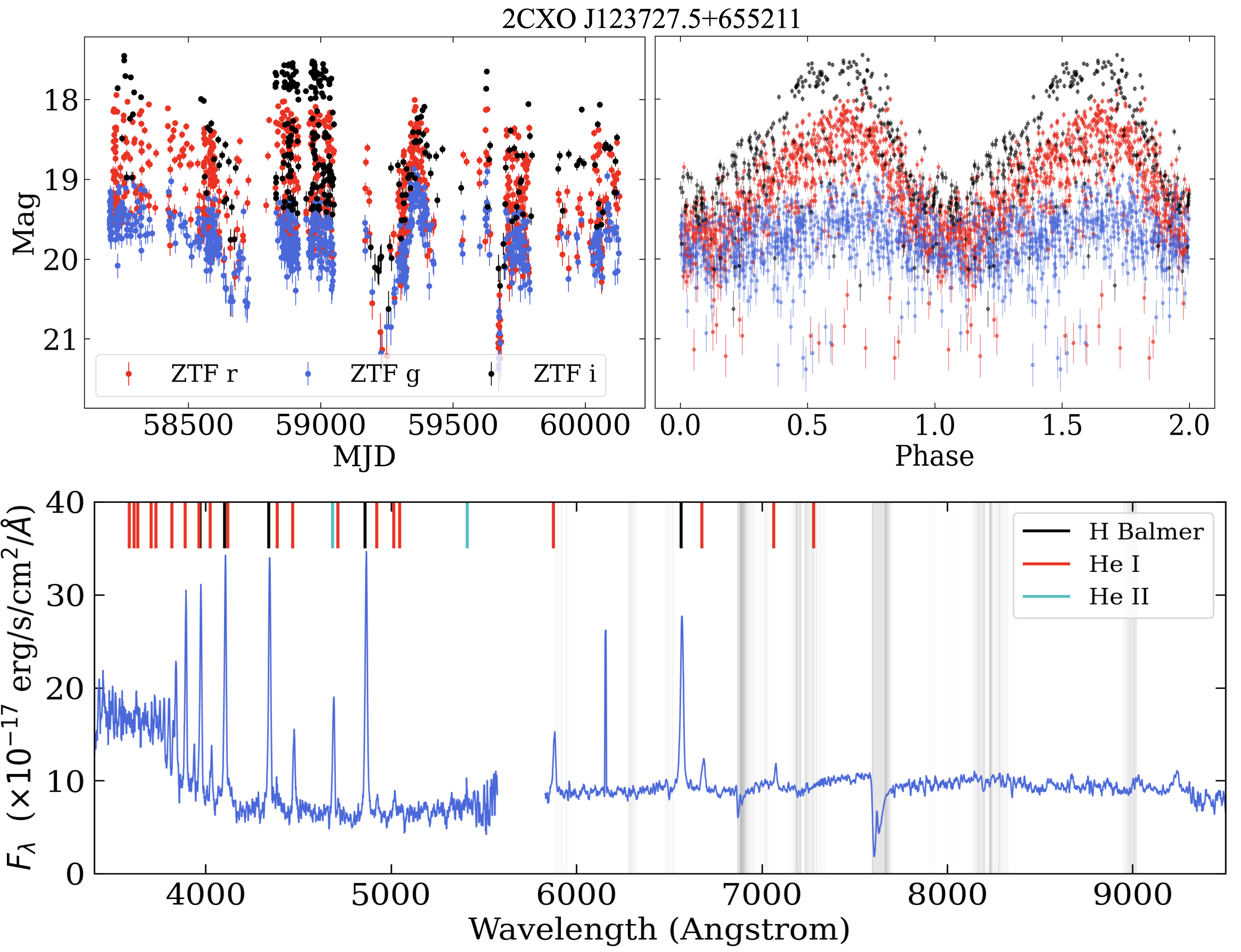}
    \caption{ ZTF long-term (top left) and phase-folded (top right) light curves in $g$, $r$ and $i$ filters and the DBSP optical spectrum of 2CXO J123727.5+655211 (bottom). {\it Grey lines} represent the Keck Telluric Line List. The optical spectrum shows Hydrogen, Helium, and high excitation He II 4686$\AA$ emission lines, giving a line ratio of $\rm He II\ / H\beta\approx 0.37$ (see Table \ref{tab:EW_lines}). Together with optical light curves, this indicates that the object is a polar (see Section \ref{sec:J1237} for more details).} 
    \label{fig:j1237_ztfspec}
\end{figure*}

\begin{figure*}
    \centering
    \includegraphics[width=0.75\textwidth]{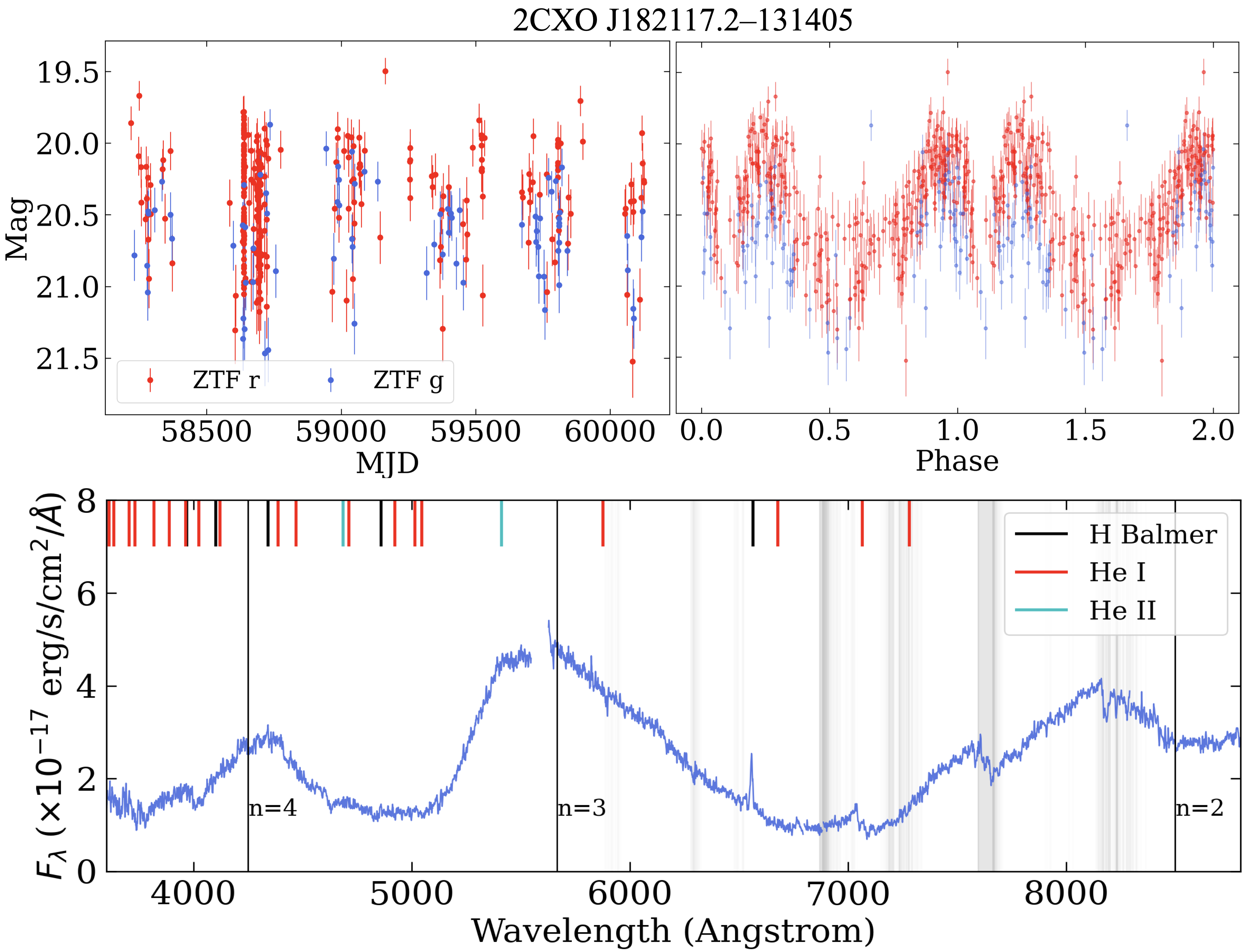}
    \caption{ZTF long-term (top left) and phase-folded (top right) light curves in $g$ and $r$ filters and the DBSP optical spectrum of 2CXO J182117.2-131405 (bottom). {\it Grey lines} represent the Keck Telluric Line List. The optical spectrum is dominated by broad humps which correspond to cyclotron harmonics, and no significant  $H\beta$ and He II 4686$\AA$ emission lines are detected. A combination of the cyclotron wavelengths seen in the spectrum and the large beaming amplitude constrains the magnetic field to be $ B\approx54-63$ MG. Optical and X-ray properties suggest that the object is a low accretion rate polar (see Section \ref{sec:J1821} for more details). }
\label{fig:J1821_ztfspec}
\end{figure*}

\subsection{Optical Spectra}
\label{sec:opt_spectra}

We performed optical spectroscopic follow-up observations for four sources in our sample, focusing on those located in the Northern Hemisphere that were accessible to our telescopes and exhibited strong optical variability and/or a high $\tt F_{X}/F_{opt}$ ratio. We used both the Double Spectrograph \citep[DBSP;][]{1982dbsp} on the 5m Hale Telescope at Palomar Observatory and the Low-Resolution Imaging Spectrometer \citep[LRIS;][]{1995lris} at the 10m Keck I Telescope on Mauna Kea. During all DBSP observations, we used the 600/4000 grism on the blue side and the 316/7500 grating on the red side. A 1.5$\arcsec$ slit was used, and the seeing on all occasions was 1.5 -- 2.0$\arcsec$. All P200/DBSP data were reduced with \texttt{DBSP-DRP}\footnote{\url{https://dbsp-drp.readthedocs.io/en/stable/index.html}}, a Python-based pipeline built upon the existing \texttt{PypeIt} package \citep{2020pypeit} and optimized for reducing spectral data. During all LRIS observations, we used the 600/4000 grism on the blue side and the 600/7500 grating on the red side. We used a 1.0$\arcsec$ slit, and the seeing on all occasions was around 0.7$\arcsec$. LRIS data were reduced with \texttt{lpipe}, an IDL-based pipeline optimized for LRIS long slit spectroscopy \citep{2019perley_lpipe}. All data were flat fielded and sky-subtracted using standard techniques, wavelength-calibrated with internal arc lamps, and flux-calibrated using a standard star. Table \ref{tab:data} shows all data acquired for the four CV candidates. Single spectra were taken at a time for each object (no coadds), with typical exposure times of 600 seconds for Keck I/LRIS and 900 seconds for P200/DBSP.

Figure \ref{fig:fov} shows {\it Chandra} X-ray images and optical images from Pan-STARRS \citep{2016panstarrs} of the same sky region for four spectroscopically confirmed CVs in our sample. Figures \ref{fig:j0440_ztfspec} - \ref{fig:J1821_ztfspec} show the optical spectra of these sources. We identified prominent emission lines and computed their equivalent widths (EWs) by fitting Gaussian profiles to them. Table \ref{tab:EW_lines} shows the EWs of selected lines and the $\rm He II(4686)\ / H\beta$ ratio\footnote{2CXO J044048.3+292434 was observed twice, and both observations show equivalent width measurements which are in agreement to within $1\sigma$. Therefore, in Table \ref{tab:EW_lines}, we only report the values from the first exposure.}. 

We conducted an extensive cross-match analysis between our 14 objects and publicly available optical data from the Sloan Digital Sky Survey (SDSS) and the Dark Energy Spectroscopic Instrument (DESI) collaboration \citep{DESI2016-Overview}. With $\rm 2\arcsec$ search radius, we found no matches between SDSS/APOGEE catalogs (APOGEE DR17 StarHorse VAC, \citet{apogeestarhorse2020}; APOGEE DR17 astroNN VAC, \citet{astroNNapogee2019}; APOGEE DR17 DistMass VAC, \citet{distmassapogee2024}; APOGEE Net VAC \citet{apogeenet2022}; MaStar Stellar Parameters VAC, \citet{Mastar2022}) and our CV sample. We cross-matched our objects with the early data release (EDR) of DESI collaboration \citep{desiedr2023}, which contains spectra of more than approximately 1 million objects \citep{desiedr2023}. We found no identification of our objects with this catalog.

\begin{table*}
\tiny

\renewcommand\arraystretch{1.25}
	\centering
	\caption{Equivalent widths ($-EW (\AA)$) of selected lines of four spectroscopically confirmed CVs. 3$\sigma$ upper limits are shown if no higher significant measurement is detected.}
    \label{tab:EW_lines}
    \setlength\tabcolsep{3.2pt}
	\begin{tabular}{lcccc} 
		\hline
    Object ID (2CXO)  & J044048.3+292434 & J044147.9--015145 & J123727.5+655211 & J182117.2--131405 \\
\textit{Line (\AA)}  & & & & \\

        \hline
		\hline
\textit{Emission features} & & & &  \\
H$\alpha$\ 6562.8 & 179 $\pm$ 8 & 435 $\pm$ 2 & 282 $\pm$ 10 &  16 $\pm$ 1 \\
H$\beta$\ 4861.3  & 95 $\pm$ 3 & 239 $\pm$ 3 & 397 $\pm$ 7 &  $<20$ \\
H$\gamma$\ 4340.5 & 62 $\pm$ 3 & 139 $\pm$ 4 & 354 $\pm$ 8 & $< 10$ \\
He I\ 5876.5   & 27 $\pm$ 6 & 64 $\pm$ 2 & 107 $\pm$ 9 &  $<12$ \\
He I\ 7065.2  & 13 $\pm$ 3 & 15 $\pm$ 2 & 35 $\pm$ 9 &  8 $\pm$ 2 \\
He II\ 4685.7  & 61 $\pm$ 4 & $< 12$ & 147 $\pm$ 7 & $< 8$ \\\hline

He II\ / H$\beta$  & $0.64\pm0.05$ & $< 0.05$ & $0.37\pm0.02$ &  -- \\
\hline
\hline
	\end{tabular}
\end{table*}

\subsection{X-ray spectra and luminosities}
\label{sec:Xspectra}

 \begin{table}
 \tiny

\renewcommand\arraystretch{1.3}
        \caption{List of CV candidates used to search for X-ray periodic signals.}
        \label{tab:xtiming}
        \centering
        \begin{tabular}{lrccc}
        \hline
        Object ID &ObsID & Bin time & Period & PF\\
        (2CXO) &  &  (s) & (s) & (\%)\\
        (1) & (2) &  (3) & (4) & (5)\\
        \hline
        \hline
        J063805--801854   & 14925 & 1,000 & $\rm 13,221 \pm 140$ & 61 \\
        J123727+655211    & 20443 & 1,000 & -- & --\\
        J173332.5--181735 & 12944 & 200 & -- & -- \\
        \hline
        \hline
        \end{tabular}
\flushleft
\tablefoot{ (1) Source name from CSC2; (2) {\it Chandra} observation identification number; (3) Bin time for X-ray light-curve; (4) Period and its error found in the {\it Chandra} data; (5) Pulse fraction. For more details, see Section \ref{sec:Xvariability}.}
\end{table}

\begin{figure}[t]
\centering
\begin{tabular}{cc}
   \includegraphics[width=1\linewidth]{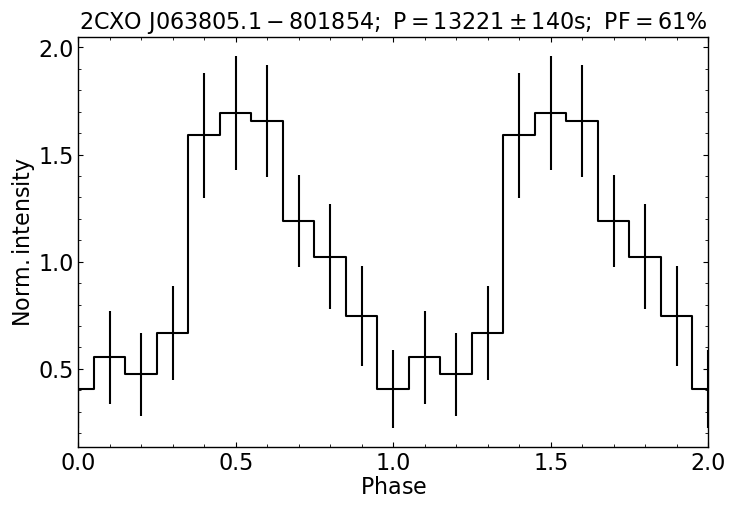}
 \end{tabular}
\caption{ {\it Chandra} X-ray pulse profile for 2CXO J063805--801854 normalized by mean count rate 0.01 $\rm cnt\ s^{-1}$.}
\label{fig:XLC}
\end{figure}

For each of the 14 sources in our sample, we queried the archival {\it Chandra} data using the observation identifier ({\tt obsid}) from source observation summary results in the CSCview application\footnote{\url{https://cxc.cfa.harvard.edu/csc2.0/gui/intro.html}}. For some objects, we additionally searched for new {\it Chandra} observations within a $\rm 10 \arcmin$ search radius using the Chaser\footnote{\url{https://cda.harvard.edu/chaser/}} webtool and included them in our analysis. The {\it Chandra} archival observations were processed by following standard {\it Chandra} Interactive Analysis of Observations (CIAO) \citep{2006SPIE.6270E..1VF}  threads\footnote{\url{http://cxc.harvard.edu/ciao/index.html}} (CIAO version 4.16.0 and CALDB version 4.11.0). The {\it chandra\_repro} tool was used to perform the initial calibration, reprocess the data, and create a level 2 event file. To extract the source and background spectra, we used the {\it specextract} tool in CIAO. A point-source aperture correction was applied for each source spectra by setting the parameter {\it correctpsf=yes}. We computed the radii of the {\it Chandra} point spread function (PSF) from the PSF map, which was constructed with {\it mkpsfmap} tool for 90\% of encircled counts fraction (ECF) and the 2.3 keV energy (parameters {\it ecf=0.9} and {\it energy=2.3} respectively). Source and background regions were centered at source positions from CSC2. The circle with the radius equal to the {\it Chandra} PSF radius was used for the source regions, and annuli with the inner and outer radii equal to two and four times the {\it Chandra} PSF radius were used for background regions. 

We performed the spectral fit using  the \texttt{XSPEC} v12.13 package \citep{1996ASPC..101...17A} in the 0.5--7 keV energy band. We binned the X-ray spectra to have at least three counts per bin and used C-statistics \citep{1979ApJ...228..939C} to get best-fit results. We used the {\tt error} command in XSPEC to compute $1\sigma$ confidence intervals for the parameters. For some sources in our sample, we estimated only the $1\sigma$ lower limit (or upper limit) of the parameters. Each spectrum was approximated using two spectral models: the power-law model (\texttt{powerlaw} in \texttt{XSPEC}) and the optically thin thermal plasma emission model (\texttt{mekal} in \texttt{XSPEC}). To take into account interstellar absorption, we used the Tuebingen-Boulder ISM absorption model (\texttt{tbabs} in \texttt{XSPEC}) and the solar elemental abundances from \cite{2000ApJ...542..914W}.

All sources in our sample are located at different distances, leading to various Galactic hydrogen column densities. We computed the distances of the sources from {\it Gaia} parallaxes, and their errors were calculated using the standard error propagation. Using the source distances, we adopted the $\rm E(B-V)$ values from the Bayestar dust map \citep{2019ApJ...887...93G}. If data from the Bayestar dust map were not available, we utilized the Galactic plane reddening map \citep{2019MNRAS.483.4277C}. We computed the hydrogen column density by assuming Cardelli's extinction law ($\rm R_{V} \approx 3.1$) and the relation $\rm N_{H (map)}\approx R_{V} \times E(B-V) \times 2.21\times10^{21}\ cm^{-2}$ \citep{1989ApJ...345..245C,2009MNRAS.400.2050G}. For sources having more than 100 counts in their X-ray spectra, we computed hydrogen column density ($N_{H(fit)}$) and metal abundance directly from the fit. We found no significant deviation between the best-fit results of $N_{H(fit)}$ and $N_{H(map)}$ computed from the Bayestar dust map. For other sources, we approximated their spectra with a fixed $N_{H(map)}$ and assuming solar metallicity. In the cases when $N_{H(map)}$ was not available at the specified distance of the source or when the distance fell outside the interval classified as reliable, we fixed the hydrogen column density to the Galactic value \citep{2016A&A...594A.116H}.

The best-fit results obtained from approximating the X-ray spectra with \texttt{powerlaw} and \texttt{mekal} models are presented in Appendix \ref{app:Xspectra} (see Table \ref{tab:xray_spectra} and Figure \ref{fig:Xspectra}). For sources with more than one {\it Chandra} observation, we analyzed the X-ray spectra for each observation separately. However, we found no significant parameter variations between these separate observations subsequently performed a joint fit of the spectra in XSPEC by combining the parameters across multiple observations. We computed absorption corrected fluxes in the 0.5--7 keV energy band for the power-law model using the {\it cflux} command and converted them to luminosities. The approximate errors of the unabsorbed X-ray luminosities were computed from the uncertainties in the distances and fluxes using standard error propagation. Table \ref{tab:new_cvs} (columns 10 and 11) shows the distances and unabsorbed X-ray luminosities in the 0.5--7 keV energy band.

\subsection{X-ray variability and timing analysis}
\label{sec:Xvariability}

We searched our sample of 14 CV candidates for X-ray variable sources using the CSC2 variability indices in the 0.5--7 keV energy band: $\tt var\_intra\_index\_b \ge 8$ (variability within a single observation) and/or $\tt var\_inter\_index\_b \ge 8$ (variability across multiple observations). We found only one source matching these criteria (2CXO J063805.1--801854). We also checked the significance of the difference between the maximum and minimum fluxes between observations using the ratio $\tt R=(F_{max}-F_{min})/(\sigma_{F_{max}}^2 + \sigma_{F_{min}}^2)^{1/2}$ and found only one source having $R \ge 3$ (2CXO J123727.5+655211).

Out of the 14 CV candidates in our sample, we performed an X-ray timing analysis for three sources with more than 300 source counts in their spectra (see Table \ref{tab:xtiming}). If sources had more than one archival {\it Chandra} observation, we used the observation with the longest exposure time for the timing analysis. We extracted the background-subtracted X-ray light curves with the {\it dmextract} tool in CIAO, using the same source and background regions as those used for the X-ray spectral analysis. Barycentric corrections were applied for the event files by using {\it axbary} tool in CIAO. We used the XRONOS sub-package in FTOOLS\footnote{\url{http://heasarc.gsfc.nasa.gov/ftools}} for timing analysis \citep{2014ascl.soft08004N}. First, we calculated the power spectrum density using a Fast Fourier Transform ({\it powspec} command in the XRONOS package) to search for periodic signal. The epoch-folding (EF) technique was then applied to constrain the period of the sources \citep{1996A&AS..117..197L} using the {\it efsearch} command in the XRONOS package. We estimated the best-fit period from the EF method by fitting a Gaussian function to the $\chi^2$ vs period diagram. The period error was estimated using the bootstrap technique similarly as described in \citet{2013AstL...39..375B}. To determine the significance of the detected period, we generated 1000 random Poisson, non-periodic light curves with mean source count rates and applied the {\it powspec} and {\it efsearch} tools on these randomly generated light curves. We computed the maximum value of the power spectrum density and the $\chi^2$ (as defined in the EF method) for each generated light curve. A $\rm 3\sigma$ upper confidence level was calculated for the distributions of power spectrum densities and $\chi^2$ values. A strong signal was considered present in the observed X-ray light curves if the results from the {\it powspec} and {\it efsearch} tools exceeded the $\rm 3\sigma$ confidence level determined from the simulations. We then used the {\it efold} tool to construct X-ray light-curves folded with the best-fit periods.

We found strong periodic X-ray signals only in one source out of three (see Table \ref{tab:xtiming}). Figure \ref{fig:XLC} shows the X-ray pulse profile normalized by the mean count rate of that source. The pulse fraction (PF) was computed using maximum and minimum count rates, where $\tt PF=(F_{max}-F_{min})/(F_{max}+F_{min})$.  We discuss the nature of the X-ray periodic source in Section \ref{sec:othersrc}.

\subsection{ Hertzsprung–Russell and X-ray luminosity -- orbital period diagrams}

\begin{figure}
    \centering
    \includegraphics[width=1.0\columnwidth]{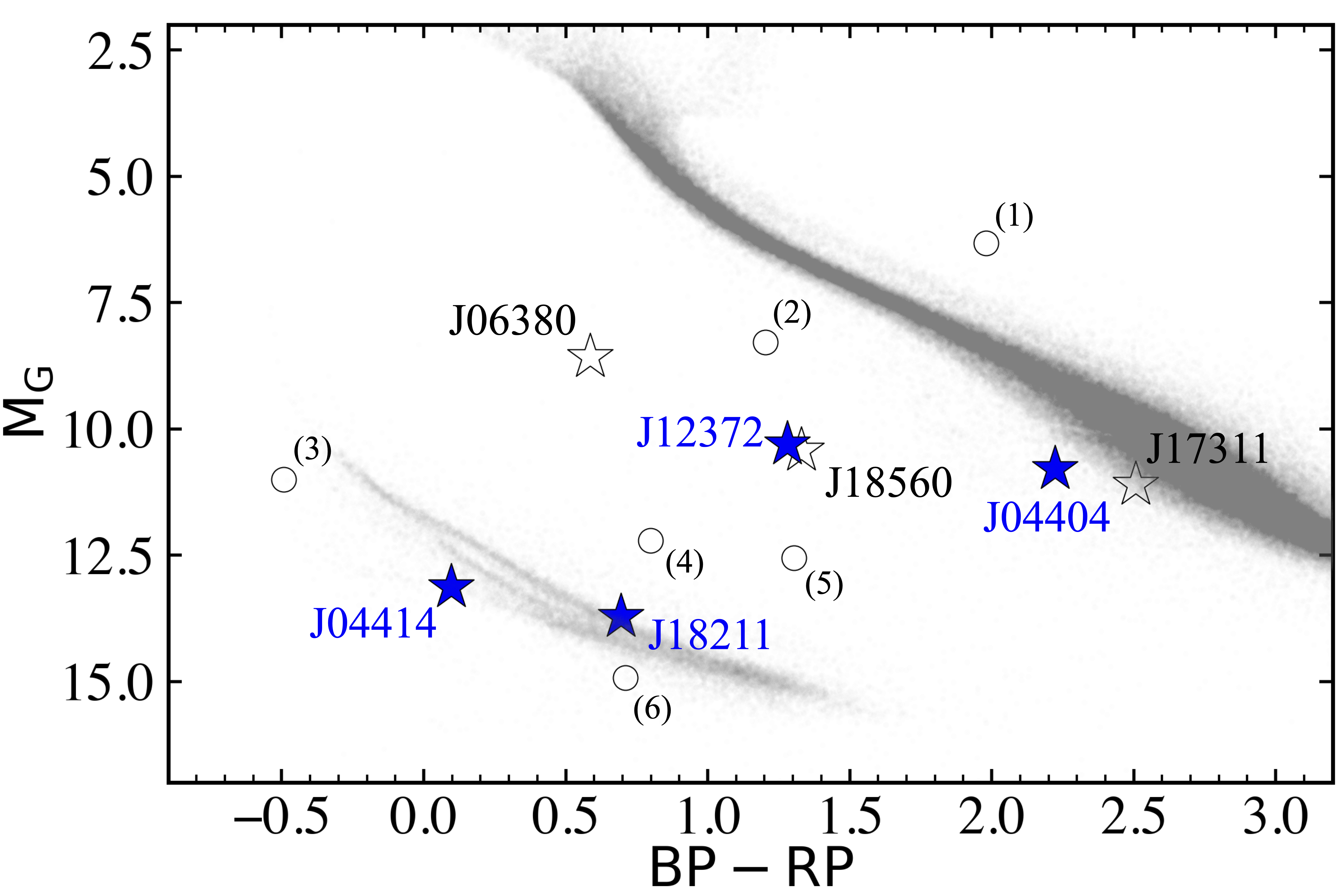}
    \caption{ Extinction-corrected Hertzsprung–Russell (HR) diagram. All {\it Gaia} sources within 100 pc with significantly measured parallaxes ($\texttt{parallax\_over\_error} > 3$) are shown in {\it grey} \citep{2023A&A...674A...1G}. {\it Star symbols} represent new, spectroscopically confirmed CVs (blue) and variable CV candidates (white) from Table \ref{tab:new_cvs}. {\it White circles} indicate CV candidates selected by a $\tt F_{X}/F_{opt}$  ratio and {\it Gaia} (BP-RP) color: (1) J165219.0--441401; (2) J173332.5--181735; (3) J094607.7--311550; (4) J155926.5--754311; (5) J190823.2+343328; (6) J173555.3--292530.}
    \label{fig:HR}
\end{figure}

\begin{figure*}
    \centering
    \includegraphics[width=0.8\textwidth]{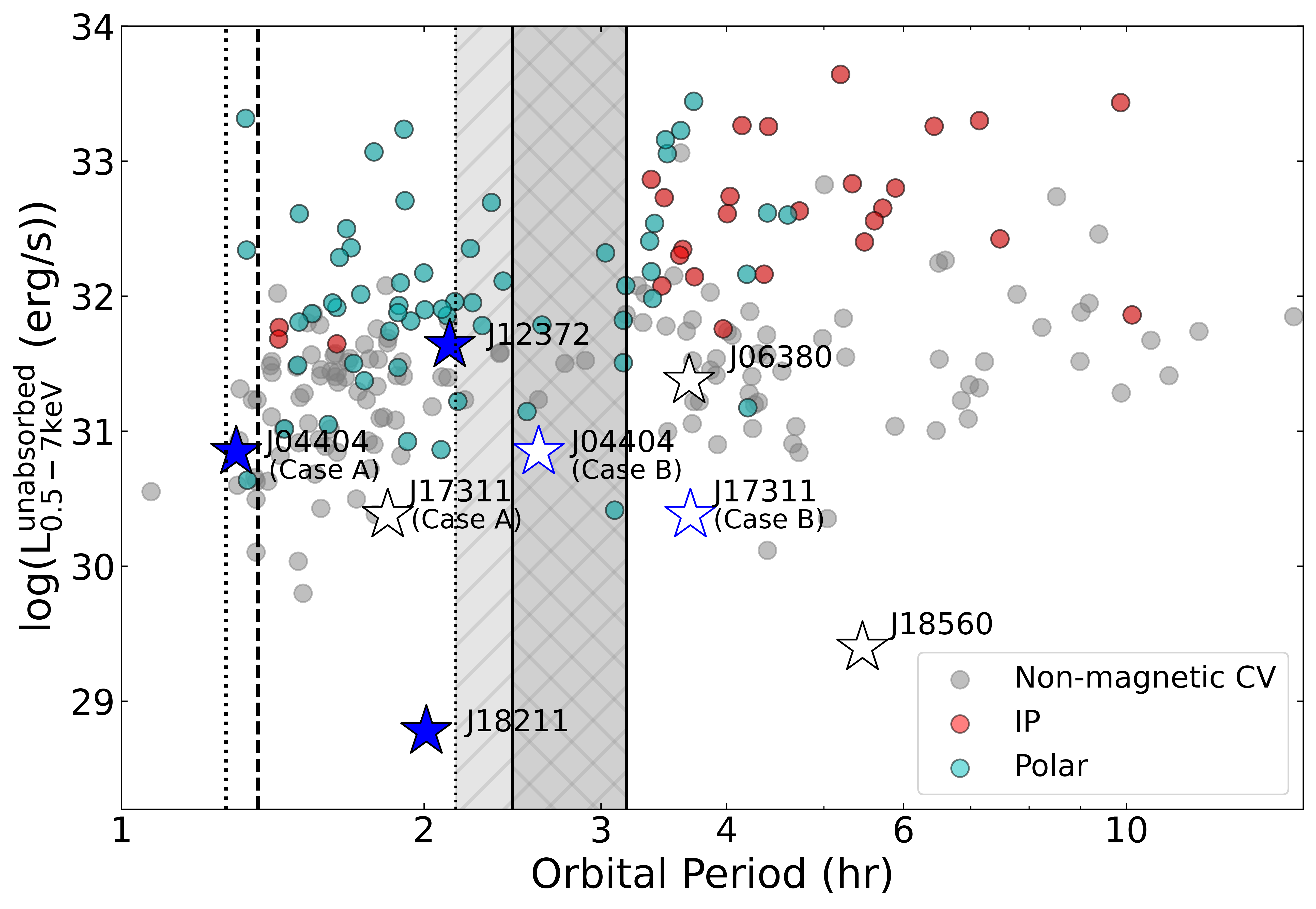}
    \caption{ Absorption-corrected X-ray luminosity (0.5--7 keV) -- orbital period diagram for known CVs from the \cite{2003ritterkolb} catalog. X-ray luminosities were computed from fluxes in the ROSAT source catalog (2RXS) and distances in {\it Gaia} DR3. {\it Circles} represent known IP (red), polars (green), and non-magnetic CVs (grey). {\it Vertical lines} indicate the period minimum for CVs from:  \cite{2009MNRAS.397.2170G} ($\rm \approx82$ min, dashed line); and \cite{2006MNRAS.373..484K} ($\rm \approx76$ min, dotted line). {\it Grey shaded regions} show the CV orbital period gap from: \cite{2024A&A...682L...7S} ($\rm \approx 2.5-3.2$ hr); and \cite{2006MNRAS.373..484K} ($\rm \approx 2.2-3.2 $ hr). {\it Star symbols} denote new, spectroscopically confirmed CVs (blue) and their candidates (white) from Table \ref{tab:new_cvs}. We present two cases (Case A and Case B) for 2CXO J044048.3+292434 and 2CXO J173118.6+224248, as we cannot distinguish between the two orbital periods with our current data. For more details, see Section \ref{sec:results}.
    }
    \label{fig:period_lx}
\end{figure*}

To compute extinction-corrected absolute G band magnitudes for CV candidates, we directly adopted the interstellar extinction $\rm A_{G}$  (at 6000 $\AA$) from Bayestar or the Galactic plane reddening maps when available. Otherwise, we converted the hydrogen column densities presented in Table  \ref{tab:xray_spectra}  for the {\tt mekal} model (see Section \ref{sec:Xspectra})  to interstellar extinction $\rm A_{G}$  using the \cite{1989ApJ...345..245C} extinction function. Figure \ref{fig:HR} shows the location of the CV candidates on the Hertzsprung–Russell (HR) diagram along with {\it Gaia} objects within 100 pc that have significantly measured parallaxes ($\texttt{parallax\_over\_error} > 3$). 

We constructed an X-ray luminosity -- orbital period diagram for six objects in our sample with periods computed from the optical and X-ray data (see Figure \ref{fig:period_lx}). To compare our objects with other known CVs, we cross-matched the CV catalog of {\it Ritter and Kolb} (Final Edition, frozen on 31 December 2015, \citet{2003ritterkolb}) with the {\it the second ROSAT all-sky survey (2RXS) catalogue} \citep{2016boller}, keeping only sources with a single match within $\rm 20\arcsec$. We then cross-matched with {\it Gaia} DR3 \citep{2023A&A...674A...1G}, keeping only sources with significant parallaxes and proper motions (5$\sigma$), which yielded 259 sources. Next, we used the WebPimms\footnote{\url{https://heasarc.gsfc.nasa.gov/cgi-bin/Tools/w3pimms/w3pimms.pl}} tool to convert ROSAT source count rates (0.1--2.4 keV) to unabsorbed X-ray luminosities in the 0.5--7 keV energy band, assuming a power-law model with a photon index of 1.7 and hydrogen column density of $\rm 3\times 10^{20}\ cm^{-2}$. To show different CV types in Figure \ref{fig:period_lx}, we used the classification scheme of the Ritter and Kolb catalog, which categorizes them as into polar (AM), intermediate polar (DQ), or otherwise as non-magnetic. In Figure \ref{fig:period_lx}, we show different period minimum values for CVs from \cite{2009MNRAS.397.2170G} ($\rm \approx82$ min) and \cite{2006MNRAS.373..484K} ($\rm \approx76$ min) as well as CV orbital period gap values from \cite{2024A&A...682L...7S} ($\rm \approx 2.5-3.2$ hr) and \cite{2006MNRAS.373..484K} ($\rm \approx 2.2-3.2 $ hr).

\section{Results and Discussion}
\label{sec:results}

Table \ref{tab:new_cvs} shows the parameters of the 14 CV candidates found in the CSC2--{\it Gaia} DR3 catalog using  the $\tt F_{X}/F_{opt}$ and the {\it Gaia} BP--RP color diagram and the period search with ZTF optical photometry (see Section \ref{sec:cv_selection}).  Figure \ref{fig:HR} shows that most of our CV candidates are located between the main sequence and the WD regions, which is consistent with the distribution of known CVs in the {\it Gaia} HR diagram \citep[e.g.,][]{2020MNRAS.492L..40A}. The X-ray luminosities of all our objects vary in the $\rm \approx 10^{29} - 10^{32}$ $\rm erg\ s^{-1}$ range, which matches with the range of X-ray luminosities of some known CVs (see Figure \ref{fig:period_lx}). X-ray spectra of our CV candidates can be approximated by a power-law model with photon indices in the $\rm \Gamma \approx 1-3$ range (or $\rm kT \sim 1-70$ keV for the optically thin thermal emission model). Such X-ray spectra are seen in some magnetic and non-magnetic CVs \citep[e.g.,][]{2021AstL...47..587G, 2022schwope, 2022schwope_b, 2023ApJ...954...63R, 2024MNRAS.528..676G}. 

Our variability search indicates that seven sources out of these 14 CV candidates are variable in nature. The X-ray luminosity -- orbital period diagram shows that most of our variable sources have X-ray luminosities and periods typically seen for magnetic and non-magnetic CVs (see Figure \ref{fig:period_lx}). Only two sources (2CXO J182117.2--131405 and  2CXO J185603.2+021259) are located in a distinct region of the phase space compared to the main sample of CVs, and exhibit low X-ray luminosity. These two objects might be a low-accretion rate  CV and an eclipsing binary\footnote{Such a binary would be an ``active binary", like RS CVn systems, which have orbital periods from as low as 5--6 hours to $\sim$10 days. In these systems, X-rays originate from the hot corona of one or both stellar components \citep[e.g.][]{1981rscvn}.}. We spectroscopically confirm four objects in our sample as new CVs. Their optical spectra show prominent emission lines or cyclotron humps, features typically observed in CVs (see Figures \ref{fig:j0440_ztfspec} - \ref{fig:J1821_ztfspec}). Below, we provide a detailed discussion of these CVs and other variable objects in our sample.

We note that some objects in our sample were previously suggested to be CV or quiescent low-mass X-ray binary (qLMXB) candidates using a multiwavelength machine-learning method, however, the results have low confidence \citep{2022ApJ...941..104Y}. Objects having {\it Gaia} optical light curves were also classified as possible CV candidates using a machine learning variability classifier in {\it Gaia} DR3 \citep{2023A&A...674A...1G}. In Appendix \ref{app:knownCVs}, we discuss three recently discovered CVs, which we independently selected using the methods described in this paper. These examples demonstrate that the $\tt F_{X}/F_{opt}$ and {\it Gaia} BP--RP color diagram, as suggested by \cite{2024PASP..136e4201R} can be a powerful method for searching and independently selecting accreting binary candidates. Without optical follow-up, it is challenging to distinguish between CVs and qLMXBs, as they display similar $\tt F_{X}/F_{opt}$ ratios and have overlapping X-ray luminosity ranges. Further spectroscopic follow-up to place constraints on the mass of the accreting object is required for the rest of the objects in our sample to clarify their nature.

\subsection{Spectroscopically Confirmed CVs}
\label{sec:individual}

Here, we discuss the nature of four objects from our final sample of CV candidates, for which we obtained optical spectra using Keck I/LRIS or Palomar/DBSP.

\subsubsection{2CXO J044048.3+292434, a new polar}
\label{sec:J0440}

We present the optical spectrum of J044048.3+292434 (hereafter J04404) in the right panel of Figure \ref{fig:j0440_ztfspec}. It shows Hydrogen Balmer and Helium emission lines along with a high-excitation He II 4686$\AA$ emission line, yielding a high line ratio  of $\rm He II\ / H\beta\approx 0.64$, which is typically seen in polars (see Table \ref{tab:EW_lines}).

J04404 shows a flux ratio of about $\rm F_{X}/F_{opt}\approx 0.6$, and an X-ray luminosity $\rm L_X \approx 7.1\times 10^{30}$ $\rm erg\ s^{-1}$ (see Table \ref{tab:new_cvs}). The X-ray spectrum can be approximated by a power-law model with an unusually shallow photon index $\rm \Gamma \approx 0.5$ (or $\rm kT \ga 55.5$ keV for an optically thin thermal emission model) (see Table \ref{tab:xray_spectra} and Figure \ref{fig:Xspectra}), a characteristic observed in some magnetic CVs \citep[e.g.,][]{2021AstL...47..587G}. J04404 is located close to the period minimum of CVs in the X-ray luminosity--period diagram (see Figure \ref{fig:period_lx}). Its optical and X-ray properties indicate that  J04404 is a polar near the CV period minimum.

In the left panels of Figure \ref{fig:j0440_ztfspec}, we present the long-term ZTF light curve of  J04404 and the phase-folded light curve at the best period (1.30 h = 78 min). It is evident that J04404 underwent a state change, commonly observed in polars, around MJD = 59000, where the average optical $r$ filter magnitude increased from $r\sim21$ to $r\sim19$ (see top left panel of Figure \ref{fig:j0440_ztfspec}). 
 
In some cases, polars can show two peaks in their optical light curve within a single orbital period due to cyclotron beaming\footnote{A two-component cyclotron model has been used to model the optical and near-infrared light curves of polars, such as in \cite{2008campbell_cyclotron}. A two-component cyclotron model leads to two peaks in a single orbital period, and can be due to any of the following: two magnetic poles of different field strength, optical depth, and/or viewing angles.}. The phase-folded optical light curve of J04404 shows $\sim 2$ mag peak-to-peak modulation, which is characteristic of cyclotron beaming following a state change in polars (see the bottom left panel of Figure \ref{fig:j0440_ztfspec}). In order to determine if the orbital period of J04404 is 1.30 hr or 2.60 hr, we utilized the fact that the Hydrogen Balmer and He II emission lines in polars typically vary by $\approx$1000 km/s within a single orbital period \citep[e.g.,][]{2023rodriguez_polars}. In the case of J04404, two spectra taken 1.4 hr apart do not show a significant radial velocity shift. In the bottom right panel of Figure \ref{fig:j0440_ztfspec}, we show both spectra of J04404, taken 1.4 hr apart. We also show a $\approx$1000 km/s shift, which would be expected if 1.4 hr were approximately half of the orbital period. However, we note that if the system were viewed nearly perpendicular to the accretion stream, such a radial velocity shift would not be observed. Time-resolved spectroscopy for 2.6 hr is necessary to confirm the orbital period of this system, leading us to currently label the orbital period as either 1.3 or 2.6 hr.

\subsubsection{2CXO J044147.9-015145, a low accretion rate non-magnetic CV}
\label{sec:J04414}

In Figure \ref{fig:J0441_ztfspec} (left panel), we present the long-term ZTF optical light curve of 2CXO J044147.9-015145 (hereafter J04414). A preliminary Lomb-Scargle period analysis reveals no periodicity in the ZTF  data, indicating that further data are required to determine the possible orbital period of this system. The optical spectrum of J04414 is shown in Figure  \ref{fig:J0441_ztfspec} (right panel). Hydrogen and Helium emission lines are present, with no detectable He II 4686\AA\ emission line, giving a $\rm 3\sigma$ upper limit for the line ratio $\rm He II\ / H\beta < 0.05$ (see Table \ref{tab:EW_lines}). J04414 shows a flux ratio of $\rm F_{X}/F_{opt}\approx 0.3$ and an X-ray luminosity $\rm L_X \approx 3.1\times 10^{29}$  $\rm erg\ s^{-1}$. The X-ray spectrum can be approximated by a power-law model with a photon index $\rm \Gamma \approx 1.8$ (or $\rm kT \approx 4.5$ keV for an optically thin thermal emission model) (see Table \ref{tab:xray_spectra} and Figure \ref{fig:Xspectra}). The optical and X-ray properties suggest that J04414 is a low accretion rate non-magnetic CV (WZ~Sge type). This classification is based on the apparent WD Balmer absorption lines and weakly double-peaked central emission lines seen in the center of its  optical spectrum. There is no sign of the donor signature in the optical spectrum. The {\it Gaia} HR diagram (see Figure \ref{fig:HR}) also supports this classification, as J04414 is located in the WD region, where WZ~Sge type systems are concentrated  \citep[e.g.,][]{2020MNRAS.492L..40A}.

\subsubsection{2CXO J123727.5+655211, a new polar}
\label{sec:J1237}

In Figure \ref{fig:j1237_ztfspec} (top panels), we present the long-term ZTF light curve and the phase-folded light curve at the best period for 2CXO J123727.5+655211 (hereafter J12372). The ZTF light curve reveals a significant period at 2.12 hr. The color dependence of cyclotron beaming is evident, with the ZTF $i$ filter showing the highest modulation (peak-to-peak $\sim 2$ mag) and the ZTF $g$ filter showing the lowest (peak-to-peak $\sim 1$ mag) (see the top left panel of Figure \ref{fig:j1237_ztfspec}). 
 
We present the optical spectrum of J12372 in Figure \ref{fig:j1237_ztfspec} (bottom panel). The spectrum shows Hydrogen, Helium, and high excitation He II 4686$\AA$ emission lines, yielding a line ratio of $\rm He II\ / H\beta\approx 0.37$ (see Table \ref{tab:EW_lines}).

J12372 shows a flux ratio $\rm F_{X}/F_{opt}\approx 1.1$ and an X-ray luminosity $\rm L_X \approx 4.4\times 10^{31}$ $\rm erg\ s^{-1}$ (see Table \ref{tab:new_cvs}). The X-ray spectrum can be approximated by a power-law model with a photon index $\rm \Gamma \approx 1.4$ (or $\rm kT \approx  38.3$ keV for an optically thin thermal emission model) (see Table \ref{tab:xray_spectra} and Figure \ref{fig:Xspectra}). X-ray timing analysis revealed no significant period in the longest exposure {\it Chandra} observation (see Table \ref{tab:xtiming}). J12372 shows significant X-ray flux variability ($\tt R\approx 6$, see Section \ref{sec:Xvariability}), with a factor of $\approx$ 2 between the {\it Chandra} observations, but no statistically significant variability between the spectral parameters was found. J12372 lies close to the CV period gap in the X-ray luminosity--period diagram (see Figure \ref{fig:period_lx}). Together with the optical light curves and spectrum, this indicates that J12372 is a polar.

\subsubsection{2CXO J182117.2-131405, low accretion rate polar}
\label{sec:J1821}

Figure \ref{fig:J1821_ztfspec} (top panels) shows the optical long-term ZTF light curve of 2CXO J182117.2--131405 (hereafter J18211) and the phase folded light curve. We determined  its best period to be 2.01 hr using the Lomb-Scargle period analysis. The phase-folded optical light curve shows cyclotron beaming  near phase $0.15$ and $\approx 1$ mag peak-to-peak modulation (see top right panel of Figure \ref{fig:J1821_ztfspec}). This modulation could be either due to cyclotron beaming or an eclipse of the WD by the donor star. High-speed photometry is needed to further investigate these scenarios.

J18211 has a low number of X-ray counts, and its spectrum can be approximately fitted by a power-law model, resulting in a large uncertainty for a photon index  $\rm \Gamma \sim 1.7$ (or $\rm kT \ga 1.6$ keV for an optically thin thermal emission model) (see Table \ref{tab:xray_spectra} and Figure \ref{fig:Xspectra}). J18211 shows a flux ratio $\rm F_{X}/F_{opt}\approx 0.15$ and a lower X-ray luminosity ($\rm L_X \approx 6\times 10^{28}$ $\rm erg\ s^{-1}$) than typically seen in CVs (see Table \ref{tab:new_cvs}). J18211 is located near the edge of the CV period gap in the X-ray luminosity--period diagram (see Figure \ref{fig:period_lx}). In this part of the diagram, the canonical picture of CV evolution predicts that detached systems begin to undergo accretion again. Hence, J18211 could be a low accretion rate polar (previously referred to as a pre-polar).

We present the optical spectrum of J18211 in Figure \ref{fig:J1821_ztfspec} (bottom panel). The spectrum is dominated by broad humps corresponding to cyclotron harmonics with no significant $H\beta$ and He II 4686$\AA$ emission lines detected. Using the equation for the electron gyration frequency around a magnetic field, we can define the wavelength of cyclotron harmonics as the following:

\begin{gather}
    \lambda_n = \frac{10710}{n}\left(\frac{100 \textrm{ MG}}{B}\right)\sin\theta\; \textup{\AA}
    \label{eq:cyclotron}
\end{gather}

With the spacing of two harmonics, it is possible to constrain the magnetic field strength, as the spacing must always be the ratio of integers. We must assume a cyclotron beaming angle, which should be close to 90$^\circ$ since the amplitude of the cyclotron harmonics is so large (An angle $\theta > 60^\circ$ is a good approximation for cyclotron beaming factors greater than 2 \citep[e.g.][]{2008campbell_cyclotron}). Detailed cyclotron modeling is beyond the scope of this paper. In this work, we fixed the magnetic field at different values and approximately matched the location of the cyclotron humps in the spectrum to different cyclotron harmonics (see Figure \ref{fig:J1821_ztfspec}). This combination of the cyclotron wavelengths observed in the spectrum of J18211 and the large beaming amplitude constrains the magnetic field to be $B\approx54-63$ MG. This magnetic field strength is typical of low accretion rate polars \citep[e.g.,][]{2021MNRAS.502.4305P}.

\subsection{Variable CV candidates}
\label{sec:othersrc}

Here, we discuss four CV candidates classified based on their X-ray and optical variability, but for which optical spectra are not yet available.

\subsubsection{ 2CXO J063805.1--801854: A magnetic CV?}

2CXO J063805.1--801854 (hereafter J06380) shows $\rm F_{X}/F_{opt}\approx 0.8$ and X-ray luminosity $\rm L_X \approx 2.4\times 10^{31}$ $\rm erg\ s^{-1}$ (see Table \ref{tab:new_cvs}). Its spectrum can be approximated by a power-law model with an unusually shallow photon index $\rm \Gamma \approx 0.2$ (or $\rm kT \ga 75$ keV for an optically thin thermal emission model) (see Table \ref{tab:xray_spectra} and Figure \ref{fig:Xspectra}). Such a shallow photon index is observed in the X-ray spectra of some magnetic CVs, such as polars or IPs \citep[e.g.,][]{2021AstL...47..587G}. X-ray timing analysis revealed a $\approx 13,200$ sec periodicity ($\rm PF=61\%$), which is consistent with the period previously found by \cite{2016MNRAS.462.4371I} in the same {\it Chandra} data (see Table \ref{tab:xtiming} and Figure \ref{fig:XLC}). There is no available optical ZTF photometry for J06380. The {\it Gaia} optical light curve shows an outbursting-type variability with $\rm \approx 3$ mag outburst lasting around $\rm \approx 100$ days and a possible recurrence time of approximately $\rm \approx 500$ days (see Figure \ref{fig:gaiaLC}). Figure \ref{fig:period_lx} shows that J06380 lies within main sample of known CVs. The X-ray periodicity (in the range of CV orbital periods), along with a high X-ray luminosity and hard spectrum, suggests that J06380 could be a magnetic CV, likely an IP.

\subsubsection{  2CXO J165219.0--441401: A possible dwarf nova?}

2CXO J165219.0--441401 (hereafter J16521) shows $\rm F_{X}/F_{opt}\approx 1.2$ and a high X-ray luminosity $\rm L_X \approx 3\times 10^{32}$ $\rm erg\ s^{-1}$ (see Table \ref{tab:new_cvs}). The X-ray spectrum can be approximated by a power-law model with a photon index $\rm \Gamma \approx 1.5$ (or $\rm kT \approx 5.8$ keV for an optically thin thermal emission model) (see Table \ref{tab:xray_spectra} and Figure \ref{fig:Xspectra}). There is no ZTF photometry for J16521 and the {\it Gaia} optical light curve shows a $\rm \approx 3$ mag outburst with a possible $\rm \approx 500$ days recurrent time (see Figure \ref{fig:gaiaLC}). This suggests that J16521 could be a dwarf nova.

\subsubsection{2CXO J173118.6+224248: A possible non-magnetic CV?}

2CXO J173118.6+224248 (hereafter J17311) shows $\rm F_{X}/F_{opt}\approx 0.3$ and an X-ray luminosity $\rm L_X \approx 2.4\times 10^{30}$ $\rm erg\ s^{-1}$ (see Table \ref{tab:new_cvs}). The optical ZTF light curves show $\approx$0.5 mag variability with a period of about $\approx$1.8 hr (see Figure \ref{fig:ZTFLC}). However, the actual period may be twice this value, possibly due to ellipsoidal modulation. The X-ray spectrum can be approximated by a power-law model with a photon index $\rm \Gamma \approx 2.2$ (or $\rm kT \approx 3.0$ keV for an optically thin thermal emission model) (see Table \ref{tab:xray_spectra} and Figure \ref{fig:Xspectra}), a characteristic observed in non-magnetic CVs. J17311 is located close to the main sequence of known CVs in the X-ray luminosity--orbital period diagram (see Figure \ref{fig:period_lx}). The X-ray properties and optical variability suggest that J17311 is a non-magnetic CV.

\subsubsection{2CXO J185603.2+021259: A possible eclipsing binary?}

2CXO J185603.2+021259 (hereafter J18560) shows $\rm F_{X}/F_{opt}\approx 0.05$ and a lower X-ray luminosity ($\rm L_X \approx 2.5\times 10^{29}$ $\rm erg\ s^{-1}$), than observed in CVs (see Table \ref{tab:new_cvs}). The ZTF optical light curve shows deep eclipses of approximately $\rm \approx 1$ mag in the $r$ filter (or approximately $\rm \approx 1.5$ mag in the $g$ filter) with a period of about 5.5 hr (see Figure \ref{fig:ZTFLC}).  Out of the eclipse sinusoidal variability with $\rm \approx 0.1$ mag is seen, possibly caused by the contribution from the donor. The X-ray spectrum can be approximated by a power-law model having a photon index $\rm \Gamma \approx 2.8$ (or $\rm kT \approx 1$ keV for an optically thin thermal plasma model) (see Table \ref{tab:xray_spectra} and Figure \ref{fig:Xspectra}). Figure \ref{fig:period_lx} shows that J18560 is located below the main CV sample, similar to J18211. These observed properties suggest that J18560 could be an eclipsing binary.

\section{Conclusions}
\label{sec:summary}

We searched for new Galactic CV candidates among 25,887 sources in the CSC2, cross-matched these with {\it Gaia} DR3, and studied their optical and X-ray properties. We summarize our findings below:

\begin{enumerate}

\item[--] We used a modern concept suggested by \cite{2024PASP..136e4201R} to classify new CVs among stellar objects based on a combination of $\tt F_{X}/F_{opt}$ and {\it Gaia} BP--RP color (so called the “X-ray Main Sequence") (see Figure \ref{fig:fxvsfopt}). We further defined a new threshold cut to compile a pure sample of CV candidates and minimize contamination from stellar objects (see Eq.\ref{eq:fx_fopt} and Section \ref{sec:cv_selection}).
\\

\item[--]  We excluded known Galactic stellar objects by cross-matching the CSC2 -- {\it Gaia} DR3 catalog with the Simbad database and other publicly available CV catalogs (see Section \ref{sec:gal_src}). Out of 13,956 unclassified Galactic objects, we identified 14 new CV candidates using two different cuts on the ``X-ray Main Sequence"  (see Table \ref{tab:new_cvs} and Section \ref{sec:cv_selection}).
\\

\item[--] All objects show X-ray luminosities in the range $L_X$ $\rm \approx 10^{29}-10^{32}$ $\rm erg\ s^{-1}$. Their X-ray spectra can be approximated by a power-law model with photon indices in the $\rm \Gamma \sim 1-3$ range or by an optically thin thermal emission model in the $\rm kT \sim 1-70$ keV range (see Figure \ref{fig:Xspectra} and Table \ref{tab:xray_spectra}). These X-ray properties are mostly typical of CVs (both magnetic and non-magnetic), but some objects might also be qLMXBs.
\\

\item[--] We performed optical follow-up observations with Keck I/LRIS and P200/DBSP and spectroscopically confirmed four objects as new CVs. Their optical spectra show prominent emission lines or cyclotron humps, typically seen in CVs (see Tables \ref{tab:data} and \ref{tab:EW_lines}). We determined orbital periods for three of these objects using ZTF photometry. Among these, we discovered two polars, one low accretion rate polar and one non-magnetic CV (WZ~Sge type) (see Section \ref{sec:individual}). 
\\

\item[--] The X-ray and optical properties of four variable CV candidates suggest that  three are potential magnetic systems or dwarf novae, while one may be an eclipsing binary (see Section \ref{sec:othersrc}). Further spectroscopic follow-up is required to confirm the nature of these objects.
\\

\end{enumerate}

Our analysis shows that a multiwavelength approach is a powerful tool for discovering new CVs and studying their properties. We illustrated that the ``X-ray Main Sequence" presented in \cite{2024PASP..136e4201R} allows us to compile a sample of CV candidates for a detailed study. This approach can also be applied to other X-ray catalogs or upcoming all-sky SRG/eROSITA data to search for accreting binary systems. Further investigation of selection effects is necessary to use the ``X-ray Main Sequence" to compile volume-limited samples of CVs and study their population.

\begin{acknowledgements}
This research has made use of data obtained from the {\it Chandra} Data Archive and software provided by the {\it Chandra} X-ray Center (CXC) in the application packages CIAO. This research has made use of the SIMBAD database, operated at CDS, Strasbourg, France.

Based on observations obtained with the Samuel Oschin Telescope 48-inch and the 60-inch Telescope at the Palomar Observatory as part of the Zwicky Transient Facility project. ZTF is supported by the National Science Foundation under Grants No. AST-1440341 and AST-2034437 and a collaboration including current partners Caltech, IPAC, the Weizmann Institute of Science, the Oskar Klein Center at Stockholm University, the University of Maryland, Deutsches Elektronen-Synchrotron and Humboldt University, the TANGO Consortium of Taiwan, the University of Wisconsin at Milwaukee, Trinity College Dublin, Lawrence Livermore National Laboratories, IN2P3, University of Warwick, Ruhr University Bochum, Northwestern University and former partners the University of Washington, Los Alamos National Laboratories, and Lawrence Berkeley National Laboratories. Operations are conducted by COO, IPAC, and UW. The ZTF forced-photometry service was funded under the Heising-Simons Foundation grant No.12540303 (PI: Graham).

We are grateful to the staffs of the Palomar and Keck Observatories for their work in helping us carry out our observations. 

This work has made use of data from the European Space Agency (ESA) mission {\it Gaia} (\url{https://www.cosmos.esa.int/gaia}), processed by the {\it Gaia} Data Processing and Analysis Consortium (DPAC, \url{https://www.cosmos.esa.int/web/gaia/dpac/consortium}). Funding for the DPAC has been provided by national institutions, in particular the institutions participating in the {\it Gaia} Multilateral Agreement.

This research made use of matplotlib, a Python library for publication quality graphics \citep{Hunter2007}; NumPy \citep{Harris2020}; Astroquery \citep{2019AJ....157...98G}; Astropy, a community-developed core Python package for Astronomy \citep{2013A&A...558A..33A, 2018AJ....156..123A}; and the VizieR catalogue access tool, CDS, Strasbourg, France. 

IG, AS, VD, NT acknowledge support from Kazan Federal University. ACR acknowledges support from the National Science Foundation via an NSF Graduate Research Fellowship. The authors thank the anonymous referee for useful and inspiring suggestions which helped to improve the manuscript.

\end{acknowledgements}

%%%%%% APPENDIX

\begin{appendix} %First appendix

\section{General object types with a Simbad and $\rm F_{X}/F_{opt}-(BP-RP)$ selection cut }
\label{appendix:PandC}

\begin{table*}
\tiny
\renewcommand\arraystretch{1.25}
\centering
\caption{Object types based on Simbad classification for 25,887 sources in the CSC2--{\it Gaia} DR3 catalog.}
\label{tab:types}
\begin{tabular}{l| l l c| r}
\hline
Main object type & Sub-type & Simbad Class & $\rm N_{src}$ & Comments \\
(1) & (2) & (3) & (4) & \\
\hline
\hline
{\bf STARS} &  &  & \\
 & Star & 1. Taxonomy of Stars & 4,779 &  General sub-type from Simbad database \\
 & Massive stars & 1.1 Massive Stars and their Remnants & 42 \\
 & YSO &  {\parbox{4cm}{\vspace{5pt} 1.2 Young Stellar Objects (Pre-Main Sequence Stars) }} & 4,823 \\
 & Main-sequence stars & 1.3 Main Sequence Stars & 37 \\
 & Evolved & 1.4 Evolved Stars & 140 \\

 & Peculiar & 1.5 Chemically Peculiar Stars & 15 \\
 & Binaries & {\parbox{4cm}{\vspace{3pt} 1.6 Interacting Binaries and close Common Proper Motion Systems }} & 638 \\
 & Low-mass stars & 1.8 Low mass Stars and substellar Objects & 101 \\

 & Properties & {\parbox{4cm}{\vspace{5pt} 1.9 Properties, variability, spectral kinematic or environment }} & 476 \\
 & Spectral properties & 1.10 Spectral properties & 239 \\
  & Kinematic & 1.11  Kinematic and Environment Properties & 205 \\
 & Sets of stars & 2. Sets of Stars & 5\\
& WD & 1.4 Evolved Stars & 13 \\
& Brown dwarf (BD) & 1.8 Low mass Stars and substellar Objects & 4 \\
 \hline
{\bf ACCRETING BINARIES} &  &  &  \\
 & CV &  {\parbox{4cm}{\vspace{5pt} 1.6 Interacting Binaries and close Common Proper Motion Systems }} & 32 &  {\parbox{4cm}{\vspace{5pt} Sub-type includes CV, Nova and Symbiotic Stars }} \\
 & X-ray Binary &   {\parbox{4cm}{\vspace{5pt} 1.6 Interacting Binaries and close Common Proper Motion Systems  and 7. General Spectral Properties }} & 50 &{\parbox{4cm}{\vspace{5pt} Sub-type includes X-ray binary (HMXB, LMXB) and ULX  }} \\
 \hline
{\bf ISM} &  &  &  \\
 & ISM & 3. Interstellar Medium & 7 \\
 \hline
{\bf GALAXY} &  &  &  \\
 & Galaxy & 4. Taxonomy of Galaxies & 47  \\
 & Sets of Galaxies & 5. Sets of Galaxies  & 6 \\
\hline
{\bf PROPERTIES} &  &  &  \\
 & Properties & 7. General Spectral Properties & 520 \\
\hline
{\bf BLENDS} &  &  &  \\
 & Blends & 8. Blends, Errors, not well Defined Objects & 2 \\
\hline
 {\bf UNKNOWN}  &  & -- & 13,706 & Object is not presented in Simbad database \\
\hline
\hline
\end{tabular}
\end{table*}

Table \ref{tab:types} shows our general object types based on the Simbad hierarchical structure. The columns are: (1) our main object type; (2) object sub-type; (3) title of Simbad hierarchical structure used to define the main object type and its sub-type; (4) number of sources per each object sub-type, including known sources and their candidates. No source in our sample had an object type from the sixth Simbad's hierarchy ("Gravitation"); therefore, it's not shown in Table \ref{tab:types}. 

We investigated purity and completeness to construct a pure sample of new CV candidates and avoid contamination from stellar objects. Based on the Simbad hierarchical structure, we created two samples of known CVs (36 sources) and STARS (4781 sources) (see Section \ref{sec:first_cut}).

\cite{2022MNRAS.510.4126S} computed purity and completeness for the red and blue galaxies classified by optical color. We applied a similar approach to define the purity and completeness as:

\begin{itemize}
\setlength\itemsep{0.5em}

\item True positive (TP) = known CVs classified as CVs based on a given threshold (i.e., known CVs with an observed X-ray-to-optical ratio exceeding the threshold cut, $\tt (F_{X}/F_{opt})_{obs} \ge (F_{X}/F_{opt})_{cut}$).

\item False positive (FP) = known STARS classified as CVs based on a given threshold (i.e., known STARS with an observed X-ray-to-optical ratio exceeding the threshold cut, $\tt (F_{X}/F_{opt})_{obs} \ge (F_{X}/F_{opt})_{cut}$).

\item False negative (FN) = known CVs classified as STARS (i.e., known CVs with an observed X-ray-to-optical ratio below the threshold cut, $\tt (F_{X}/F_{opt})_{obs} < (F_{X}/F_{opt})_{cut}$).

\end{itemize}

Purity (P) is computed as the fraction of true identification to all identifications, 
\begin{gather}
  \tt P = TP / (TP + FP),
    \label{eq:purity}
\end{gather}
and the completeness (C) is calculated as the fraction of true detections to all that should have been classified, 
\begin{gather}
  \tt C = TP / (TP + FN).
    \label{eq:completeness}
\end{gather}
To find the optimal combination of linear parameters (A, B) for Eq.\ref{eq:fx_fopt}, we created a grid of parameters with the 0.1 steps in the range $\rm 0 \la {\tt A} \la 3$ and $\rm -8 \la {\tt B} \la 3$ and computed the purity and completeness. 

\begin{figure}
    \centering
    \includegraphics[width=0.99\columnwidth]{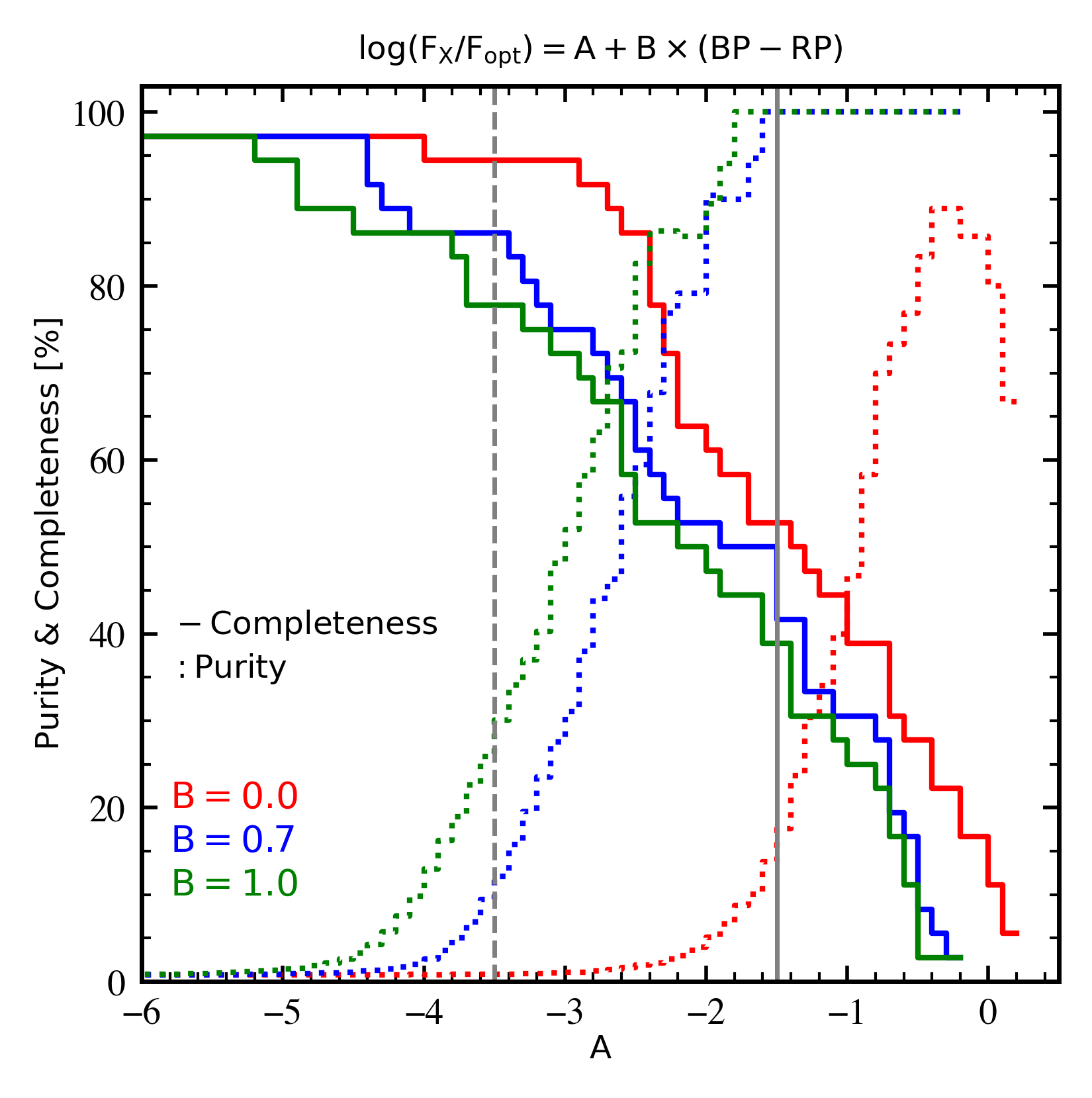}
    \caption{Purity (dotted line) and Completeness (solid line) computed for different $\tt (A, B)$ parameters from Eq. \ref{eq:fx_fopt}. {\it The vertical solid line} corresponds to  ${\tt A} = -1.5$ (and ${\tt B}=0.7$), which is used to select a pure sample of CV candidates in this paper. {\it The vertical dashed line} shows ${\tt A} = -3.5$ (and ${\tt B}=1$) for an empirical cut from \cite{2024PASP..136e4201R}. }
    \label{fig:purity_comp}
\end{figure}

Figure \ref{fig:purity_comp} shows the purity and completeness as a function of linear parameters $\tt (A, B)$. For illustrative purposes, we show the purity and completeness only for three $\tt B$ parameters. The $\tt B=0$ case corresponds to the canonical selection cut based only on the X-ray-to-optical flux ratio without additional information from {\it Gaia} colors. For example, the cut $\tt F_{X}/F_{opt} = 0.1$ (where ${\tt A} = -1$ and ${\tt B}=0$) gives $\approx$27\% purity and $\approx$48\% completeness. The drop of the purity above the $\tt A \ga -0.2$ is caused by a low number of sources and is not statistically significant. The \cite{2024PASP..136e4201R} empirical cut (${\tt A} = -3.5$ and ${\tt B}=1$) shows $\approx$25\% purity and $\approx$78\% completeness. In our study, we chose to use ${\tt A} = -1.5$ and ${\tt B}=0.7$, which gives $\approx$100 \% purity and $\approx$52\% completeness.

We note that our threshold is defined based only on a small sample of confirmed CVs (36 objects). To study the CV population and create a volume-limited sample, the $\tt F_{X}/F_{opt}$ and {\it Gaia} BP--RP color diagram should be further investigated for selection effects with a large sample of confirmed CVs and stellar objects. A detailed investigation is beyond the scope of this paper.

\section{Recently discovered CVs}
\label{app:knownCVs}

\begin{table*}
\tiny
\caption{ List of previously known CVs independently found in our search of the CSC2--{\it Gaia} DR3 catalog. }
\label{tab:app_new_cvs}
\renewcommand\arraystretch{1.5}
\setlength\tabcolsep{3.pt}
\centering
\begin{tabular}{lccccccccllcc}
\hline
Object ID & R.A. & DEC. & $\rm R_{95}$ & {\it Gaia} DR3 & G & BP--RP & $\tt F_{X}/F_{opt}$ 
& $\tt Var.$ 
& d 
& $\rm L_X$ 
& Obj. 
& Opt.  
\\

(2CXO) &  &  & ($\arcsec$) & source\_id & (mag) & (mag) & 
& 
& (pc) 
& ($10^{30}$ $\rm erg\ s^{-1}$) 
& Type 
& Spec.  
\\

(1) & (2) & (3) & (4) & (5) & (6) & (7) & (8)& (9)& (10)& (11)  
& (12) 
& (13)  
\\

\hline
\hline

J024131.0+593630 & 02 41 31.1 & +59 36 30.5 & 0.72 & 464373929923792384 & 19.84 & 1.47 & 1.71 
& Z/Ga 
& $541\pm121$ 
& $14.78\pm{6.70}$ 
& 1 
& \checkmark  
\\

J143435.3+334049 & 14 34 35.3 & +33 40 49.5 & 3.72 & 1287753107989718784 & 20.03 & 0.23 & 0.30 
& -- 
& $343\pm50$ 
& $0.48\pm{0.16}$ 
& 4 
& \checkmark  
\\

J174133.7--284034 & 17 41 33.8 & --28 40 34.5 & 0.73 & 4060179580171407360 & 18.75 & 0.99 & 0.97 
& -- 
& $784\pm161$ 
& $44.12\pm{18.23}$ 
& 3 
& 
\\

\hline
\hline

\end{tabular}
\flushleft
\tablefoot{ Same notes as in Table \ref{tab:new_cvs}. (9) "Z" --outbursting source based on ZTF photometry (see Section \ref{sec:second_cut}).} 
\end{table*}

\begin{figure*}
    \centering
    \includegraphics[width=1.0\textwidth]{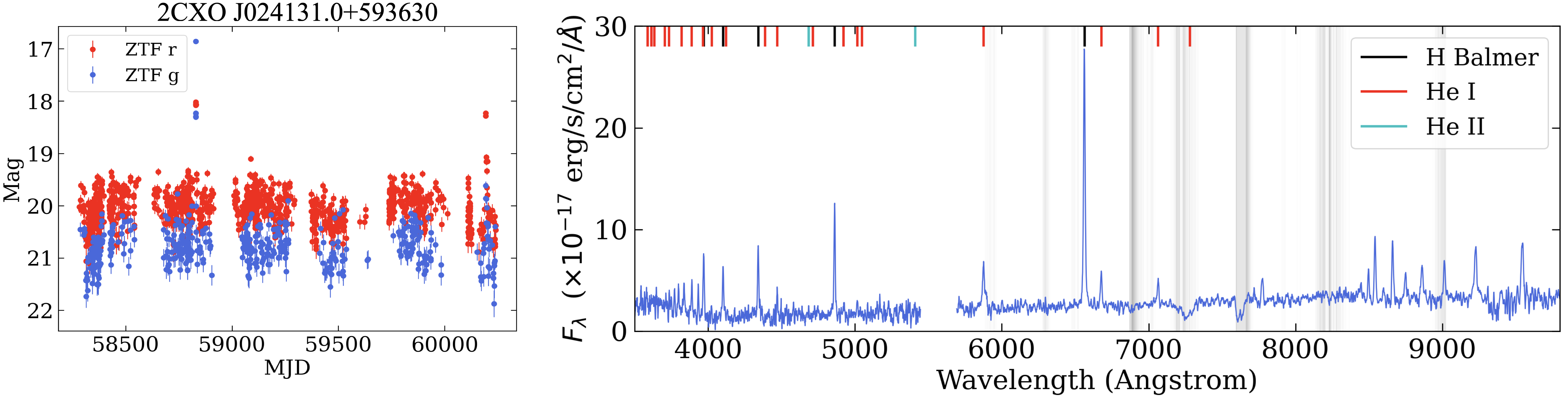}
    \caption{ZTF long-term light curves in $g$ and $r$ filters (left) and DBSP optical spectrum of 2CXO J024131.0+593630 (right). {\it Grey lines} represent the Keck Telluric Line List. The optical spectrum shows single peaked Hydrogen Balmer emission lines and a weak He II 4686$\AA$ emission line, giving a low line ratio $\rm He II\ / H\beta \approx0.12$  (see Table \ref{app:EW_lines}). Our analysis, based on optical and X-ray properties, suggests that J0241 is a new CV, likely non-magnetic (see Section \ref{sec:J0241} for more details).} 
    \label{fig:J0241_ztfspec}
\end{figure*}

\begin{figure*}
    \centering
    \includegraphics[width=1.0\textwidth]{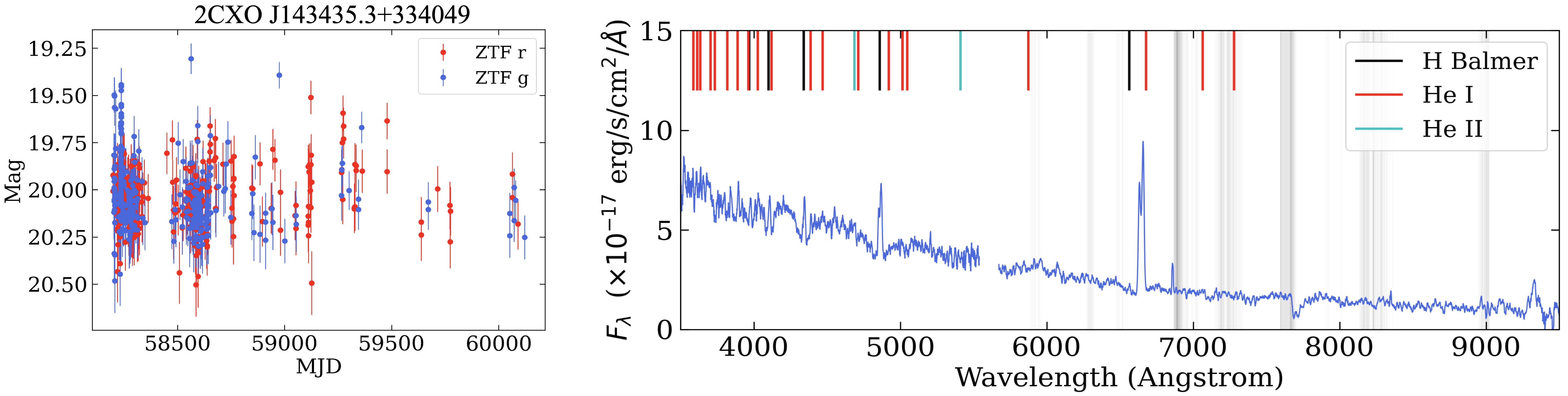}
    \caption{ ZTF long-term (left) in $g$ and $r$ filters and the DBSP optical spectrum of 2CXO J143435.3+334049 (right). {\it Grey lines} represent the Keck Telluric Line List. The optical spectrum shows WD Balmer absorption lines with no sign of the donor signature in the optical spectrum and double-peaked, asymmetric H$\alpha$ and H$\beta$ emission lines. There is no detection of the high-excitation He II 4686$\AA$ emission line providing an upper limit for a line ratio  $\rm He II\ / H\beta < 0.15$ (see Table \ref{app:EW_lines}). X-ray and optical properties suggest that the object is a low-accretion rate CV (WZ~Sge type) (see Section \ref{sec:J1434} for more details).}
\label{fig:J1434_ztfspec}
\end{figure*}

\begin{table}
\tiny

\renewcommand\arraystretch{1.25}
	\centering
	\caption{Equivalent widths of selected lines of two previously known CVs.}
    \label{app:EW_lines}
    \setlength\tabcolsep{3.2pt}
	\begin{tabular}{lcc} 
		\hline
    Object ID (2CXO) & J024131.0+593630 & J143435.3+334049 \\
\textit{Line (\AA)}  & & \\

        \hline
		\hline
\textit{Emission features} & & \\
H$\alpha$\ 6562.8 & 336 $\pm$ 3 & 222 $\pm$ 6 \\
H$\beta$\ 4861.3 & 114 $\pm$ 2 & 55 $\pm$ 4 \\
H$\gamma$\ 4340.5 & 86 $\pm$ 2 & $<12$ \\
He I\ 5876.5  & 91 $\pm$ 2 & $<10$  \\
He I\ 7065.2  & 26 $\pm$ 3 & 9 $\pm$ 2  \\
He II\ 4685.7  & 14 $\pm$ 4$^{*}$  & $<8$\\\hline

He II\ / H$\beta$  & $0.12\pm0.04$$^{*}$ & $< 0.15$  \\
\hline
\hline
	\end{tabular}
 \flushleft
\tablefoot{ ($*$) -- The measurement is statistically significant, but visual inspection suggests the line is not present. 3$\sigma$ upper limits are shown if no higher significant measurement is detected. }
\end{table}

We selected three objects from the CSC2--{\it Gaia} DR3 catalog using the X-ray main sequence as CV candidates (see Figure \ref{fig:fxvsfopt} and Table \ref{tab:app_new_cvs}).  In the course of this study, two of these objects (2CXO J024131.0+593630 and 2CXO J143435.3+334049) were recently discovered as CVs with SDSS spectra and ZTF light curves \citep{2023MNRAS.525.3597I}. Another object, 2CXO J174133.7–284034\footnote{We included this system in our analysis because it appears as an X-ray source ("X") in Simbad, not explicitly as a CV. }, was confirmed as a CV via optical spectroscopy by \cite{2017MNRAS.470.4512W}. However, we included this object in our analyses to search for possible X-ray variability through timing and spectral analyses. Here, we individually discuss the X-ray and optical properties of these objects.

\subsection{2CXO J024131.0+593630, a non-magnetic CV}
\label{sec:J0241}

In the left panel of Figure \ref{fig:J0241_ztfspec}, we present the long-term optical ZTF light curve of 2CXO J024131.0+593630 (hereafter J02413), which shows two significant ($\approx 2$ mag in $r$ filter and $\approx 3$ mag in $g$ filter) short outbursts. A preliminary Lomb-Scargle period analysis reveals no periodicity in the ZTF and {\it Gaia} data, indicating that further observations are needed to determine the possible orbital period of this system.

We observed J02413 with P200/DBSP on 6 December 2023 for 900s. The optical spectrum of J02413 is presented in the right panel of Figure \ref{fig:J0241_ztfspec}. Single-peaked Hydrogen Balmer emission lines are clearly visible, and a weak He II 4686$\AA$ emission line is also present, giving a low line ratio $\rm He II\ / H\beta \approx0.12$  (see Table \ref{app:EW_lines}). However, as noted in Table \ref{app:EW_lines}, the detection of He II 4686$\AA$ does not pass a visual inspection, so the value of the ratio should be interpreted as an upper limit. We also note that the detection (or non-detection) of He II 4686$\AA$ does not affect our final interpretation of this object. 

J02413 shows a high flux ratio $\rm F_{X}/F_{opt}\approx 1.7$ and an X-ray luminosity $\rm L_X \approx 1.5\times 10^{31}$ $\rm erg\ s^{-1}$ (see Table \ref{tab:app_new_cvs}). The X-ray spectrum of J02413 can be approximated by a power-law model with $\rm \Gamma \approx 1.8$ (or $\rm kT \approx 5.5$ keV for an optically thin thermal emission model) (see Table \ref{tab:xray_spectra} and Figure \ref{fig:Xspectra}). We found no significant period for J02413 from the X-ray timing analysis of the longest exposure {\it Chandra} observation (ObsID = 7152). J02413 also does not show X-ray flux variability and or changes in spectral parameters between observations. Our analysis, using optical and X-ray properties, suggests that J02413 is a CV, likely non-magnetic. This agrees with \citet{2023MNRAS.525.3597I} conclusion, where J02413 is identified as a new non-magnetic CV (U Gem type). The spectrum of this object is not shown in \cite{2023MNRAS.525.3597I}, making our spectrum a useful addition to the literature.

\subsection{2CXO J143435.3+334049, low-accretion rate CV}
\label{sec:J1434}

In Figure \ref{fig:J1434_ztfspec} (left panel), we present the long-term ZTF light curve of 2CXO J143435.3+334049 (hereafter J14343). A preliminary Lomb-Scargle period analysis reveals no periodicity in the ZTF  data, indicating that further observations are needed to determine the possible orbital period of this system.

We observed J14343 with P200/DBSP on 21 January 2023 for 900s. The optical spectrum of J14343 is presented in Figure \ref{fig:J1434_ztfspec} (right panel). The WD Balmer absorption lines are seen with double-peaked emission lines at the center, though the peaks appear to be asymmetric in height and width for both the H$\alpha$ and H$\beta$ lines. We do not detect the high-excitation He II 4686$\AA$ emission line providing an upper limit for a line ratio  $\rm He II\ / H\beta < 0.15$ (see Table \ref{app:EW_lines}). There is no sign of the donor signature in the optical spectrum. The spectrum of this object in Figure \ref{fig:J1434_ztfspec} is different from the one shown in \cite{2023MNRAS.525.3597I}, with the Balmer emission lines here being noticeably asymmetric. This could warrant future study of this system to further understand the nature of this source.

J14343 shows a flux ratio $\rm F_{X}/F_{opt}\approx 0.3$  and a lower X-ray luminosity ($\rm L_X \approx 4.8\times 10^{29}$ $\rm erg\ s^{-1}$) than typically seen in CVs (see Table \ref{tab:app_new_cvs}). We have not detected statistically significant flux and spectral parameter variability between {\it Chandra} observations. The X-ray spectrum can be approximated by a power-law model with a photon index $\rm \Gamma \approx 1.7$ (or $\rm kT \ga 5.2$ keV for optically thin thermal emission model) (see Table \ref{tab:xray_spectra} and Figure \ref{fig:Xspectra}). X-ray and optical properties suggest that J14343 is a low-accretion rate CV (WZ~Sge type), similar to what \citet{2023MNRAS.525.3597I} concluded in their analysis.

\subsection{2CXO J174133.7–284034, a non-magnetic CV}

2CXO J174133.7--284034 (hereafter J17413 shows a flux ratio $\rm F_{X}/F_{opt}\approx 1.0$ and an X-ray luminosity $\rm L_X \approx 4.4\times 10^{31}$ $\rm erg\ s^{-1}$ (see Table \ref{tab:app_new_cvs}). The X-ray spectrum of J17413 can be approximated by a power-law model with $\rm \Gamma \approx 1.8$ (or $\rm kT \approx 5.6$ keV for an optically thin thermal emission model) (see Table \ref{tab:xray_spectra} and Figure \ref{fig:Xspectra}). We have not detected statistically significant flux and spectral parameter variability between {\it Chandra} observations. We found no significant period for J17413 with the X-ray timing analysis for the longest exposure {\it Chandra} observation (ObsID = 9565). \citet{2014ApJS..214...10B} observed optical variability ($\approx 1$ mag) with no periodicity. \cite{2017MNRAS.470.4512W} observed prominent emission lines and confirmed via optical spectroscopy that J17413 is a CV. Our X-ray analysis supports  \cite{2014ApJS..214...10B} and \cite{2017MNRAS.470.4512W}  conclusions and also suggests that J17413 is likely a non-magnetic system.

\section{X-ray spectra figures}
\label{app:Xspectra}

\begin{figure*}[h!]
\centering
\setlength{\tabcolsep}{1pt}
    \centering
    \includegraphics[width=0.3\linewidth]{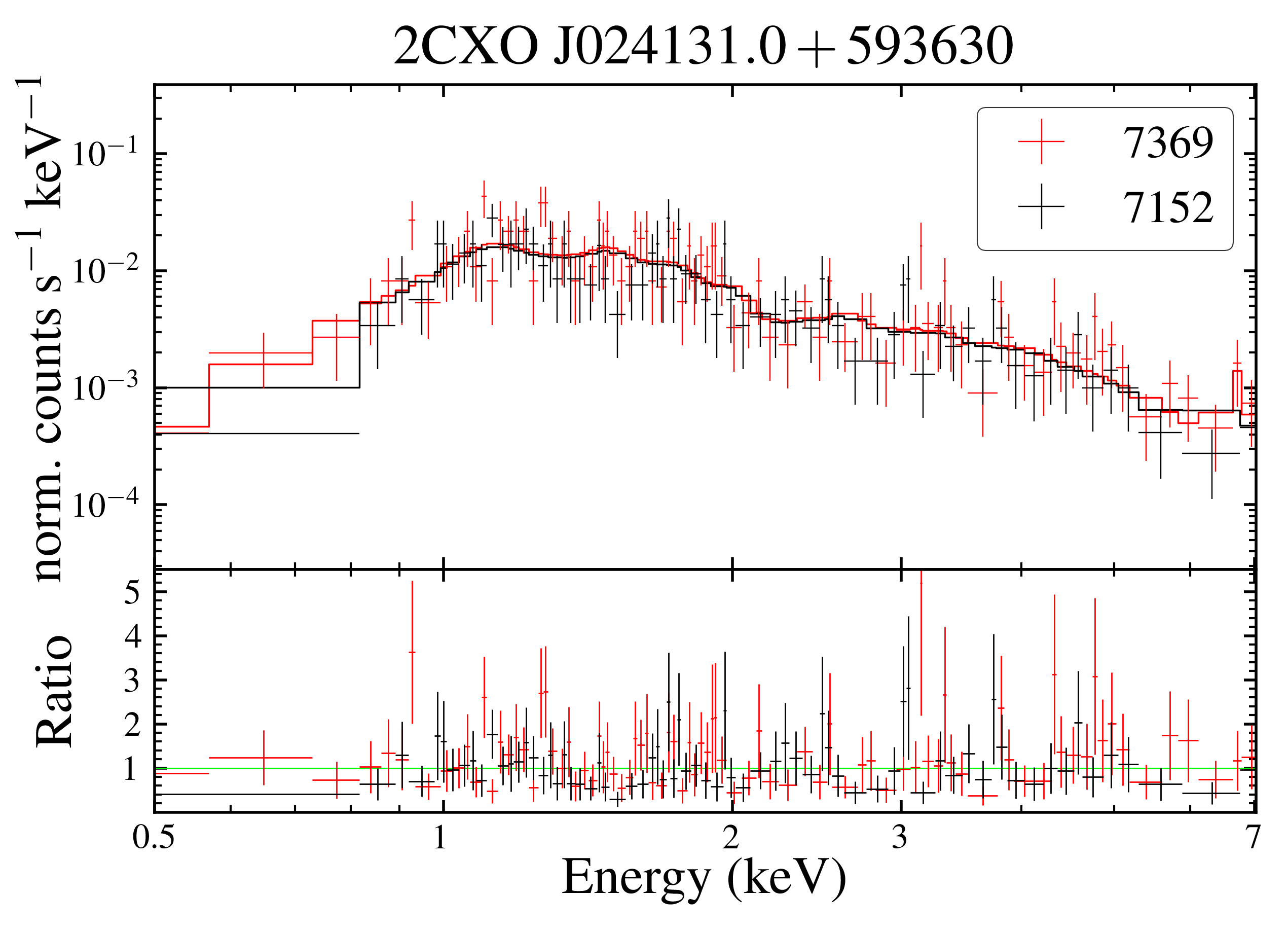}
    \includegraphics[width=0.3\linewidth]{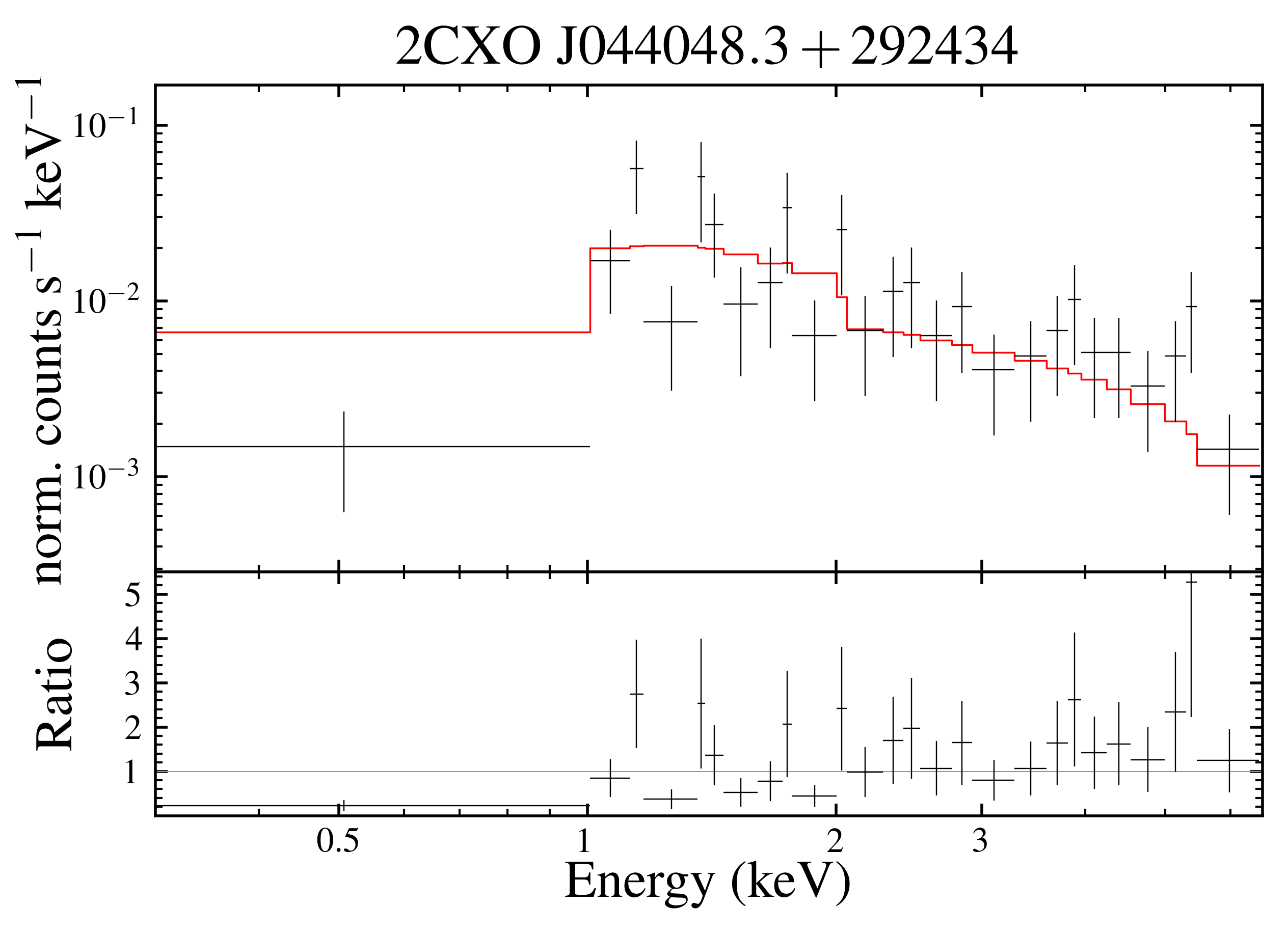}
    \includegraphics[width=0.3\linewidth]{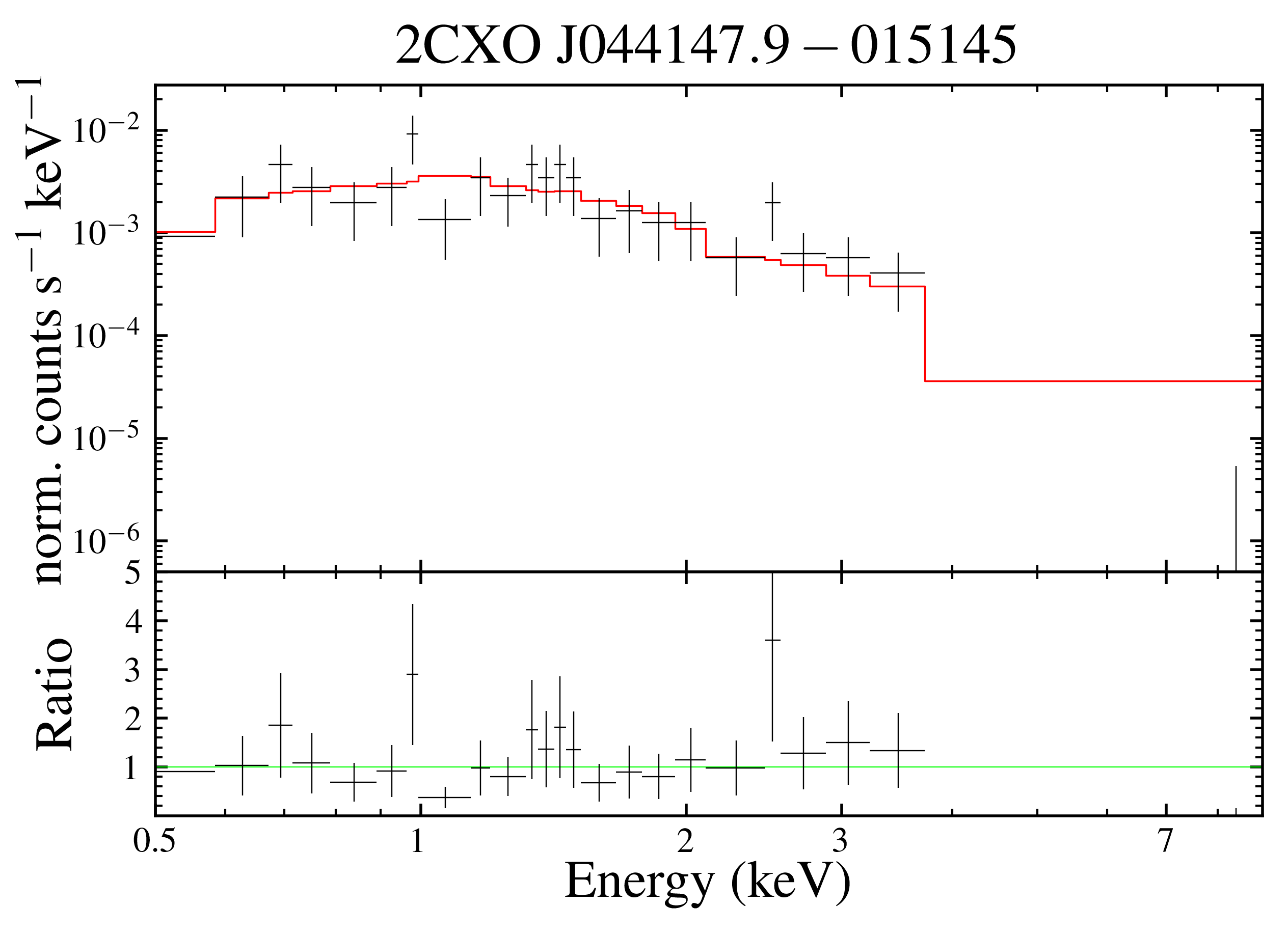}
    \includegraphics[width=0.3\linewidth]{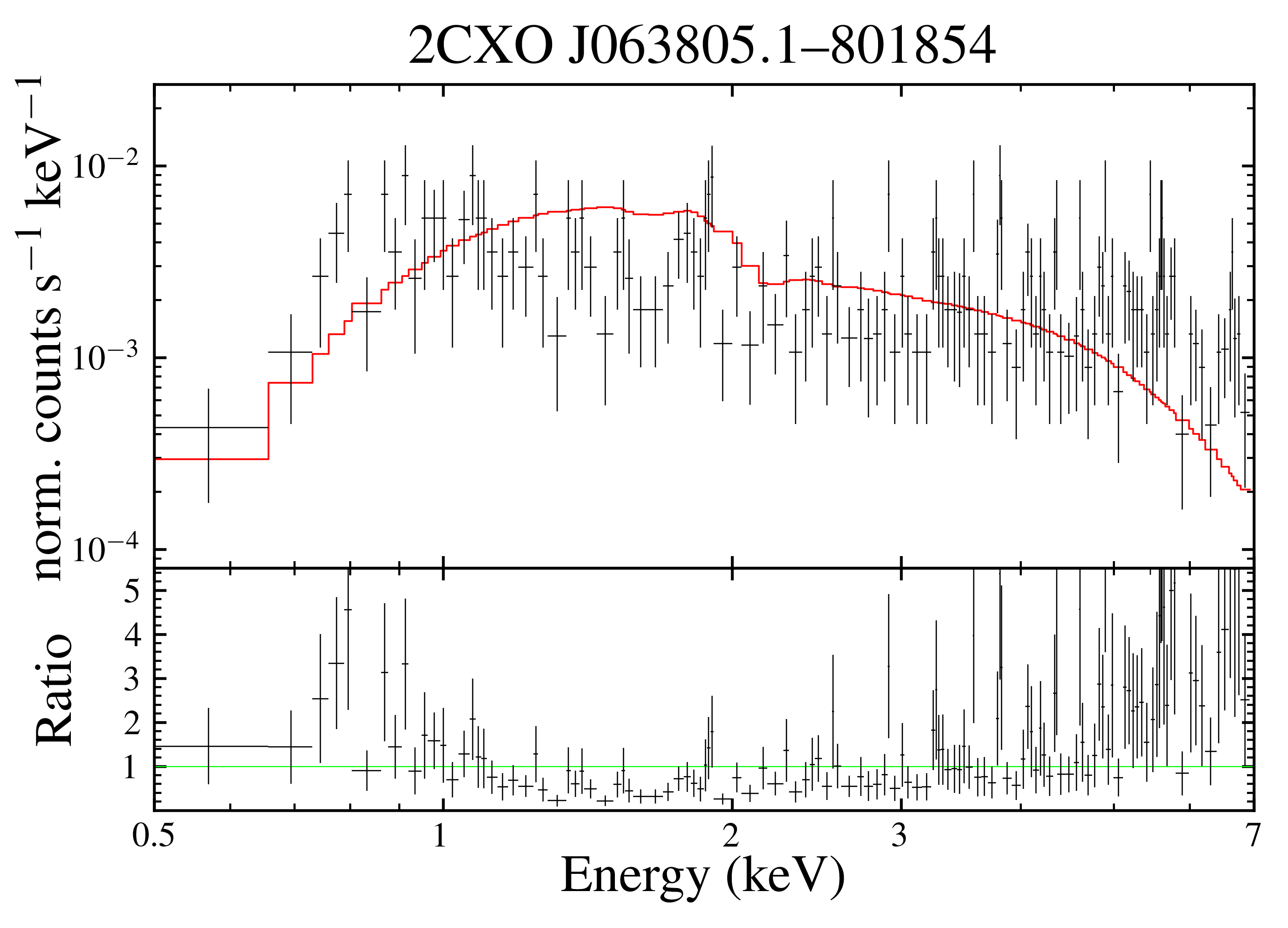}
    \includegraphics[width=0.3\linewidth]{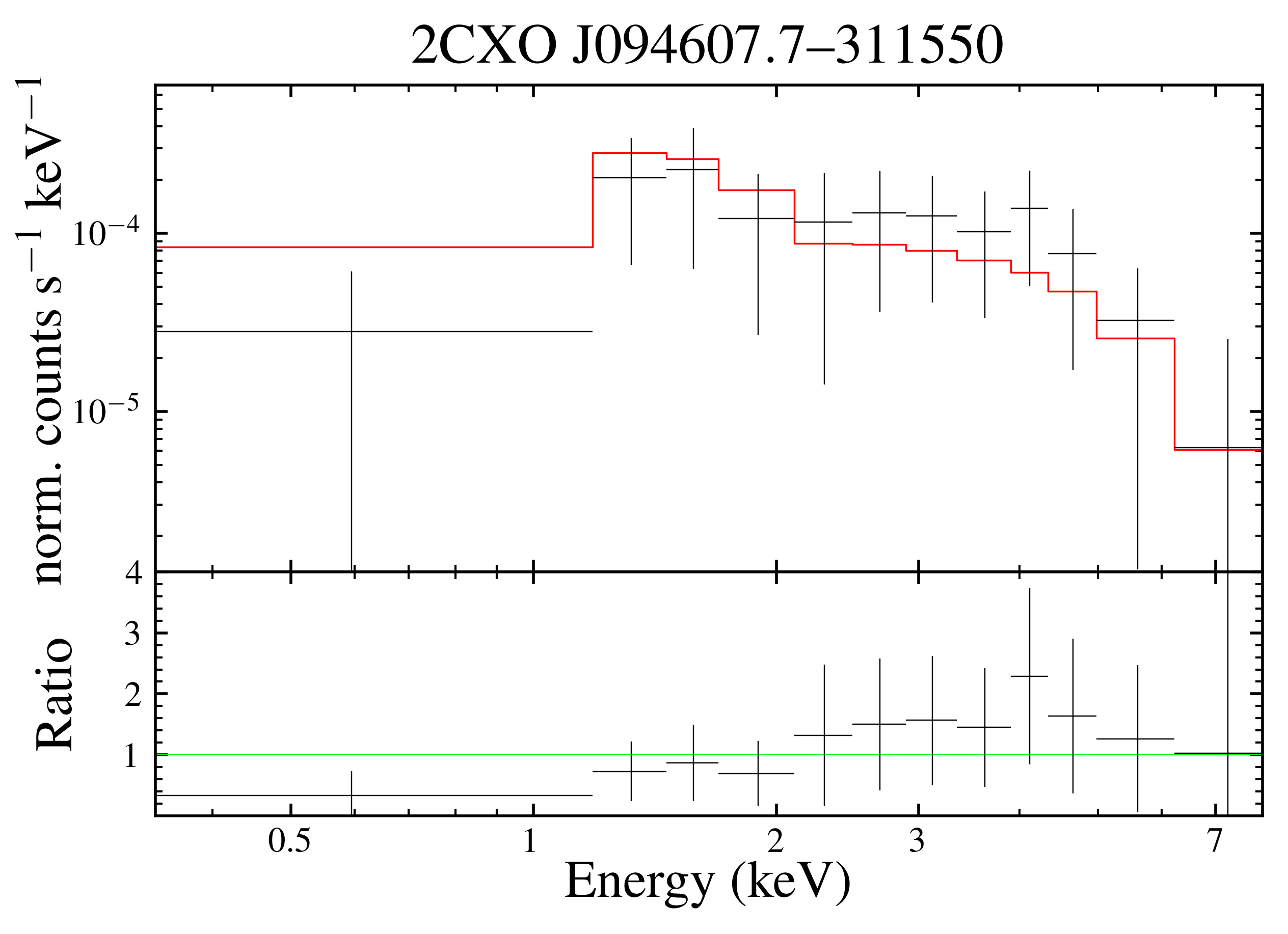}
    \includegraphics[width=0.3\linewidth]{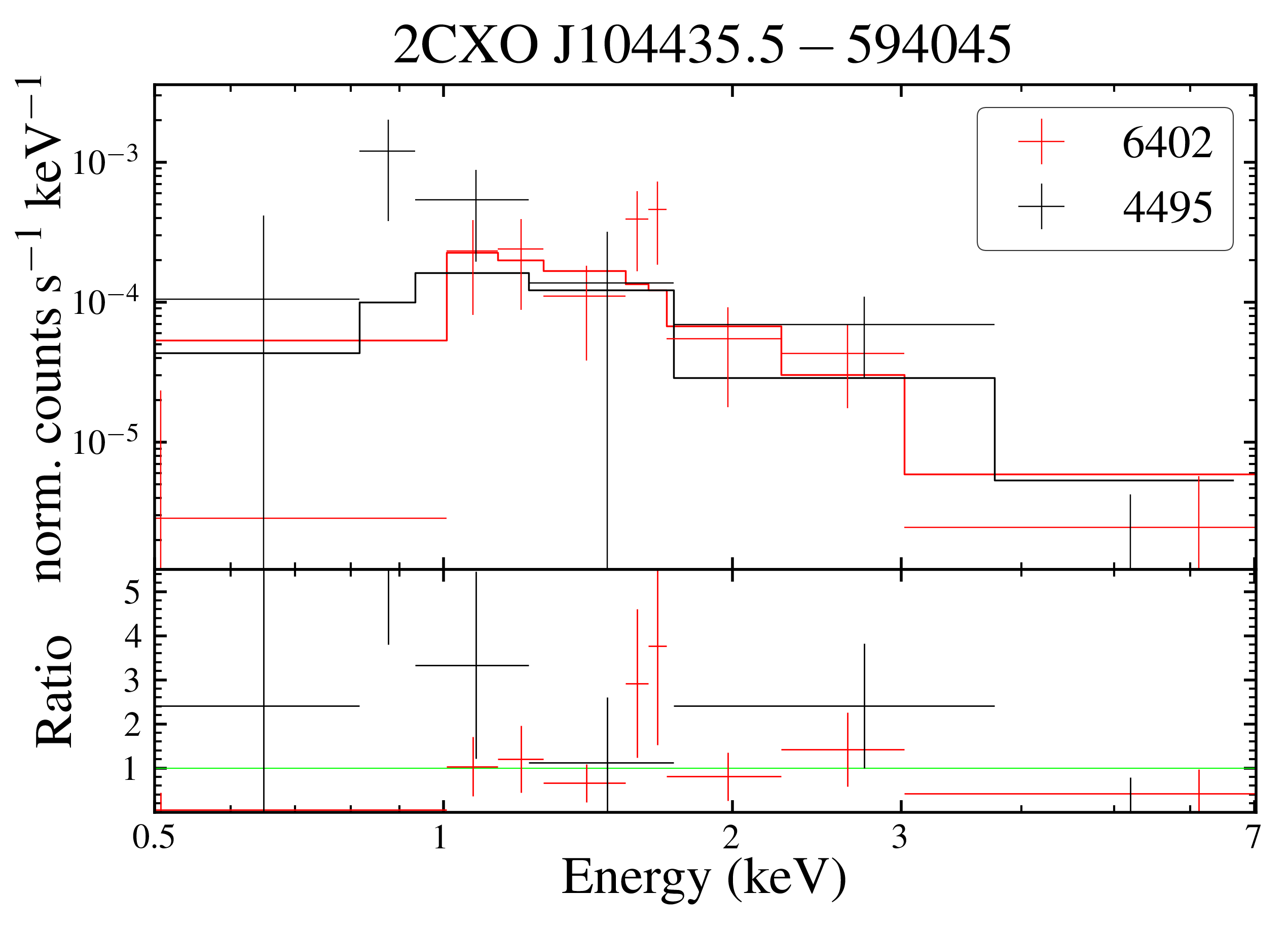}
    \includegraphics[width=0.3\linewidth]{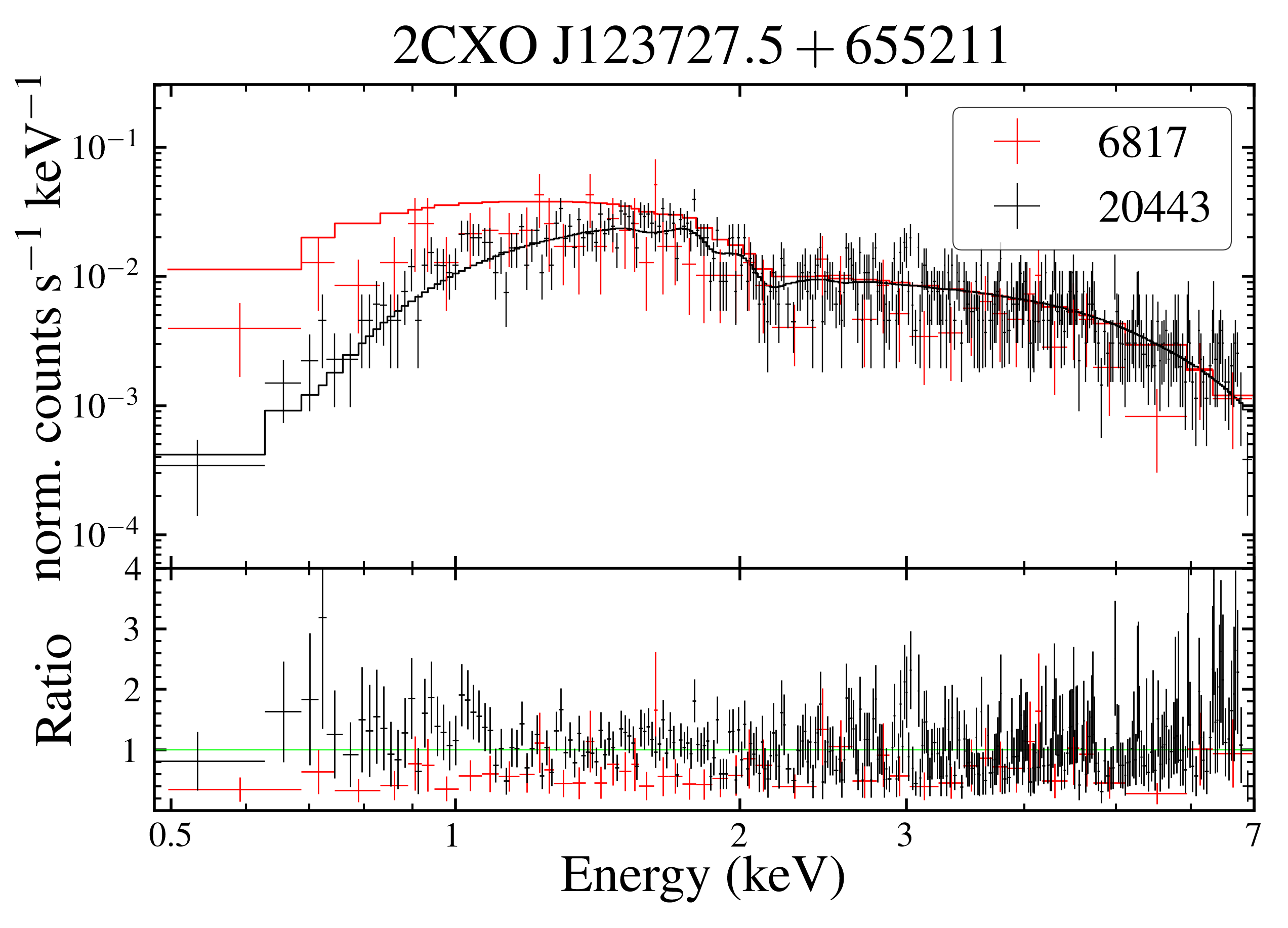}
    \includegraphics[width=0.3\linewidth]{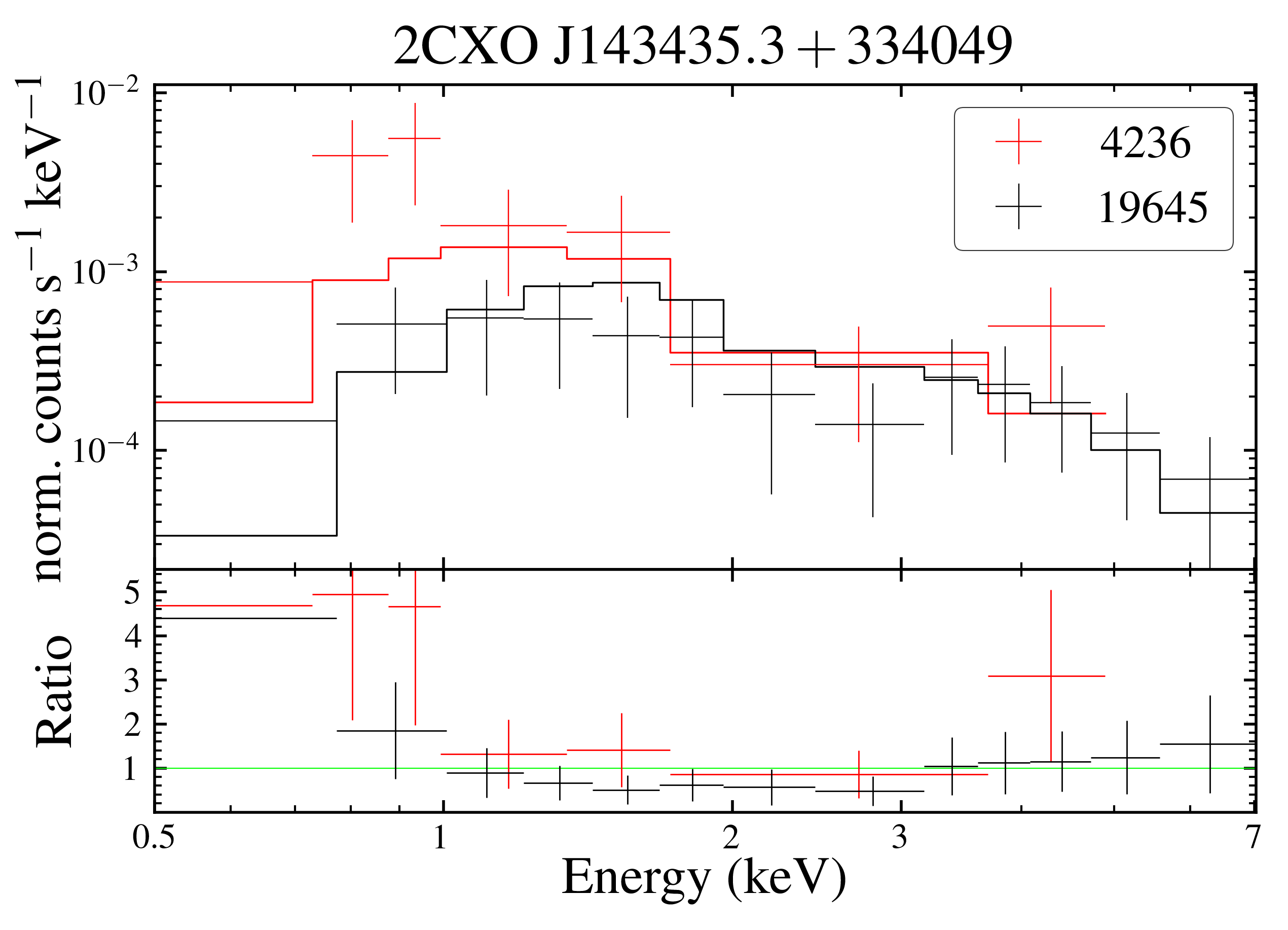}
    \includegraphics[width=0.3\linewidth]{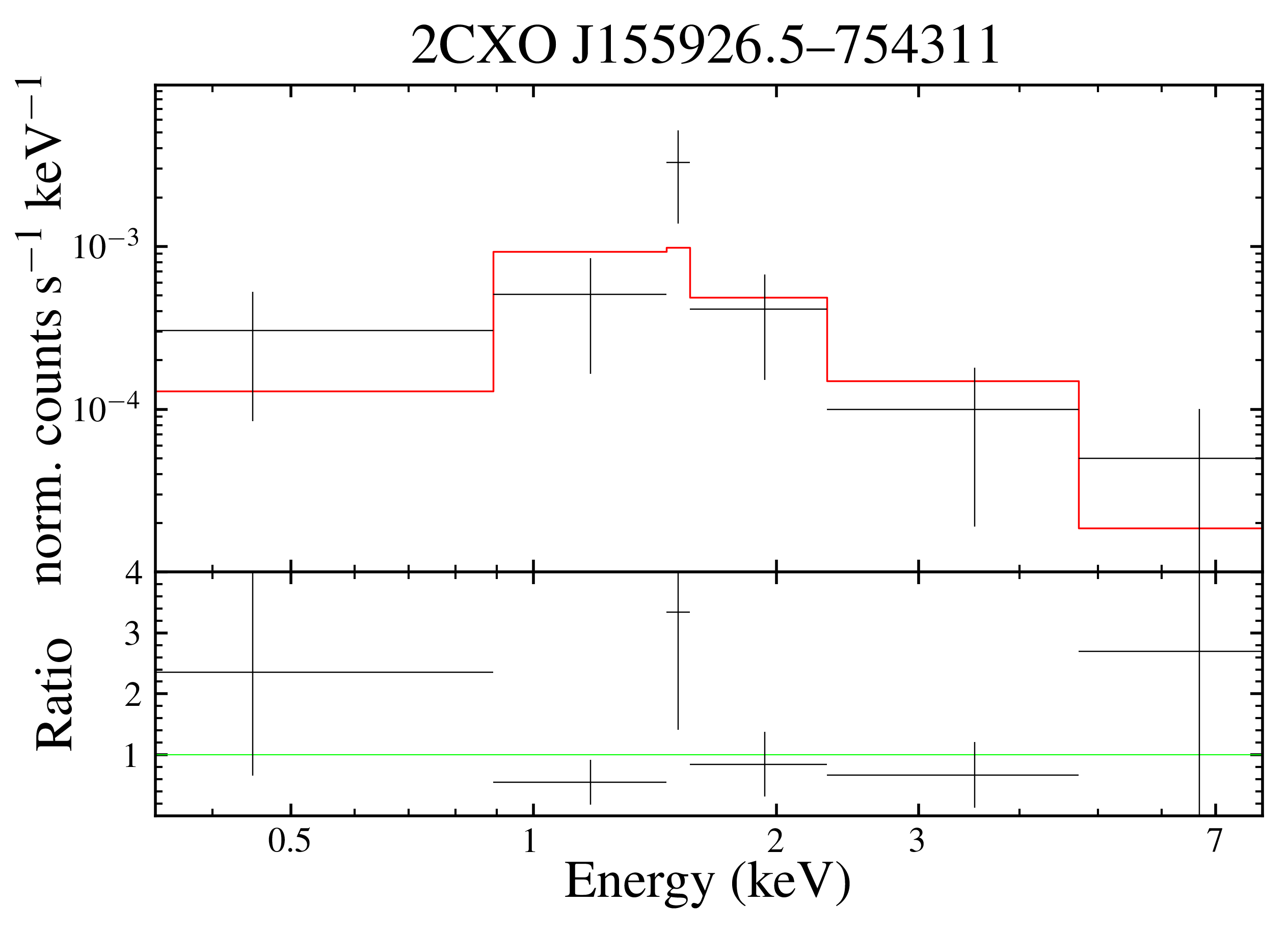}
    \includegraphics[width=0.3\linewidth]{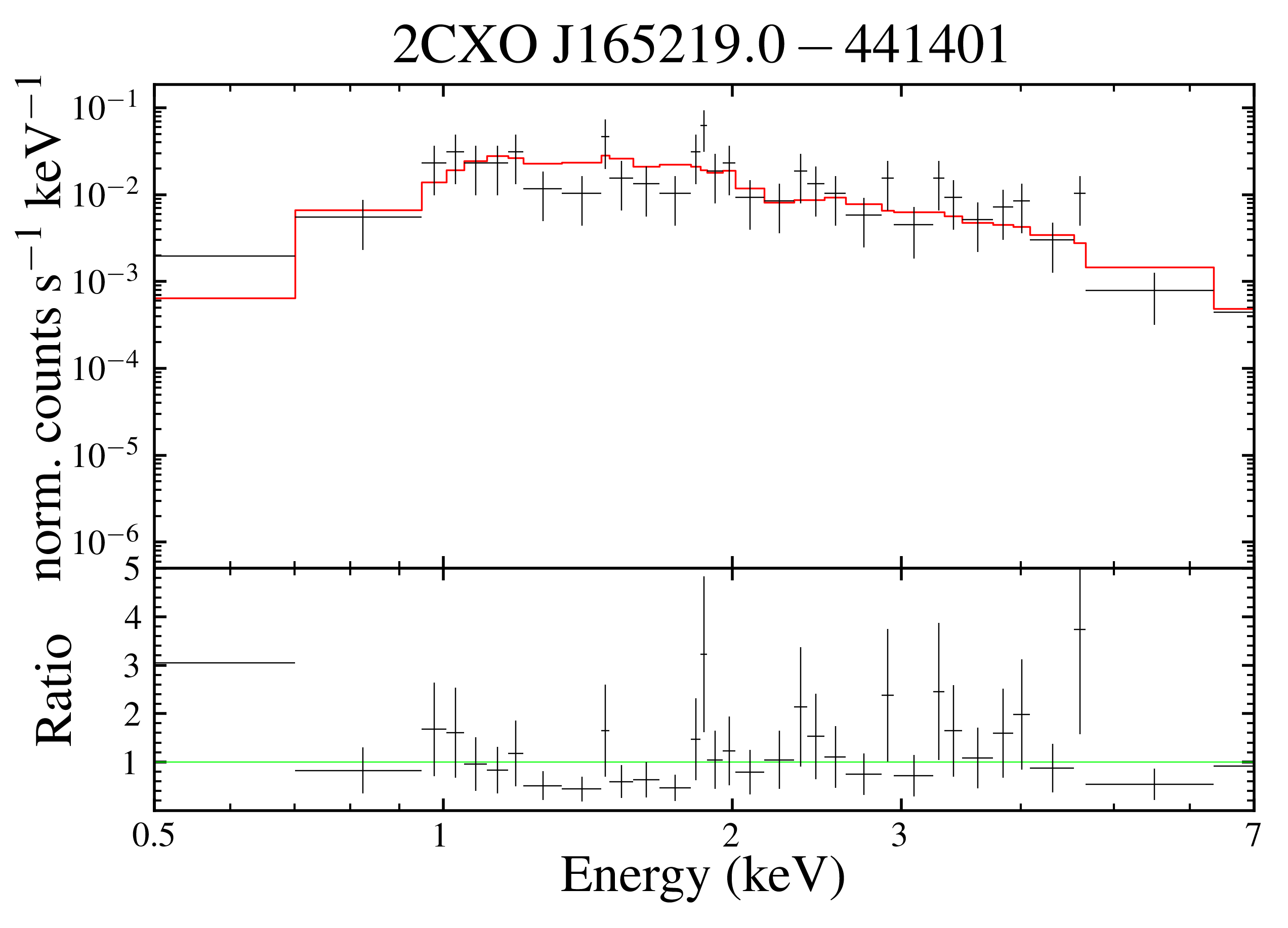}
    \includegraphics[width=0.3\linewidth]{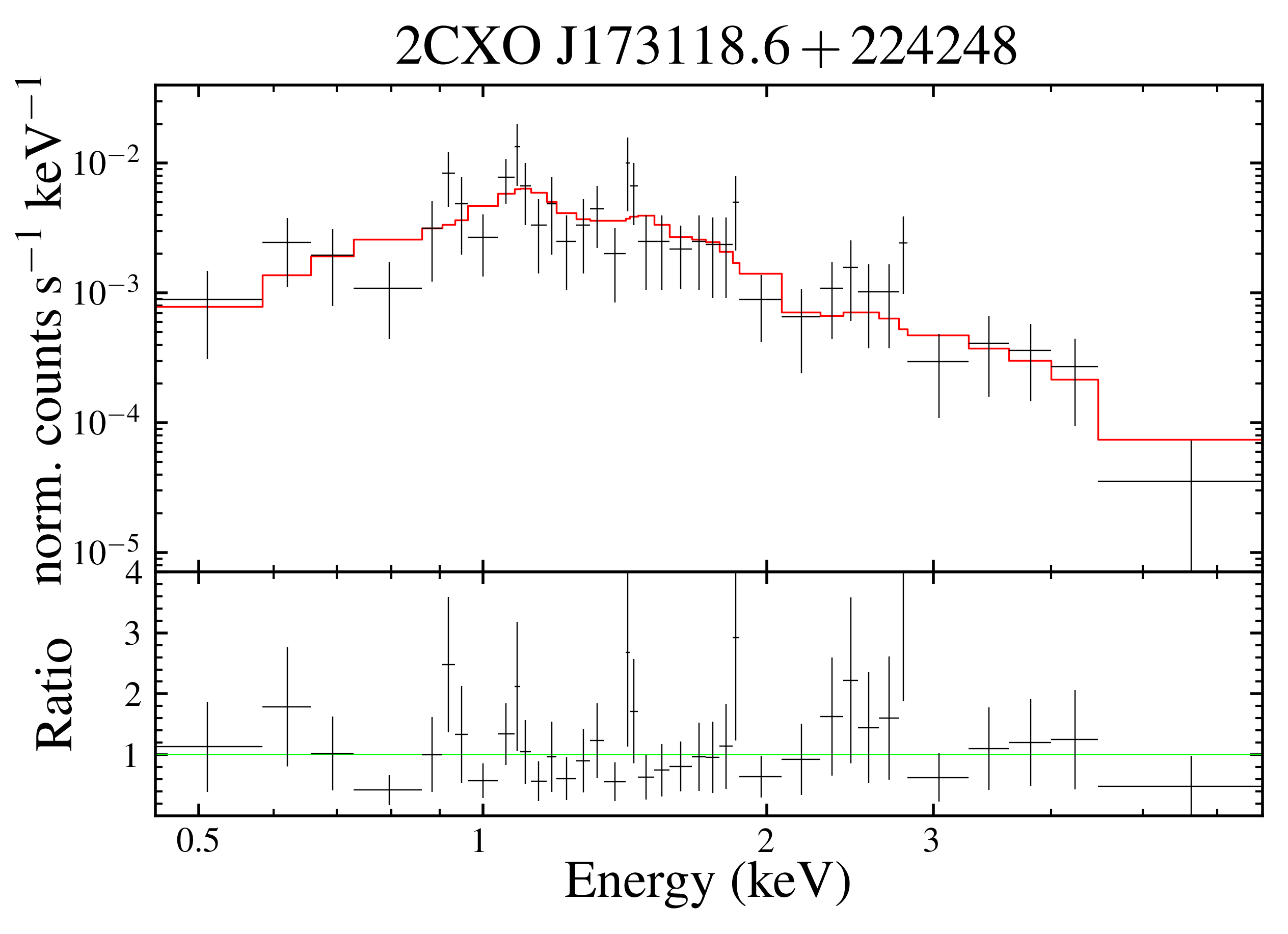}
    \includegraphics[width=0.3\linewidth]{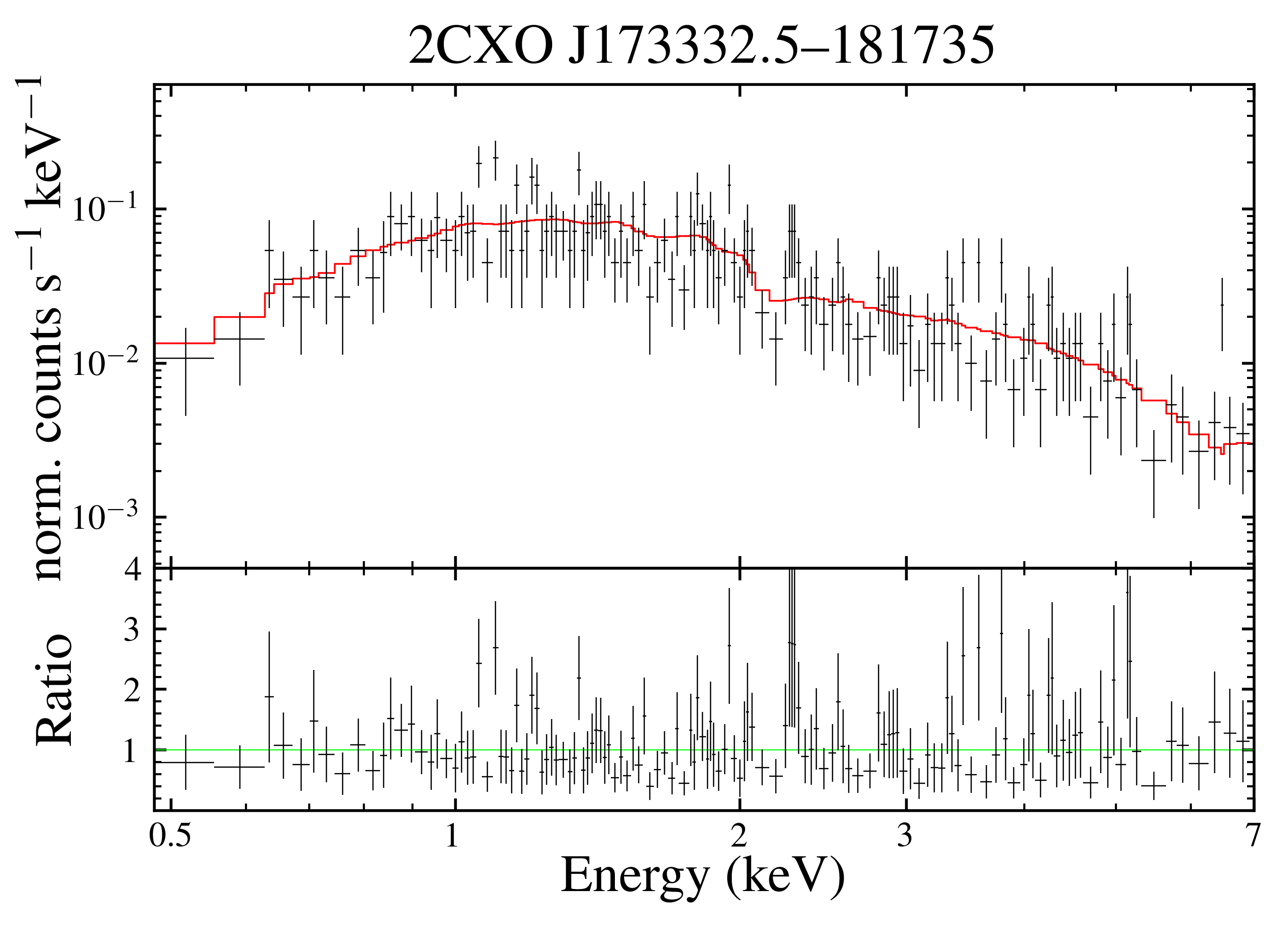}
    \includegraphics[width=0.3\linewidth]{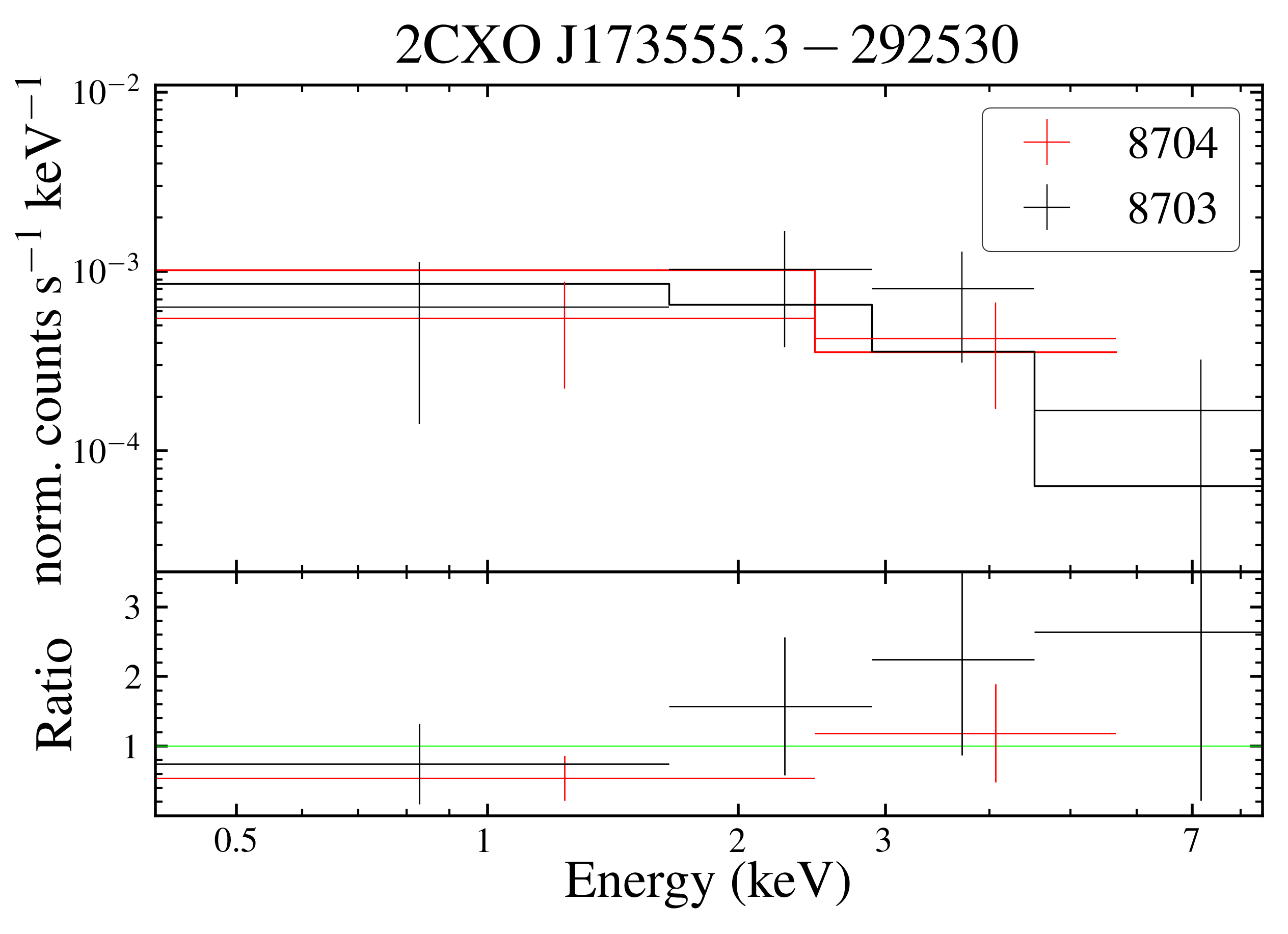}
    \includegraphics[width=0.3\linewidth]{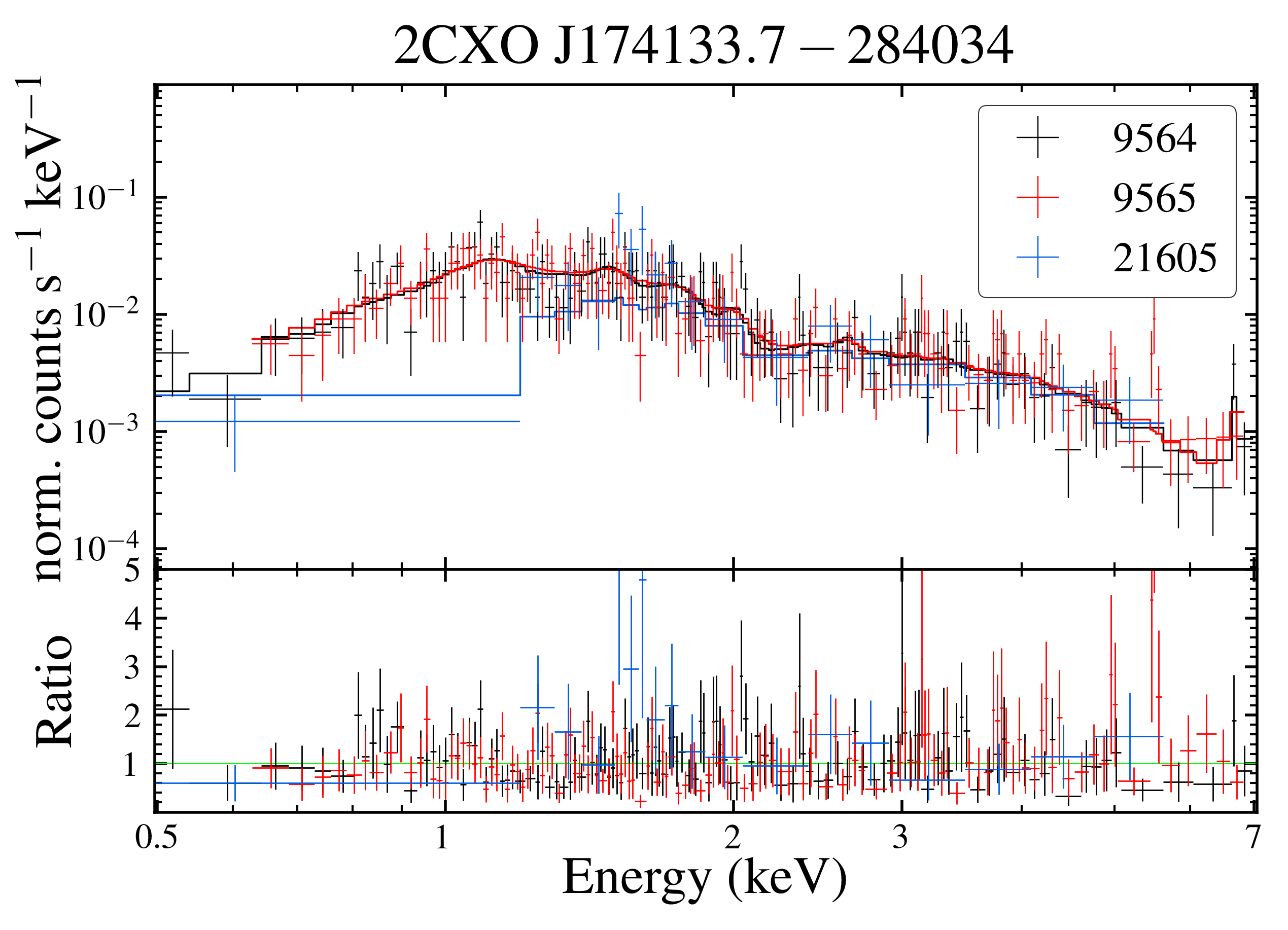}
    \includegraphics[width=0.3\linewidth]{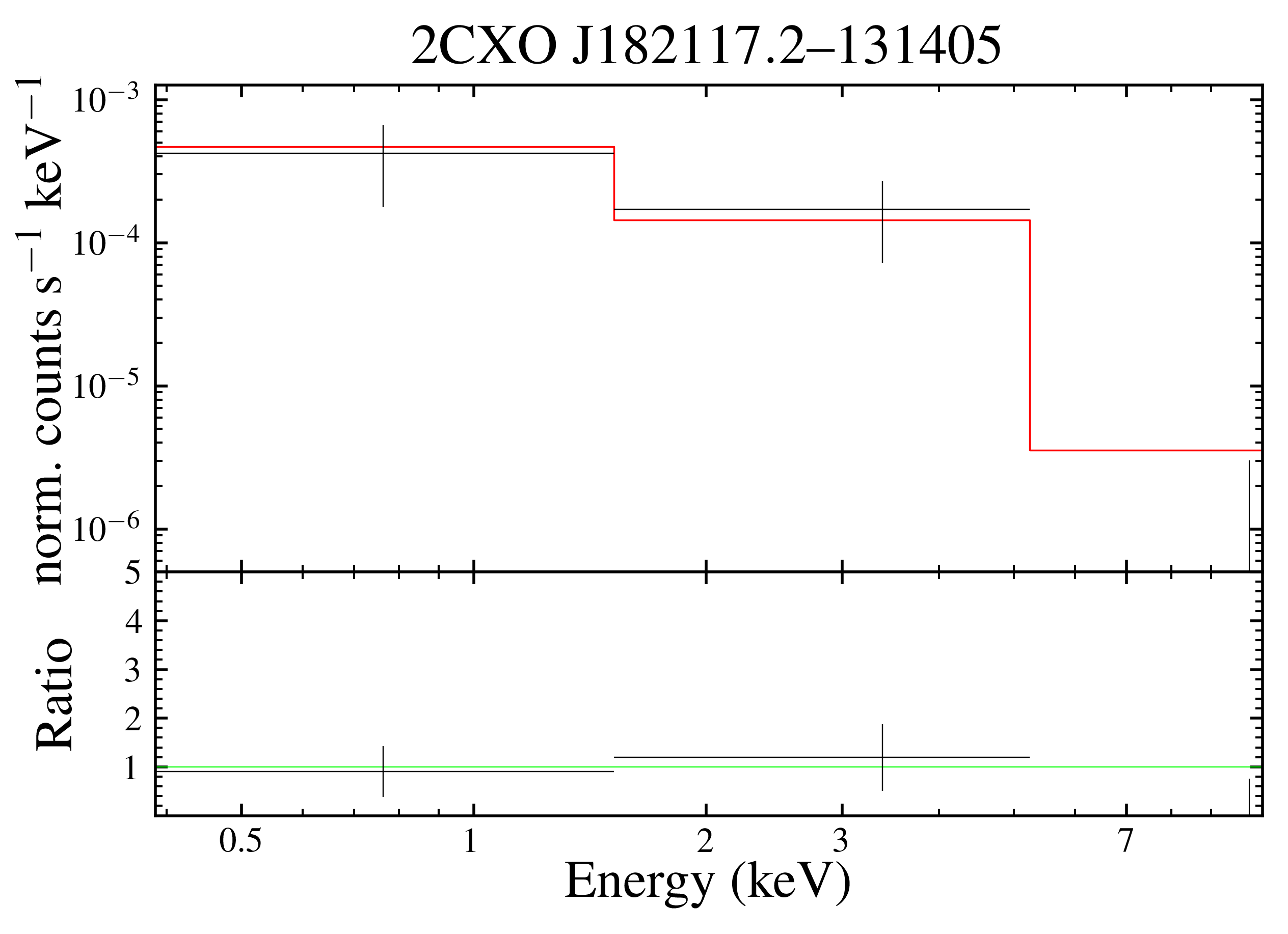}
    \includegraphics[width=0.3\linewidth]{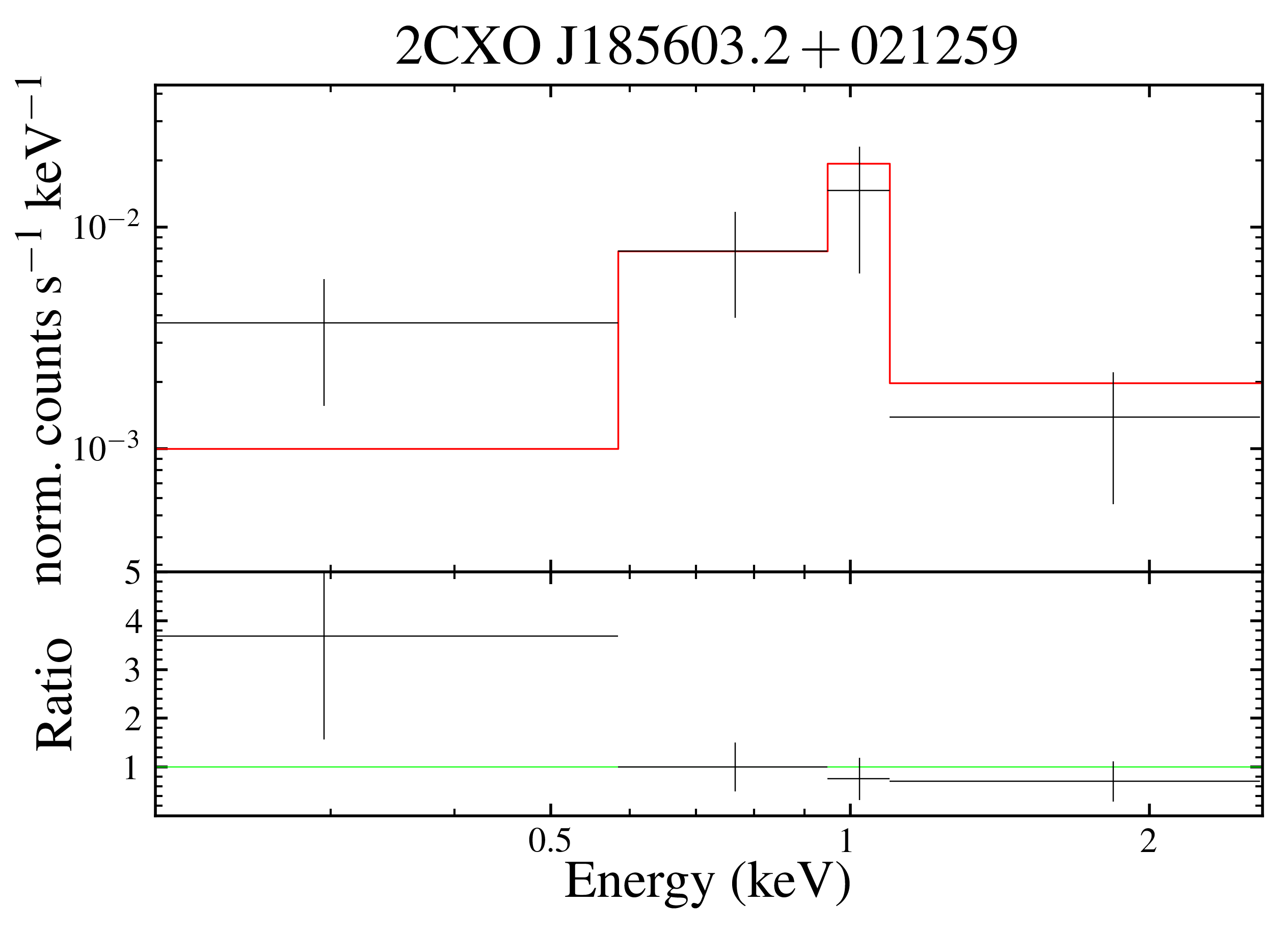}
    \includegraphics[width=0.3\linewidth]{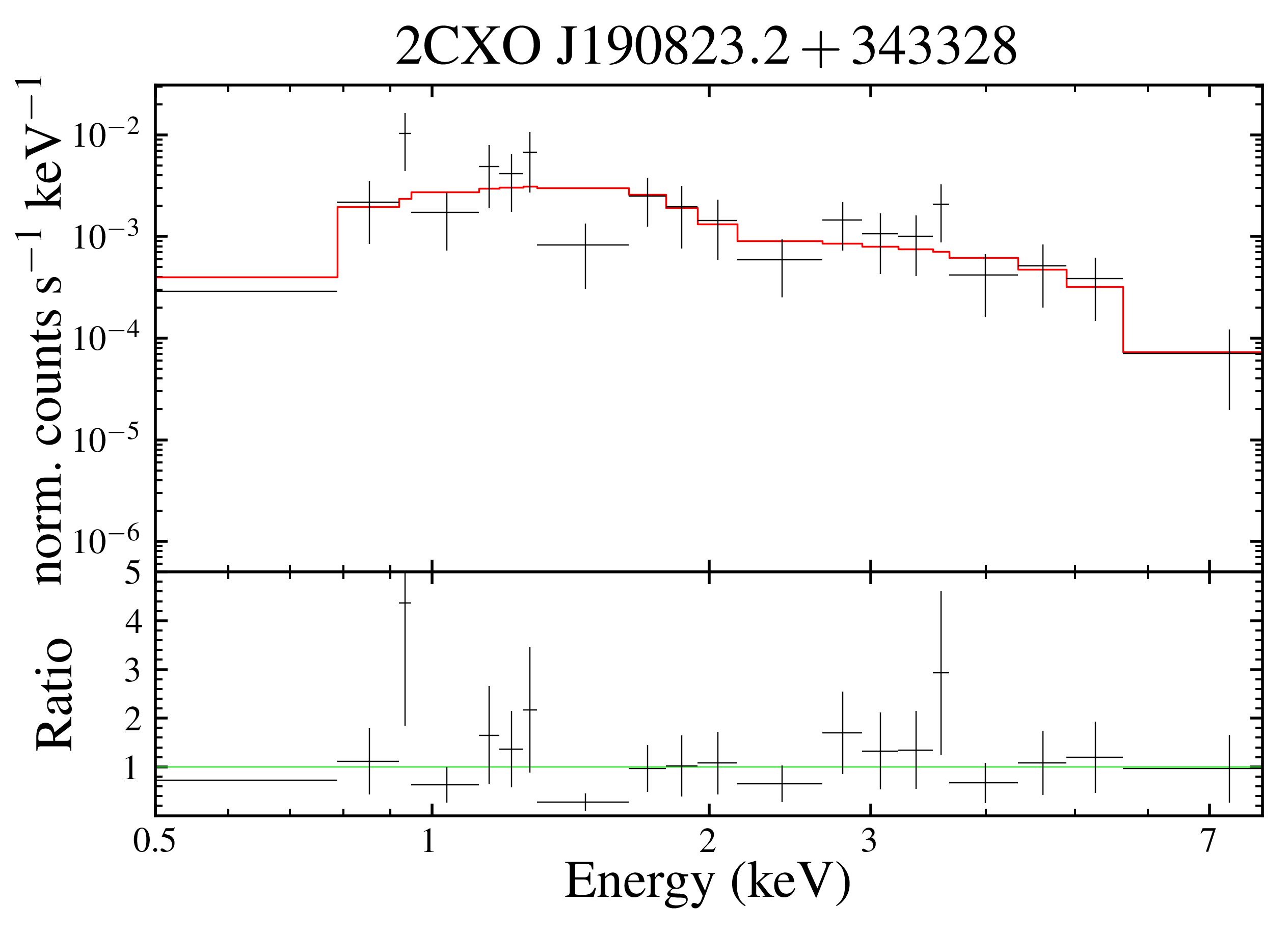} 
    \rule{0.3\linewidth}{0pt}

    \caption{ The {\it Chandra} X-ray spectra of objects from Tables \ref{tab:new_cvs} and \ref{tab:app_new_cvs} (top panel). {\it The solid lines} show the best-fit {\tt mekal} model. {\it The bottom panel} shows the ratio of the data divided by the model in each energy channel. }

    \label{fig:Xspectra}
\end{figure*}

Figure \ref{fig:Xspectra} shows {\it Chandra} X-ray spectra of all objects from Tables \ref{tab:new_cvs} and \ref{tab:app_new_cvs}. Table \ref{tab:xray_spectra} shows the results of the approximation of X-ray spectra with two spectral models: power-law (\texttt{powerlaw} in \texttt{XSPEC} with photon index parameter, $\Gamma$) and  optically thin thermal plasma emission model (\texttt{mekal} in \texttt{XSPEC} with temperature $\rm kT$, and metal abundance $[Z]$ parameters).

\begin{table*}
\tiny
\caption{Results of approximation of X-ray spectra of objects from Tables \ref{tab:new_cvs} and \ref{tab:app_new_cvs} by different models.}
\label{tab:xray_spectra}
\renewcommand\arraystretch{1.5}
\setlength\tabcolsep{4pt}
\centering
\begin{tabular}{l|c|c|c|c|c|c}
    \hline \hline
    \textbf{Name (2CXO)} & J024131.0+593630 & J044048.3+292434 & J044147.9--015145 & J063805.1--801854 & J094607.7--311550 & J104435.5--594045 \\ 
    \hline
    \textbf{ObsID} & 7152 + 7369 & 12677 & 11676 & 14925 & 15383 & 4495 + 6402 \\    \hline  \hline
    \texttt{tbabs x powerlaw} & & & & & &\\
    $N_H$($\times 10^{22} \, \text{cm}^{-2}$) & $0.42\pm0.11$ & $0.18^{b}$ (fixed) & $0.04^{b}$  (fixed) & $\rm \lesssim 0.01$ & $0.03$ (fixed) & $0.05$ (fixed) \\
    \rule{0cm}{0.35cm}$\Gamma$ & $1.78\pm0.14$ & $0.54\pm0.19$ & $1.78^{+0.23}_{-0.22}$ & $0.22^{+0.05}_{-0.08}$ & $0.48^{+0.40}_{-0.41}$ & $1.77^{+0.38}_{-0.37}$\\
    \rule{0cm}{0.35cm}C-stat/dof & 153.25/153 & 29.27/24 & 22.09/21 & 196.92/132 & 1.45/10 & 45.08/35 \\ \hline 
    \texttt{tbabs x mekal} & & & & & &\\
    $N_H$($\times 10^{22} \, \text{cm}^{-2}$) & $0.35^{+0.09}_{-0.08}$ & $0.18^{b}$ (fixed) & $0.04^{b}$ (fixed) & $0.45^{+0.08}_{-0.07}$ & $0.03$ (fixed) & $0.05$ (fixed) \\
    \rule{0cm}{0.35cm}kT(keV) & $5.54^{+1.12}_{-0.83}$ & $\rm \gtrsim 55.54$ & $4.54^{+3.47}_{-1.46}$ & $\rm \gtrsim 74.80$ & $\rm \gtrsim 24.92$ & $3.50^{+4.09}_{-1.29}$\\
    \rule{0cm}{0.35cm}$Z(Z_\odot)$ & $2.90^{+1.91}_{-1.21}$ & $1.0$ (fixed) & $1.0$ (fixed) & $1.0$ (fixed) & $1.0$ (fixed) & $1.0$ (fixed) \\
    \rule{0cm}{0.35cm}C-stat/dof & 144.18/152 & 42.83/24 & 21.55/21 & 314.20/132 & 4.93/10 & 43.49/35 \\

    \hline \hline 
    \multicolumn{7}{c}{}  \\
    \hline  \hline

    \textbf{Name} (2CXO)& J123727.5+655211 & J143435.3+334049 & J155926.5--754311 & J165219.0--441401 & J173118.6+224248 & J173332.5--181735 \\ 
    \hline
    \textbf{ObsID} & 6817 + 20443$^c$ & 4236 + 19645$^c$ & 14386 & 8145 & 3281 & 12944 \\ \hline  \hline  
    \texttt{tbabs x powerlaw}& & & & & &\\
    $N_H$($\times 10^{22} \, \text{cm}^{-2}$) & $0.08^{+0.06}_{-0.05}$ & $0.009^{a}$ (fixed) & $0.06^{a}$ (fixed) & $0.79^{+0.32}_{-0.29}$ & $0.17^{+0.15}_{-0.13}$ & $0.27^{+0.07}_{-0.06}$ \\
    \rule{0cm}{0.35cm}$\Gamma$ & $1.36\pm0.06$ & $1.71^{+0.32}_{-0.31}$ & $1.95^{+0.77}_{-0.67}$ & $1.53^{+0.35}_{-0.33}$ & $2.23^{+0.34}_{-0.31}$ & $1.36\pm0.12$ \\
    \rule{0cm}{0.35cm}C-stat/dof & 394.68/367 & 16.67/16 & 5.70/4 & 26.22/29 & 34.85/35 & 146.50/138 \\ \hline 
    \texttt{tbabs x mekal} & & & & & &\\
    $N_H$($\times 10^{22} \, \text{cm}^{-2}$) & $\rm \lesssim 0.08$ & $0.009^{a}$ (fixed) & $0.06^{a}$ (fixed) & $0.89^{+0.32}_{-0.27}$ & $\rm \lesssim 0.12$ & $0.26^{+0.05}_{-0.06}$ \\
    \rule{0cm}{0.35cm}kT(keV) & $38.28^{+31.40}_{-13.15}$ & $\rm \gtrsim 5.16$ & $\rm \gtrsim 2.00$ & $5.78^{+6.48}_{-2.01}$ & $2.97^{+0.88}_{-0.75}$ & $20.82^{+51.70}_{-8.61}$ \\
    \rule{0cm}{0.35cm}$Z(Z_\odot)$ & $\rm \lesssim 1.46$ & $1.0$ (fixed) & $1.0$ (fixed) & $4.16^{+10.75}_{-3.12}$ & $1.24^{+1.92}_{-0.90}$ & $\rm \lesssim 15.59$ \\
    \rule{0cm}{0.35cm}C-stat/dof & 397.16/366 & 18.52/16 & 5.89/4 & 23.93/28 & 30.63/34 & 146.34/137 \\ 

    \hline \hline 
    \multicolumn{7}{c}{}  \\
    \hline  \hline

    \textbf{Name (2CXO)} & J173555.3--292530 & \multicolumn{2}{c|}{J174133.7--284034} & J182117.2--131405 & J185603.2+021259 & J190823.2+343328  \\ 
    \hline
    \textbf{ObsID} & 8703 + 8704 & \multicolumn{2}{c|}{9564 + 9565 + 21605$^c$} & 7561 & 9613 & 13659  \\ \hline \hline
    \texttt{tbabs x powerlaw} & & \multicolumn{2}{c|}{} & & & \\
    $N_H$($\times 10^{22} \, \text{cm}^{-2}$) & $0.002$ (fixed) & \multicolumn{2}{c|}{$0.21\pm0.06$} & $0.03$ (fixed) & $0.005$ (fixed) & $0.01$ (fixed)  \\
    \rule{0cm}{0.35cm}$\Gamma$ & $0.58\pm0.51$ & \multicolumn{2}{c|}{$1.78^{+0.10}_{-0.09}$} & $1.74^{+0.88}_{-0.79}$ & $2.79^{+0.64}_{-0.63}$ & $1.15^{+0.25}_{-0.24}$  \\
    \rule{0cm}{0.35cm}C-stat/dof & 4.02/6 & \multicolumn{2}{c|}{275.22/273} & 1.34/1 & 1.98/2 & 19.04/17 \\ \hline
    \texttt{tbabs x mekal} & & \multicolumn{2}{c|}{} & & &\\
    $N_H$($\times 10^{22} \, \text{cm}^{-2}$) & $0.002$ (fixed) & \multicolumn{2}{c|}{$0.13\pm0.04$} & $0.03$ (fixed) & $0.005$ (fixed) & $0.01$ (fixed) \\
    \rule{0cm}{0.35cm}kT(keV) & $\rm \gtrsim 14.69$ & \multicolumn{2}{c|}{$5.62^{+0.77}_{-0.62}$} & $\rm \gtrsim 1.60$ & $1.04^{+1.22}_{-0.27}$ & $\rm \gtrsim 18.01$ \\
    \rule{0cm}{0.35cm}$Z(Z_\sun)$ & $1.0$ (fixed) & \multicolumn{2}{c|}{$2.61^{+1.21}_{-0.83}$} & $1.0$ (fixed) & $1.0$ (fixed) & $1.0$ (fixed) \\
    \rule{0cm}{0.35cm}C-stat/dof & 5.65/6 & \multicolumn{2}{c|}{247.52/270} & 1.24/1 & 4.12/2 & 19.28/17 \\ 
    \hline \hline

\end{tabular}
\flushleft
\tablefoot{($a$) -- $N_H$ is fixed at Galactic hydrogen column density; ($b$) -- Color excess $\rm E(B-V)$  from the Bayestar dust map lies outside the reliable distance interval, so we fixed $N_H$ at Galactic hydrogen column density; ($c$) -- New {\it Chandra} observation is used in the analysis, which is not presented in CSC2.
}
\end{table*}

\end{appendix}

\bibliographystyle{aa}
\bibliography{CVsCSC}

\begin{thebibliography}{88}
\expandafter\ifx\csname natexlab\endcsname\relax\def\natexlab#1{#1}\fi

\bibitem[{{Abril} {et~al.}(2020){Abril}, {Schmidtobreick}, {Ederoclite}, \& {L{\'o}pez-Sanjuan}}]{2020MNRAS.492L..40A}
{Abril}, J., {Schmidtobreick}, L., {Ederoclite}, A., \& {L{\'o}pez-Sanjuan}, C. 2020, \mnras, 492, L40

\bibitem[{{Arnaud}(1996)}]{1996ASPC..101...17A}
{Arnaud}, K.~A. 1996, in Astronomical Society of the Pacific Conference Series, Vol. 101, Astronomical Data Analysis Software and Systems V, ed. G.~H. {Jacoby} \& J.~{Barnes}, 17

\bibitem[{{Astropy Collaboration} {et~al.}(2018){Astropy Collaboration}, {Price-Whelan}, {Sip{\H{o}}cz}, {G{\"u}nther}, {Lim}, {Crawford}, {Conseil}, {Shupe}, {Craig}, {Dencheva}, {Ginsburg}, {VanderPlas}, {Bradley}, {P{\'e}rez-Su{\'a}rez}, {de Val-Borro}, {Aldcroft}, {Cruz}, {Robitaille}, {Tollerud}, {Ardelean}, {Babej}, {Bach}, {Bachetti}, {Bakanov}, {Bamford}, {Barentsen}, {Barmby}, {Baumbach}, {Berry}, {Biscani}, {Boquien}, {Bostroem}, {Bouma}, {Brammer}, {Bray}, {Breytenbach}, {Buddelmeijer}, {Burke}, {Calderone}, {Cano Rodr{\'\i}guez}, {Cara}, {Cardoso}, {Cheedella}, {Copin}, {Corrales}, {Crichton}, {D'Avella}, {Deil}, {Depagne}, {Dietrich}, {Donath}, {Droettboom}, {Earl}, {Erben}, {Fabbro}, {Ferreira}, {Finethy}, {Fox}, {Garrison}, {Gibbons}, {Goldstein}, {Gommers}, {Greco}, {Greenfield}, {Groener}, {Grollier}, {Hagen}, {Hirst}, {Homeier}, {Horton}, {Hosseinzadeh}, {Hu}, {Hunkeler}, {Ivezi{\'c}}, {Jain}, {Jenness}, {Kanarek}, {Kendrew}, {Kern}, {Kerzendorf}, {Khvalko}, {King}, {Kirkby}, {Kulkarni},
  {Kumar}, {Lee}, {Lenz}, {Littlefair}, {Ma}, {Macleod}, {Mastropietro}, {McCully}, {Montagnac}, {Morris}, {Mueller}, {Mumford}, {Muna}, {Murphy}, {Nelson}, {Nguyen}, {Ninan}, {N{\"o}the}, {Ogaz}, {Oh}, {Parejko}, {Parley}, {Pascual}, {Patil}, {Patil}, {Plunkett}, {Prochaska}, {Rastogi}, {Reddy Janga}, {Sabater}, {Sakurikar}, {Seifert}, {Sherbert}, {Sherwood-Taylor}, {Shih}, {Sick}, {Silbiger}, {Singanamalla}, {Singer}, {Sladen}, {Sooley}, {Sornarajah}, {Streicher}, {Teuben}, {Thomas}, {Tremblay}, {Turner}, {Terr{\'o}n}, {van Kerkwijk}, {de la Vega}, {Watkins}, {Weaver}, {Whitmore}, {Woillez}, {Zabalza}, \& {Astropy Contributors}}]{2018AJ....156..123A}
{Astropy Collaboration}, {Price-Whelan}, A.~M., {Sip{\H{o}}cz}, B.~M., {et~al.} 2018, \aj, 156, 123

\bibitem[{{Astropy Collaboration} {et~al.}(2013){Astropy Collaboration}, {Robitaille}, {Tollerud}, {Greenfield}, {Droettboom}, {Bray}, {Aldcroft}, {Davis}, {Ginsburg}, {Price-Whelan}, {Kerzendorf}, {Conley}, {Crighton}, {Barbary}, {Muna}, {Ferguson}, {Grollier}, {Parikh}, {Nair}, {Unther}, {Deil}, {Woillez}, {Conseil}, {Kramer}, {Turner}, {Singer}, {Fox}, {Weaver}, {Zabalza}, {Edwards}, {Azalee Bostroem}, {Burke}, {Casey}, {Crawford}, {Dencheva}, {Ely}, {Jenness}, {Labrie}, {Lim}, {Pierfederici}, {Pontzen}, {Ptak}, {Refsdal}, {Servillat}, \& {Streicher}}]{2013A&A...558A..33A}
{Astropy Collaboration}, {Robitaille}, T.~P., {Tollerud}, E.~J., {et~al.} 2013, \aap, 558, A33

\bibitem[{{Bellm} {et~al.}(2019){Bellm}, {Kulkarni}, {Graham}, {Dekany}, {Smith}, {Riddle}, {Masci}, {Helou}, {Prince}, {Adams}, {Barbarino}, {Barlow}, {Bauer}, {Beck}, {Belicki}, {Biswas}, {Blagorodnova}, {Bodewits}, {Bolin}, {Brinnel}, {Brooke}, {Bue}, {Bulla}, {Burruss}, {Cenko}, {Chang}, {Connolly}, {Coughlin}, {Cromer}, {Cunningham}, {De}, {Delacroix}, {Desai}, {Duev}, {Eadie}, {Farnham}, {Feeney}, {Feindt}, {Flynn}, {Franckowiak}, {Frederick}, {Fremling}, {Gal-Yam}, {Gezari}, {Giomi}, {Goldstein}, {Golkhou}, {Goobar}, {Groom}, {Hacopians}, {Hale}, {Henning}, {Ho}, {Hover}, {Howell}, {Hung}, {Huppenkothen}, {Imel}, {Ip}, {Ivezi{\'c}}, {Jackson}, {Jones}, {Juric}, {Kasliwal}, {Kaspi}, {Kaye}, {Kelley}, {Kowalski}, {Kramer}, {Kupfer}, {Landry}, {Laher}, {Lee}, {Lin}, {Lin}, {Lunnan}, {Giomi}, {Mahabal}, {Mao}, {Miller}, {Monkewitz}, {Murphy}, {Ngeow}, {Nordin}, {Nugent}, {Ofek}, {Patterson}, {Penprase}, {Porter}, {Rauch}, {Rebbapragada}, {Reiley}, {Rigault}, {Rodriguez}, {van Roestel}, {Rusholme}, {van
  Santen}, {Schulze}, {Shupe}, {Singer}, {Soumagnac}, {Stein}, {Surace}, {Sollerman}, {Szkody}, {Taddia}, {Terek}, {Van Sistine}, {van Velzen}, {Vestrand}, {Walters}, {Ward}, {Ye}, {Yu}, {Yan}, \& {Zolkower}}]{bellm2019}
{Bellm}, E.~C., {Kulkarni}, S.~R., {Graham}, M.~J., {et~al.} 2019, \pasp, 131, 018002

\bibitem[{{Boldin} {et~al.}(2013){Boldin}, {Tsygankov}, \& {Lutovinov}}]{2013AstL...39..375B}
{Boldin}, P.~A., {Tsygankov}, S.~S., \& {Lutovinov}, A.~A. 2013, Astronomy Letters, 39, 375

\bibitem[{{Boller} {et~al.}(2016){Boller}, {Freyberg}, {Tr{\"u}mper}, {Haberl}, {Voges}, \& {Nandra}}]{2016boller}
{Boller}, T., {Freyberg}, M.~J., {Tr{\"u}mper}, J., {et~al.} 2016, \aap, 588, A103

\bibitem[{{Britt} {et~al.}(2014){Britt}, {Hynes}, {Johnson}, {Baldwin}, {Jonker}, {Nelemans}, {Torres}, {Maccarone}, {Steeghs}, {Greiss}, {Heinke}, {Bassa}, {Collazzi}, {Villar}, {Gabb}, \& {Gossen}}]{2014ApJS..214...10B}
{Britt}, C.~T., {Hynes}, R.~I., {Johnson}, C.~B., {et~al.} 2014, \apjs, 214, 10

\bibitem[{{Campbell} {et~al.}(2008){Campbell}, {Harrison}, {Schwope}, \& {Howell}}]{2008campbell_cyclotron}
{Campbell}, R.~K., {Harrison}, T.~E., {Schwope}, A.~D., \& {Howell}, S.~B. 2008, \apj, 672, 531

\bibitem[{{Cardelli} {et~al.}(1989){Cardelli}, {Clayton}, \& {Mathis}}]{1989ApJ...345..245C}
{Cardelli}, J.~A., {Clayton}, G.~C., \& {Mathis}, J.~S. 1989, \apj, 345, 245

\bibitem[{{Cash}(1979)}]{1979ApJ...228..939C}
{Cash}, W. 1979, \apj, 228, 939

\bibitem[{{Chambers} {et~al.}(2016){Chambers}, {Magnier}, {Metcalfe}, {Flewelling}, {Huber}, {Waters}, {Denneau}, {Draper}, {Farrow}, {Finkbeiner}, {Holmberg}, {Koppenhoefer}, {Price}, {Rest}, {Saglia}, {Schlafly}, {Smartt}, {Sweeney}, {Wainscoat}, {Burgett}, {Chastel}, {Grav}, {Heasley}, {Hodapp}, {Jedicke}, {Kaiser}, {Kudritzki}, {Luppino}, {Lupton}, {Monet}, {Morgan}, {Onaka}, {Shiao}, {Stubbs}, {Tonry}, {White}, {Ba{\~n}ados}, {Bell}, {Bender}, {Bernard}, {Boegner}, {Boffi}, {Botticella}, {Calamida}, {Casertano}, {Chen}, {Chen}, {Cole}, {Deacon}, {Frenk}, {Fitzsimmons}, {Gezari}, {Gibbs}, {Goessl}, {Goggia}, {Gourgue}, {Goldman}, {Grant}, {Grebel}, {Hambly}, {Hasinger}, {Heavens}, {Heckman}, {Henderson}, {Henning}, {Holman}, {Hopp}, {Ip}, {Isani}, {Jackson}, {Keyes}, {Koekemoer}, {Kotak}, {Le}, {Liska}, {Long}, {Lucey}, {Liu}, {Martin}, {Masci}, {McLean}, {Mindel}, {Misra}, {Morganson}, {Murphy}, {Obaika}, {Narayan}, {Nieto-Santisteban}, {Norberg}, {Peacock}, {Pier}, {Postman}, {Primak}, {Rae}, {Rai},
  {Riess}, {Riffeser}, {Rix}, {R{\"o}ser}, {Russel}, {Rutz}, {Schilbach}, {Schultz}, {Scolnic}, {Strolger}, {Szalay}, {Seitz}, {Small}, {Smith}, {Soderblom}, {Taylor}, {Thomson}, {Taylor}, {Thakar}, {Thiel}, {Thilker}, {Unger}, {Urata}, {Valenti}, {Wagner}, {Walder}, {Walter}, {Watters}, {Werner}, {Wood-Vasey}, \& {Wyse}}]{2016panstarrs}
{Chambers}, K.~C., {Magnier}, E.~A., {Metcalfe}, N., {et~al.} 2016, arXiv e-prints, arXiv:1612.05560

\bibitem[{{Chen} {et~al.}(2019){Chen}, {Huang}, {Yuan}, {Wang}, {Fan}, {Xiang}, {Zhang}, {Tian}, \& {Liu}}]{2019MNRAS.483.4277C}
{Chen}, B.~Q., {Huang}, Y., {Yuan}, H.~B., {et~al.} 2019, \mnras, 483, 4277

\bibitem[{{Cordova} {et~al.}(1981){Cordova}, {Jensen}, \& {Nugent}}]{1981MNRAS.196....1C}
{Cordova}, F.~A., {Jensen}, K.~A., \& {Nugent}, J.~J. 1981, \mnras, 196, 1

\bibitem[{{Cordova} \& {Mason}(1984)}]{1984MNRAS.206..879C}
{Cordova}, F.~A. \& {Mason}, K.~O. 1984, \mnras, 206, 879

\bibitem[{{Dekany} {et~al.}(2020){Dekany}, {Smith}, {Riddle}, {Feeney}, {Porter}, {Hale}, {Zolkower}, {Belicki}, {Kaye}, {Henning}, {Walters}, {Cromer}, {Delacroix}, {Rodriguez}, {Reiley}, {Mao}, {Hover}, {Murphy}, {Burruss}, {Baker}, {Kowalski}, {Reif}, {Mueller}, {Bellm}, {Graham}, \& {Kulkarni}}]{dekanyztf}
{Dekany}, R., {Smith}, R.~M., {Riddle}, R., {et~al.} 2020, \pasp, 132, 038001

\bibitem[{{DESI Collaboration} {et~al.}(2023){DESI Collaboration}, {Adame}, {Aguilar}, {Ahlen}, {Alam}, {Aldering}, {Alexander}, {Alfarsy}, {Allende Prieto}, {Alvarez}, {Alves}, {Anand}, {Andrade-Oliveira}, {Armengaud}, {Asorey}, {Avila}, {Aviles}, {Bailey}, {Balaguera-Antol{\'\i}nez}, {Ballester}, {Baltay}, {Bault}, {Bautista}, {Behera}, {Beltran}, {BenZvi}, {Beraldo e Silva}, {Bermejo-Climent}, {Berti}, {Besuner}, {Beutler}, {Bianchi}, {Blake}, {Blum}, {Bolton}, {Brieden}, {Brodzeller}, {Brooks}, {Brown}, {Buckley-Geer}, {Burtin}, {Cabayol-Garcia}, {Cai}, {Canning}, {Cardiel-Sas}, {Carnero Rosell}, {Castander}, {Cervantes-Cota}, {Chabanier}, {Chaussidon}, {Chaves-Montero}, {Chen}, {Chuang}, {Claybaugh}, {Cole}, {Cooper}, {Cuceu}, {Davis}, {Dawson}, {de Belsunce}, {de la Cruz}, {de la Macorra}, {de Mattia}, {Demina}, {Demirbozan}, {DeRose}, {Dey}, {Dey}, {Dhungana}, {Ding}, {Ding}, {Doel}, {Doshi}, {Douglass}, {Edge}, {Eftekharzadeh}, {Eisenstein}, {Elliott}, {Escoffier}, {Fagrelius}, {Fan}, {Fanning},
  {Fawcett}, {Ferraro}, {Ereza}, {Flaugher}, {Font-Ribera}, {Forero-S{\'a}nchez}, {Forero-Romero}, {Frenk}, {G{\"a}nsicke}, {Garc{\'\i}a}, {Garc{\'\i}a-Bellido}, {Garcia-Quintero}, {Garrison}, {Gil-Mar{\'\i}n}, {Golden-Marx}, {Gontcho}, {Gonzalez-Morales}, {Gonzalez-Perez}, {Gordon}, {Graur}, {Green}, {Gruen}, {Guy}, {Hadzhiyska}, {Hahn}, {Han}, {Hanif}, {Herrera-Alcantar}, {Honscheid}, {Hou}, {Howlett}, {Huterer}, {Ir{\v{s}}i{\v{c}}}, {Ishak}, {Jacques}, {Jana}, {Jiang}, {Jimenez}, {Jing}, {Joudaki}, {Jullo}, {Juneau}, {Kizhuprakkat}, {Kara{\c{c}}ayl{\i}}, {Karim}, {Kehoe}, {Kent}, {Khederlarian}, {Kim}, {Kirkby}, {Kisner}, {Kitaura}, {Kneib}, {Koposov}, {Kov{\'a}cs}, {Kremin}, {Krolewski}, {L'Huillier}, {Lambert}, {Lamman}, {Lan}, {Landriau}, {Lang}, {Lange}, {Lasker}, {Le Guillou}, {Leauthaud}, {Levi}, {Li}, {Linder}, {Lyons}, {Magneville}, {Manera}, {Manser}, {Margala}, {Martini}, {McDonald}, {Medina}, {Medina-Varela}, {Meisner}, {Mena-Fern{\'a}ndez}, {Meneses-Rizo}, {Mezcua}, {Miquel}, {Montero-Camacho},
  {Moon}, {Moore}, {Moustakas}, {Mueller}, {Mundet}, {Mu{\~n}oz-Guti{\'e}rrez}, {Myers}, {Nadathur}, {Napolitano}, {Neveux}, {Newman}, {Nie}, {Nikutta}, {Niz}, {Norberg}, {Noriega}, {Paillas}, {Palanque-Delabrouille}, {Palmese}, {Zhiwei}, {Parkinson}, {Penmetsa}, {Percival}, {P{\'e}rez-Fern{\'a}ndez}, {P{\'e}rez-R{\`a}fols}, {Pieri}, {Poppett}, {Porredon}, {Pothier}, {Prada}, {Pucha}, {Raichoor}, {Ram{\'\i}rez-P{\'e}rez}, {Ramirez-Solano}, {Rashkovetskyi}, {Ravoux}, {Rocher}, {Rockosi}, {Ross}, {Rossi}, {Ruggeri}, {Ruhlmann-Kleider}, {Sabiu}, {Said}, {Saintonge}, {Samushia}, {Sanchez}, {Saulder}, {Schaan}, {Schlafly}, {Schlegel}, {Scholte}, {Schubnell}, {Seo}, {Shafieloo}, {Sharples}, {Sheu}, {Silber}, {Sinigaglia}, {Siudek}, {Slepian}, {Smith}, {Sprayberry}, {Stephey}, {Su{\'a}rez-P{\'e}rez}, {Sun}, {Tan}, {Tarl{\'e}}, {Tojeiro}, {Ure{\~n}a-L{\'o}pez}, {Vaisakh}, {Valcin}, {Valdes}, {Valluri}, {Vargas-Maga{\~n}a}, {Variu}, {Verde}, {Walther}, {Wang}, {Wang}, {Weaver}, {Weaverdyck}, {Wechsler}, {White},
  {Xie}, {Yang}, {Y{\`e}che}, {Yu}, {Yuan}, {Zhang}, {Zhang}, {Zhao}, {Zheng}, {Zhou}, {Zhou}, {Zou}, {Zou}, \& {Zu}}]{desiedr2023}
{DESI Collaboration}, {Adame}, A.~G., {Aguilar}, J., {et~al.} 2023, arXiv e-prints, arXiv:2306.06308

\bibitem[{{DESI Collaboration} {et~al.}(2016){DESI Collaboration}, {Aghamousa}, {Aguilar}, {Ahlen}, {Alam}, {Allen}, {Allende Prieto}, {Annis}, {Bailey}, {Balland}, {Ballester}, {Baltay}, {Beaufore}, {Bebek}, {Beers}, {Bell}, {Bernal}, {Besuner}, {Beutler}, {Blake}, {Bleuler}, {Blomqvist}, {Blum}, {Bolton}, {Briceno}, {Brooks}, {Brownstein}, {Buckley-Geer}, {Burden}, {Burtin}, {Busca}, {Cahn}, {Cai}, {Cardiel-Sas}, {Carlberg}, {Carton}, {Casas}, {Castander}, {Cervantes-Cota}, {Claybaugh}, {Close}, {Coker}, {Cole}, {Comparat}, {Cooper}, {Cousinou}, {Crocce}, {Cuby}, {Cunningham}, {Davis}, {Dawson}, {de la Macorra}, {De Vicente}, {Delubac}, {Derwent}, {Dey}, {Dhungana}, {Ding}, {Doel}, {Duan}, {Ealet}, {Edelstein}, {Eftekharzadeh}, {Eisenstein}, {Elliott}, {Escoffier}, {Evatt}, {Fagrelius}, {Fan}, {Fanning}, {Farahi}, {Farihi}, {Favole}, {Feng}, {Fernandez}, {Findlay}, {Finkbeiner}, {Fitzpatrick}, {Flaugher}, {Flender}, {Font-Ribera}, {Forero-Romero}, {Fosalba}, {Frenk}, {Fumagalli}, {Gaensicke}, {Gallo},
  {Garcia-Bellido}, {Gaztanaga}, {Pietro Gentile Fusillo}, {Gerard}, {Gershkovich}, {Giannantonio}, {Gillet}, {Gonzalez-de-Rivera}, {Gonzalez-Perez}, {Gott}, {Graur}, {Gutierrez}, {Guy}, {Habib}, {Heetderks}, {Heetderks}, {Heitmann}, {Hellwing}, {Herrera}, {Ho}, {Holland}, {Honscheid}, {Huff}, {Hutchinson}, {Huterer}, {Hwang}, {Illa Laguna}, {Ishikawa}, {Jacobs}, {Jeffrey}, {Jelinsky}, {Jennings}, {Jiang}, {Jimenez}, {Johnson}, {Joyce}, {Jullo}, {Juneau}, {Kama}, {Karcher}, {Karkar}, {Kehoe}, {Kennamer}, {Kent}, {Kilbinger}, {Kim}, {Kirkby}, {Kisner}, {Kitanidis}, {Kneib}, {Koposov}, {Kovacs}, {Koyama}, {Kremin}, {Kron}, {Kronig}, {Kueter-Young}, {Lacey}, {Lafever}, {Lahav}, {Lambert}, {Lampton}, {Landriau}, {Lang}, {Lauer}, {Le Goff}, {Le Guillou}, {Le Van Suu}, {Lee}, {Lee}, {Leitner}, {Lesser}, {Levi}, {L'Huillier}, {Li}, {Liang}, {Lin}, {Linder}, {Loebman}, {Luki{\'c}}, {Ma}, {MacCrann}, {Magneville}, {Makarem}, {Manera}, {Manser}, {Marshall}, {Martini}, {Massey}, {Matheson}, {McCauley}, {McDonald},
  {McGreer}, {Meisner}, {Metcalfe}, {Miller}, {Miquel}, {Moustakas}, {Myers}, {Naik}, {Newman}, {Nichol}, {Nicola}, {Nicolati da Costa}, {Nie}, {Niz}, {Norberg}, {Nord}, {Norman}, {Nugent}, {O'Brien}, {Oh}, {Olsen}, {Padilla}, {Padmanabhan}, {Padmanabhan}, {Palanque-Delabrouille}, {Palmese}, {Pappalardo}, {P{\^a}ris}, {Park}, {Patej}, {Peacock}, {Peiris}, {Peng}, {Percival}, {Perruchot}, {Pieri}, {Pogge}, {Pollack}, {Poppett}, {Prada}, {Prakash}, {Probst}, {Rabinowitz}, {Raichoor}, {Ree}, {Refregier}, {Regal}, {Reid}, {Reil}, {Rezaie}, {Rockosi}, {Roe}, {Ronayette}, {Roodman}, {Ross}, {Ross}, {Rossi}, {Rozo}, {Ruhlmann-Kleider}, {Rykoff}, {Sabiu}, {Samushia}, {Sanchez}, {Sanchez}, {Schlegel}, {Schneider}, {Schubnell}, {Secroun}, {Seljak}, {Seo}, {Serrano}, {Shafieloo}, {Shan}, {Sharples}, {Sholl}, {Shourt}, {Silber}, {Silva}, {Sirk}, {Slosar}, {Smith}, {Smoot}, {Som}, {Song}, {Sprayberry}, {Staten}, {Stefanik}, {Tarle}, {Sien Tie}, {Tinker}, {Tojeiro}, {Valdes}, {Valenzuela}, {Valluri}, {Vargas-Magana},
  {Verde}, {Walker}, {Wang}, {Wang}, {Weaver}, {Weaverdyck}, {Wechsler}, {Weinberg}, {White}, {Yang}, {Yeche}, {Zhang}, {Zhao}, {Zheng}, {Zhou}, {Zhou}, {Zhu}, {Zou}, \& {Zu}}]{DESI2016-Overview}
{DESI Collaboration}, {Aghamousa}, A., {Aguilar}, J., {et~al.} 2016, arXiv e-prints, arXiv:1611.00036

\bibitem[{{Downes} {et~al.}(2001){Downes}, {Webbink}, {Shara}, {Ritter}, {Kolb}, \& {Duerbeck}}]{2001PASP..113..764D}
{Downes}, R.~A., {Webbink}, R.~F., {Shara}, M.~M., {et~al.} 2001, \pasp, 113, 764

\bibitem[{{El-Badry} {et~al.}(2022){El-Badry}, {Conroy}, {Fuller}, {Kiman}, {van Roestel}, {Rodriguez}, \& {Burdge}}]{2022MNRAS.517.4916E}
{El-Badry}, K., {Conroy}, C., {Fuller}, J., {et~al.} 2022, \mnras, 517, 4916

\bibitem[{{Evans} {et~al.}(2010){Evans}, {Primini}, {Glotfelty}, {Anderson}, {Bonaventura}, {Chen}, {Davis}, {Doe}, {Evans}, {Fabbiano}, {Galle}, {Gibbs}, {Grier}, {Hain}, {Hall}, {Harbo}, {He}, {Houck}, {Karovska}, {Kashyap}, {Lauer}, {McCollough}, {McDowell}, {Miller}, {Mitschang}, {Morgan}, {Mossman}, {Nichols}, {Nowak}, {Plummer}, {Refsdal}, {Rots}, {Siemiginowska}, {Sundheim}, {Tibbetts}, {Van Stone}, {Winkelman}, \& {Zografou}}]{2010ApJS..189...37E}
{Evans}, I.~N., {Primini}, F.~A., {Glotfelty}, K.~J., {et~al.} 2010, \apjs, 189, 37

\bibitem[{{Faulkner}(1971)}]{1971ApJ...170L..99F}
{Faulkner}, J. 1971, \apjl, 170, L99

\bibitem[{{Fruscione} {et~al.}(2006){Fruscione}, {McDowell}, {Allen}, {Brickhouse}, {Burke}, {Davis}, {Durham}, {Elvis}, {Galle}, {Harris}, {Huenemoerder}, {Houck}, {Ishibashi}, {Karovska}, {Nicastro}, {Noble}, {Nowak}, {Primini}, {Siemiginowska}, {Smith}, \& {Wise}}]{2006SPIE.6270E..1VF}
{Fruscione}, A., {McDowell}, J.~C., {Allen}, G.~E., {et~al.} 2006, in Society of Photo-Optical Instrumentation Engineers (SPIE) Conference Series, Vol. 6270, Observatory Operations: Strategies, Processes, and Systems, ed. D.~R. {Silva} \& R.~E. {Doxsey}, 62701V

\bibitem[{{Gaia Collaboration} {et~al.}(2023){Gaia Collaboration}, {Vallenari}, {Brown}, {Prusti}, {de Bruijne}, {Arenou}, {Babusiaux}, {Biermann}, {Creevey}, {Ducourant}, {Evans}, {Eyer}, {Guerra}, {Hutton}, {Jordi}, {Klioner}, {Lammers}, {Lindegren}, {Luri}, {Mignard}, {Panem}, {Pourbaix}, {Randich}, {Sartoretti}, {Soubiran}, {Tanga}, {Walton}, {Bailer-Jones}, {Bastian}, {Drimmel}, {Jansen}, {Katz}, {Lattanzi}, {van Leeuwen}, {Bakker}, {Cacciari}, {Casta{\~n}eda}, {De Angeli}, {Fabricius}, {Fouesneau}, {Fr{\'e}mat}, {Galluccio}, {Guerrier}, {Heiter}, {Masana}, {Messineo}, {Mowlavi}, {Nicolas}, {Nienartowicz}, {Pailler}, {Panuzzo}, {Riclet}, {Roux}, {Seabroke}, {Sordo}, {Th{\'e}venin}, {Gracia-Abril}, {Portell}, {Teyssier}, {Altmann}, {Andrae}, {Audard}, {Bellas-Velidis}, {Benson}, {Berthier}, {Blomme}, {Burgess}, {Busonero}, {Busso}, {C{\'a}novas}, {Carry}, {Cellino}, {Cheek}, {Clementini}, {Damerdji}, {Davidson}, {de Teodoro}, {Nu{\~n}ez Campos}, {Delchambre}, {Dell'Oro}, {Esquej},
  {Fern{\'a}ndez-Hern{\'a}ndez}, {Fraile}, {Garabato}, {Garc{\'\i}a-Lario}, {Gosset}, {Haigron}, {Halbwachs}, {Hambly}, {Harrison}, {Hern{\'a}ndez}, {Hestroffer}, {Hodgkin}, {Holl}, {Jan{\ss}en}, {Jevardat de Fombelle}, {Jordan}, {Krone-Martins}, {Lanzafame}, {L{\"o}ffler}, {Marchal}, {Marrese}, {Moitinho}, {Muinonen}, {Osborne}, {Pancino}, {Pauwels}, {Recio-Blanco}, {Reyl{\'e}}, {Riello}, {Rimoldini}, {Roegiers}, {Rybizki}, {Sarro}, {Siopis}, {Smith}, {Sozzetti}, {Utrilla}, {van Leeuwen}, {Abbas}, {{\'A}brah{\'a}m}, {Abreu Aramburu}, {Aerts}, {Aguado}, {Ajaj}, {Aldea-Montero}, {Altavilla}, {{\'A}lvarez}, {Alves}, {Anders}, {Anderson}, {Anglada Varela}, {Antoja}, {Baines}, {Baker}, {Balaguer-N{\'u}{\~n}ez}, {Balbinot}, {Balog}, {Barache}, {Barbato}, {Barros}, {Barstow}, {Bartolom{\'e}}, {Bassilana}, {Bauchet}, {Becciani}, {Bellazzini}, {Berihuete}, {Bernet}, {Bertone}, {Bianchi}, {Binnenfeld}, {Blanco-Cuaresma}, {Blazere}, {Boch}, {Bombrun}, {Bossini}, {Bouquillon}, {Bragaglia}, {Bramante}, {Breedt},
  {Bressan}, {Brouillet}, {Brugaletta}, {Bucciarelli}, {Burlacu}, {Butkevich}, {Buzzi}, {Caffau}, {Cancelliere}, {Cantat-Gaudin}, {Carballo}, {Carlucci}, {Carnerero}, {Carrasco}, {Casamiquela}, {Castellani}, {Castro-Ginard}, {Chaoul}, {Charlot}, {Chemin}, {Chiaramida}, {Chiavassa}, {Chornay}, {Comoretto}, {Contursi}, {Cooper}, {Cornez}, {Cowell}, {Crifo}, {Cropper}, {Crosta}, {Crowley}, {Dafonte}, {Dapergolas}, {David}, {David}, {de Laverny}, {De Luise}, {De March}, {De Ridder}, {de Souza}, {de Torres}, {del Peloso}, {del Pozo}, {Delbo}, {Delgado}, {Delisle}, {Demouchy}, {Dharmawardena}, {Di Matteo}, {Diakite}, {Diener}, {Distefano}, {Dolding}, {Edvardsson}, {Enke}, {Fabre}, {Fabrizio}, {Faigler}, {Fedorets}, {Fernique}, {Fienga}, {Figueras}, {Fournier}, {Fouron}, {Fragkoudi}, {Gai}, {Garcia-Gutierrez}, {Garcia-Reinaldos}, {Garc{\'\i}a-Torres}, {Garofalo}, {Gavel}, {Gavras}, {Gerlach}, {Geyer}, {Giacobbe}, {Gilmore}, {Girona}, {Giuffrida}, {Gomel}, {Gomez}, {Gonz{\'a}lez-N{\'u}{\~n}ez},
  {Gonz{\'a}lez-Santamar{\'\i}a}, {Gonz{\'a}lez-Vidal}, {Granvik}, {Guillout}, {Guiraud}, {Guti{\'e}rrez-S{\'a}nchez}, {Guy}, {Hatzidimitriou}, {Hauser}, {Haywood}, {Helmer}, {Helmi}, {Sarmiento}, {Hidalgo}, {Hilger}, {H{\l}adczuk}, {Hobbs}, {Holland}, {Huckle}, {Jardine}, {Jasniewicz}, {Jean-Antoine Piccolo}, {Jim{\'e}nez-Arranz}, {Jorissen}, {Juaristi Campillo}, {Julbe}, {Karbevska}, {Kervella}, {Khanna}, {Kontizas}, {Kordopatis}, {Korn}, {K{\'o}sp{\'a}l}, {Kostrzewa-Rutkowska}, {Kruszy{\'n}ska}, {Kun}, {Laizeau}, {Lambert}, {Lanza}, {Lasne}, {Le Campion}, {Lebreton}, {Lebzelter}, {Leccia}, {Leclerc}, {Lecoeur-Taibi}, {Liao}, {Licata}, {Lindstr{\o}m}, {Lister}, {Livanou}, {Lobel}, {Lorca}, {Loup}, {Madrero Pardo}, {Magdaleno Romeo}, {Managau}, {Mann}, {Manteiga}, {Marchant}, {Marconi}, {Marcos}, {Marcos Santos}, {Mar{\'\i}n Pina}, {Marinoni}, {Marocco}, {Marshall}, {Martin Polo}, {Mart{\'\i}n-Fleitas}, {Marton}, {Mary}, {Masip}, {Massari}, {Mastrobuono-Battisti}, {Mazeh}, {McMillan}, {Messina}, {Michalik},
  {Millar}, {Mints}, {Molina}, {Molinaro}, {Moln{\'a}r}, {Monari}, {Mongui{\'o}}, {Montegriffo}, {Montero}, {Mor}, {Mora}, {Morbidelli}, {Morel}, {Morris}, {Muraveva}, {Murphy}, {Musella}, {Nagy}, {Noval}, {Oca{\~n}a}, {Ogden}, {Ordenovic}, {Osinde}, {Pagani}, {Pagano}, {Palaversa}, {Palicio}, {Pallas-Quintela}, {Panahi}, {Payne-Wardenaar}, {Pe{\~n}alosa Esteller}, {Penttil{\"a}}, {Pichon}, {Piersimoni}, {Pineau}, {Plachy}, {Plum}, {Poggio}, {Pr{\v{s}}a}, {Pulone}, {Racero}, {Ragaini}, {Rainer}, {Raiteri}, {Rambaux}, {Ramos}, {Ramos-Lerate}, {Re Fiorentin}, {Regibo}, {Richards}, {Rios Diaz}, {Ripepi}, {Riva}, {Rix}, {Rixon}, {Robichon}, {Robin}, {Robin}, {Roelens}, {Rogues}, {Rohrbasser}, {Romero-G{\'o}mez}, {Rowell}, {Royer}, {Ruz Mieres}, {Rybicki}, {Sadowski}, {S{\'a}ez N{\'u}{\~n}ez}, {Sagrist{\`a} Sell{\'e}s}, {Sahlmann}, {Salguero}, {Samaras}, {Sanchez Gimenez}, {Sanna}, {Santove{\~n}a}, {Sarasso}, {Schultheis}, {Sciacca}, {Segol}, {Segovia}, {S{\'e}gransan}, {Semeux}, {Shahaf}, {Siddiqui}, {Siebert},
  {Siltala}, {Silvelo}, {Slezak}, {Slezak}, {Smart}, {Snaith}, {Solano}, {Solitro}, {Souami}, {Souchay}, {Spagna}, {Spina}, {Spoto}, {Steele}, {Steidelm{\"u}ller}, {Stephenson}, {S{\"u}veges}, {Surdej}, {Szabados}, {Szegedi-Elek}, {Taris}, {Taylor}, {Teixeira}, {Tolomei}, {Tonello}, {Torra}, {Torra}, {Torralba Elipe}, {Trabucchi}, {Tsounis}, {Turon}, {Ulla}, {Unger}, {Vaillant}, {van Dillen}, {van Reeven}, {Vanel}, {Vecchiato}, {Viala}, {Vicente}, {Voutsinas}, {Weiler}, {Wevers}, {Wyrzykowski}, {Yoldas}, {Yvard}, {Zhao}, {Zorec}, {Zucker}, \& {Zwitter}}]{2023A&A...674A...1G}
{Gaia Collaboration}, {Vallenari}, A., {Brown}, A.~G.~A., {et~al.} 2023, \aap, 674, A1

\bibitem[{{Galiullin} {et~al.}(2024){Galiullin}, {Rodriguez}, {Kulkarni}, {Sunyaev}, {Gilfanov}, {Bikmaev}, {Yungelson}, {van Roestel}, {G{\"a}nsicke}, {Khamitov}, {Szkody}, {El-Badry}, {Suslikov}, {Prince}, {Buntov}, {Caiazzo}, {Gorbachev}, {Graham}, {Gumerov}, {Irtuganov}, {Laher}, {Medvedev}, {Riddle}, {Rusholme}, {Sakhibullin}, {Sklyanov}, \& {Vanderbosch}}]{2024MNRAS.528..676G}
{Galiullin}, I., {Rodriguez}, A.~C., {Kulkarni}, S.~R., {et~al.} 2024, \mnras, 528, 676

\bibitem[{{Galiullin} \& {Gilfanov}(2021)}]{2021AstL...47..587G}
{Galiullin}, I.~I. \& {Gilfanov}, M.~R. 2021, Astronomy Letters, 47, 587

\bibitem[{{G{\"a}nsicke} {et~al.}(2009){G{\"a}nsicke}, {Dillon}, {Southworth}, {Thorstensen}, {Rodr{\'\i}guez-Gil}, {Aungwerojwit}, {Marsh}, {Szkody}, {Barros}, {Casares}, {de Martino}, {Groot}, {Hakala}, {Kolb}, {Littlefair}, {Mart{\'\i}nez-Pais}, {Nelemans}, \& {Schreiber}}]{2009MNRAS.397.2170G}
{G{\"a}nsicke}, B.~T., {Dillon}, M., {Southworth}, J., {et~al.} 2009, \mnras, 397, 2170

\bibitem[{{Ginsburg} {et~al.}(2019){Ginsburg}, {Sip{\H{o}}cz}, {Brasseur}, {Cowperthwaite}, {Craig}, {Deil}, {Guillochon}, {Guzman}, {Liedtke}, {Lian Lim}, {Lockhart}, {Mommert}, {Morris}, {Norman}, {Parikh}, {Persson}, {Robitaille}, {Segovia}, {Singer}, {Tollerud}, {de Val-Borro}, {Valtchanov}, {Woillez}, {Astroquery Collaboration}, \& {a subset of astropy Collaboration}}]{2019AJ....157...98G}
{Ginsburg}, A., {Sip{\H{o}}cz}, B.~M., {Brasseur}, C.~E., {et~al.} 2019, \aj, 157, 98

\bibitem[{{Graham} {et~al.}(2019){Graham}, {Kulkarni}, {Bellm}, {Adams}, {Barbarino}, {Blagorodnova}, {Bodewits}, {Bolin}, {Brady}, {Cenko}, {Chang}, {Coughlin}, {De}, {Eadie}, {Farnham}, {Feindt}, {Franckowiak}, {Fremling}, {Gezari}, {Ghosh}, {Goldstein}, {Golkhou}, {Goobar}, {Ho}, {Huppenkothen}, {Ivezi{\'c}}, {Jones}, {Juric}, {Kaplan}, {Kasliwal}, {Kelley}, {Kupfer}, {Lee}, {Lin}, {Lunnan}, {Mahabal}, {Miller}, {Ngeow}, {Nugent}, {Ofek}, {Prince}, {Rauch}, {van Roestel}, {Schulze}, {Singer}, {Sollerman}, {Taddia}, {Yan}, {Ye}, {Yu}, {Barlow}, {Bauer}, {Beck}, {Belicki}, {Biswas}, {Brinnel}, {Brooke}, {Bue}, {Bulla}, {Burruss}, {Connolly}, {Cromer}, {Cunningham}, {Dekany}, {Delacroix}, {Desai}, {Duev}, {Feeney}, {Flynn}, {Frederick}, {Gal-Yam}, {Giomi}, {Groom}, {Hacopians}, {Hale}, {Helou}, {Henning}, {Hover}, {Hillenbrand}, {Howell}, {Hung}, {Imel}, {Ip}, {Jackson}, {Kaspi}, {Kaye}, {Kowalski}, {Kramer}, {Kuhn}, {Landry}, {Laher}, {Mao}, {Masci}, {Monkewitz}, {Murphy}, {Nordin}, {Patterson}, {Penprase},
  {Porter}, {Rebbapragada}, {Reiley}, {Riddle}, {Rigault}, {Rodriguez}, {Rusholme}, {van Santen}, {Shupe}, {Smith}, {Soumagnac}, {Stein}, {Surace}, {Szkody}, {Terek}, {Van Sistine}, {van Velzen}, {Vestrand}, {Walters}, {Ward}, {Zhang}, \& {Zolkower}}]{graham2019}
{Graham}, M.~J., {Kulkarni}, S.~R., {Bellm}, E.~C., {et~al.} 2019, \pasp, 131, 078001

\bibitem[{{Green} {et~al.}(2019){Green}, {Schlafly}, {Zucker}, {Speagle}, \& {Finkbeiner}}]{2019ApJ...887...93G}
{Green}, G.~M., {Schlafly}, E., {Zucker}, C., {Speagle}, J.~S., \& {Finkbeiner}, D. 2019, \apj, 887, 93

\bibitem[{{G{\"u}ver} \& {{\"O}zel}(2009)}]{2009MNRAS.400.2050G}
{G{\"u}ver}, T. \& {{\"O}zel}, F. 2009, \mnras, 400, 2050

\bibitem[{{Haberl} \& {Motch}(1995)}]{1995A&A...297L..37H}
{Haberl}, F. \& {Motch}, C. 1995, \aap, 297, L37

\bibitem[{{Hameury}(2020)}]{2020AdSpR..66.1004H}
{Hameury}, J.~M. 2020, Advances in Space Research, 66, 1004

\bibitem[{Harris {et~al.}(2020)Harris, Millman, van~der Walt, Gommers, Virtanen, Cournapeau, Wieser, Taylor, Berg, Smith, Kern, Picus, Hoyer, van Kerkwijk, Brett, Haldane, del R{'{\i}}o, Wiebe, Peterson, G{'{e}}rard-Marchant, Sheppard, Reddy, Weckesser, Abbasi, Gohlke, \& Oliphant}]{Harris2020}
Harris, C.~R., Millman, K.~J., van~der Walt, S.~J., {et~al.} 2020, Nature, 585, 357

\bibitem[{{HI4PI Collaboration} {et~al.}(2016){HI4PI Collaboration}, {Ben Bekhti}, {Fl{\"o}er}, {Keller}, {Kerp}, {Lenz}, {Winkel}, {Bailin}, {Calabretta}, {Dedes}, {Ford}, {Gibson}, {Haud}, {Janowiecki}, {Kalberla}, {Lockman}, {McClure-Griffiths}, {Murphy}, {Nakanishi}, {Pisano}, \& {Staveley-Smith}}]{2016A&A...594A.116H}
{HI4PI Collaboration}, {Ben Bekhti}, N., {Fl{\"o}er}, L., {et~al.} 2016, \aap, 594, A116

\bibitem[{Hunter(2007)}]{Hunter2007}
Hunter, J.~D. 2007, Computing in Science \& Engineering, 9, 90

\bibitem[{{Inight} {et~al.}(2023){Inight}, {G{\"a}nsicke}, {Schwope}, {Anderson}, {Badenes}, {Breedt}, {Chandra}, {Davies}, {Gentile Fusillo}, {Green}, {Hermes}, {Huamani}, {Hwang}, {Knauff}, {Kurpas}, {Long}, {Malanushenko}, {Morrison}, {Quiroz C.}, {Ramos}, {Roman-Lopes}, {Schreiber}, {Standke}, {St{\"u}tz}, {Thorstensen}, {Toloza}, {Tovmassian}, \& {Zakamska}}]{2023MNRAS.525.3597I}
{Inight}, K., {G{\"a}nsicke}, B.~T., {Schwope}, A., {et~al.} 2023, \mnras, 525, 3597

\bibitem[{{Israel} {et~al.}(2016){Israel}, {Esposito}, {Rodr{\'\i}guez Castillo}, \& {Sidoli}}]{2016MNRAS.462.4371I}
{Israel}, G.~L., {Esposito}, P., {Rodr{\'\i}guez Castillo}, G.~A., \& {Sidoli}, L. 2016, \mnras, 462, 4371

\bibitem[{{Ivanova} {et~al.}(2013){Ivanova}, {Justham}, {Chen}, {De Marco}, {Fryer}, {Gaburov}, {Ge}, {Glebbeek}, {Han}, {Li}, {Lu}, {Marsh}, {Podsiadlowski}, {Potter}, {Soker}, {Taam}, {Tauris}, {van den Heuvel}, \& {Webbink}}]{2013A&ARv..21...59I}
{Ivanova}, N., {Justham}, S., {Chen}, X., {et~al.} 2013, \aapr, 21, 59

\bibitem[{{Jackim} {et~al.}(2020){Jackim}, {Szkody}, {Hazelton}, \& {Benson}}]{2020RNAAS...4..219J}
{Jackim}, R., {Szkody}, P., {Hazelton}, B., \& {Benson}, N.~C. 2020, Research Notes of the American Astronomical Society, 4, 219

\bibitem[{{Khamitov} {et~al.}(2023){Khamitov}, {Bikmaev}, {Gilfanov}, {Sunyaev}, {Medvedev}, \& {Gorbachev}}]{2023AstL...49..271K}
{Khamitov}, I.~M., {Bikmaev}, I.~F., {Gilfanov}, M.~R., {et~al.} 2023, Astronomy Letters, 49, 271

\bibitem[{{Knigge}(2006)}]{2006MNRAS.373..484K}
{Knigge}, C. 2006, \mnras, 373, 484

\bibitem[{{Knigge} {et~al.}(2011){Knigge}, {Baraffe}, \& {Patterson}}]{2011ApJS..194...28K}
{Knigge}, C., {Baraffe}, I., \& {Patterson}, J. 2011, \apjs, 194, 28

\bibitem[{{Kupfer} {et~al.}(2023){Kupfer}, {Korol}, {Littenberg}, {Shah}, {Savalle}, {Groot}, {Marsh}, {Le Jeune}, {Nelemans}, {Petiteau}, {Ramsay}, {Steeghs}, \& {Babak}}]{2023arXiv230212719K}
{Kupfer}, T., {Korol}, V., {Littenberg}, T.~B., {et~al.} 2023, arXiv e-prints, arXiv:2302.12719

\bibitem[{{Larsson}(1996)}]{1996A&AS..117..197L}
{Larsson}, S. 1996, \aaps, 117, 197

\bibitem[{{Lazarz} {et~al.}(2022){Lazarz}, {Yan}, {Wilhelm}, {Chen}, {Hill}, {Holtzman}, {Imig}, {Maraston}, {M{\'e}sz{\'a}ros}, {Stringfellow}, {Thomas}, {Beers}, {Bizyaev}, {Drory}, {Lane}, \& {Nitschelm}}]{Mastar2022}
{Lazarz}, D., {Yan}, R., {Wilhelm}, R., {et~al.} 2022, \aap, 668, A21

\bibitem[{{Mackereth} {et~al.}(2019){Mackereth}, {Bovy}, {Leung}, {Schiavon}, {Trick}, {Chaplin}, {Cunha}, {Feuillet}, {Majewski}, {Martig}, {Miglio}, {Nidever}, {Pinsonneault}, {Aguirre}, {Sobeck}, {Tayar}, \& {Zasowski}}]{astroNNapogee2019}
{Mackereth}, J.~T., {Bovy}, J., {Leung}, H.~W., {et~al.} 2019, \mnras, 489, 176

\bibitem[{{Makarov} \& {Secrest}(2022)}]{2022ApJ...933...28M}
{Makarov}, V.~V. \& {Secrest}, N.~J. 2022, \apj, 933, 28

\bibitem[{{Maoz} {et~al.}(2014){Maoz}, {Mannucci}, \& {Nelemans}}]{2014ARA&A..52..107M}
{Maoz}, D., {Mannucci}, F., \& {Nelemans}, G. 2014, \araa, 52, 107

\bibitem[{{Masci} {et~al.}(2019){Masci}, {Laher}, {Rusholme}, {Shupe}, {Groom}, {Surace}, {Jackson}, {Monkewitz}, {Beck}, {Flynn}, {Terek}, {Landry}, {Hacopians}, {Desai}, {Howell}, {Brooke}, {Imel}, {Wachter}, {Ye}, {Lin}, {Cenko}, {Cunningham}, {Rebbapragada}, {Bue}, {Miller}, {Mahabal}, {Bellm}, {Patterson}, {Juri{\'c}}, {Golkhou}, {Ofek}, {Walters}, {Graham}, {Kasliwal}, {Dekany}, {Kupfer}, {Burdge}, {Cannella}, {Barlow}, {Van Sistine}, {Giomi}, {Fremling}, {Blagorodnova}, {Levitan}, {Riddle}, {Smith}, {Helou}, {Prince}, \& {Kulkarni}}]{masci_ztf}
{Masci}, F.~J., {Laher}, R.~R., {Rusholme}, B., {et~al.} 2019, \pasp, 131, 018003

\bibitem[{{Merloni} {et~al.}(2024){Merloni}, {Lamer}, {Liu}, {Ramos-Ceja}, {Brunner}, {Bulbul}, {Dennerl}, {Doroshenko}, {Freyberg}, {Friedrich}, {Gatuzz}, {Georgakakis}, {Haberl}, {Igo}, {Kreykenbohm}, {Liu}, {Maitra}, {Malyali}, {Mayer}, {Nandra}, {Predehl}, {Robrade}, {Salvato}, {Sanders}, {Stewart}, {Tub{\'\i}n-Arenas}, {Weber}, {Wilms}, {Arcodia}, {Artis}, {Aschersleben}, {Avakyan}, {Aydar}, {Bahar}, {Balzer}, {Becker}, {Berger}, {Boller}, {Bornemann}, {Br{\"u}ggen}, {Brusa}, {Buchner}, {Burwitz}, {Camilloni}, {Clerc}, {Comparat}, {Coutinho}, {Czesla}, {Dannhauer}, {Dauner}, {Dauser}, {Dietl}, {Dolag}, {Dwelly}, {Egg}, {Ehl}, {Freund}, {Friedrich}, {Gaida}, {Garrel}, {Ghirardini}, {Gokus}, {Gr{\"u}nwald}, {Grandis}, {Grotova}, {Gruen}, {Gueguen}, {H{\"a}mmerich}, {Hamaus}, {Hasinger}, {Haubner}, {Homan}, {Ider Chitham}, {Joseph}, {Joyce}, {K{\"o}nig}, {Kaltenbrunner}, {Khokhriakova}, {Kink}, {Kirsch}, {Kluge}, {Knies}, {Krippendorf}, {Krumpe}, {Kurpas}, {Li}, {Liu}, {Locatelli}, {Lorenz}, {M{\"u}ller},
  {Magaudda}, {Mannes}, {McCall}, {Meidinger}, {Michailidis}, {Migkas}, {Mu{\~n}oz-Giraldo}, {Musiimenta}, {Nguyen-Dang}, {Ni}, {Olechowska}, {Ota}, {Pacaud}, {Pasini}, {Perinati}, {Pires}, {Pommranz}, {Ponti}, {Poppenhaeger}, {P{\"u}hlhofer}, {Rau}, {Reh}, {Reiprich}, {Roster}, {Saeedi}, {Santangelo}, {Sasaki}, {Schmitt}, {Schneider}, {Schrabback}, {Schuster}, {Schwope}, {Seppi}, {Serim}, {Shreeram}, {Sokolova-Lapa}, {Starck}, {Stelzer}, {Stierhof}, {Suleimanov}, {Tenzer}, {Traulsen}, {Tr{\"u}mper}, {Tsuge}, {Urrutia}, {Veronica}, {Waddell}, {Willer}, {Wolf}, {Yeung}, {Zainab}, {Zangrandi}, {Zhang}, {Zhang}, \& {Zheng}}]{2024A&A...682A..34M}
{Merloni}, A., {Lamer}, G., {Liu}, T., {et~al.} 2024, \aap, 682, A34

\bibitem[{{Nasa High Energy Astrophysics Science Archive Research Center (Heasarc)}(2014)}]{2014ascl.soft08004N}
{Nasa High Energy Astrophysics Science Archive Research Center (Heasarc)}. 2014, {HEAsoft: Unified Release of FTOOLS and XANADU}, Astrophysics Source Code Library, record ascl:1408.004

\bibitem[{{Oke} {et~al.}(1995){Oke}, {Cohen}, {Carr}, {Cromer}, {Dingizian}, {Harris}, {Labrecque}, {Lucinio}, {Schaal}, {Epps}, \& {Miller}}]{1995lris}
{Oke}, J.~B., {Cohen}, J.~G., {Carr}, M., {et~al.} 1995, \pasp, 107, 375

\bibitem[{{Oke} \& {Gunn}(1982)}]{1982dbsp}
{Oke}, J.~B. \& {Gunn}, J.~E. 1982, \pasp, 94, 586

\bibitem[{{Paczy{\'n}ski}(1967)}]{1967AcA....17..287P}
{Paczy{\'n}ski}, B. 1967, \actaa, 17, 287

\bibitem[{{Paczynski}(1976)}]{1976IAUS...73...75P}
{Paczynski}, B. 1976, in Structure and Evolution of Close Binary Systems, ed. P.~{Eggleton}, S.~{Mitton}, \& J.~{Whelan}, Vol.~73, 75

\bibitem[{{Parsons} {et~al.}(2021){Parsons}, {G{\"a}nsicke}, {Schreiber}, {Marsh}, {Ashley}, {Breedt}, {Littlefair}, \& {Meusinger}}]{2021MNRAS.502.4305P}
{Parsons}, S.~G., {G{\"a}nsicke}, B.~T., {Schreiber}, M.~R., {et~al.} 2021, \mnras, 502, 4305

\bibitem[{{Patterson}(1998)}]{1998PASP..110.1132P}
{Patterson}, J. 1998, \pasp, 110, 1132

\bibitem[{{Perley}(2019)}]{2019perley_lpipe}
{Perley}, D.~A. 2019, \pasp, 131, 084503

\bibitem[{{Predehl} {et~al.}(2021){Predehl}, {Andritschke}, {Arefiev}, {Babyshkin}, {Batanov}, {Becker}, {B{\"o}hringer}, {Bogomolov}, {Boller}, {Borm}, {Bornemann}, {Br{\"a}uninger}, {Br{\"u}ggen}, {Brunner}, {Brusa}, {Bulbul}, {Buntov}, {Burwitz}, {Burkert}, {Clerc}, {Churazov}, {Coutinho}, {Dauser}, {Dennerl}, {Doroshenko}, {Eder}, {Emberger}, {Eraerds}, {Finoguenov}, {Freyberg}, {Friedrich}, {Friedrich}, {F{\"u}rmetz}, {Georgakakis}, {Gilfanov}, {Granato}, {Grossberger}, {Gueguen}, {Gureev}, {Haberl}, {H{\"a}lker}, {Hartner}, {Hasinger}, {Huber}, {Ji}, {Kienlin}, {Kink}, {Korotkov}, {Kreykenbohm}, {Lamer}, {Lomakin}, {Lapshov}, {Liu}, {Maitra}, {Meidinger}, {Menz}, {Merloni}, {Mernik}, {Mican}, {Mohr}, {M{\"u}ller}, {Nandra}, {Nazarov}, {Pacaud}, {Pavlinsky}, {Perinati}, {Pfeffermann}, {Pietschner}, {Ramos-Ceja}, {Rau}, {Reiffers}, {Reiprich}, {Robrade}, {Salvato}, {Sanders}, {Santangelo}, {Sasaki}, {Scheuerle}, {Schmid}, {Schmitt}, {Schwope}, {Shirshakov}, {Steinmetz}, {Stewart}, {Str{\"u}der},
  {Sunyaev}, {Tenzer}, {Tiedemann}, {Tr{\"u}mper}, {Voron}, {Weber}, {Wilms}, \& {Yaroshenko}}]{2021A&A...647A...1P}
{Predehl}, P., {Andritschke}, R., {Arefiev}, V., {et~al.} 2021, \aap, 647, A1

\bibitem[{{Pretorius} \& {Knigge}(2012)}]{2012MNRAS.419.1442P}
{Pretorius}, M.~L. \& {Knigge}, C. 2012, \mnras, 419, 1442

\bibitem[{{Prochaska} {et~al.}(2020){Prochaska}, {Hennawi}, {Westfall}, {Cooke}, {Wang}, {Hsyu}, {Davies}, {Farina}, \& {Pelliccia}}]{2020pypeit}
{Prochaska}, J., {Hennawi}, J., {Westfall}, K., {et~al.} 2020, The Journal of Open Source Software, 5, 2308

\bibitem[{{Queiroz} {et~al.}(2020){Queiroz}, {Anders}, {Chiappini}, {Khalatyan}, {Santiago}, {Steinmetz}, {Valentini}, {Miglio}, {Bossini}, {Barbuy}, {Minchev}, {Minniti}, {Garc{\'\i}a Hern{\'a}ndez}, {Schultheis}, {Beaton}, {Beers}, {Bizyaev}, {Brownstein}, {Cunha}, {Fern{\'a}ndez-Trincado}, {Frinchaboy}, {Lane}, {Majewski}, {Nataf}, {Nitschelm}, {Pan}, {Roman-Lopes}, {Sobeck}, {Stringfellow}, \& {Zamora}}]{apogeestarhorse2020}
{Queiroz}, A.~B.~A., {Anders}, F., {Chiappini}, C., {et~al.} 2020, \aap, 638, A76

\bibitem[{{Ramsay} {et~al.}(2018){Ramsay}, {Green}, {Marsh}, {Kupfer}, {Breedt}, {Korol}, {Groot}, {Knigge}, {Nelemans}, {Steeghs}, {Woudt}, \& {Aungwerojwit}}]{2018A&A...620A.141R}
{Ramsay}, G., {Green}, M.~J., {Marsh}, T.~R., {et~al.} 2018, \aap, 620, A141

\bibitem[{{Rappaport} {et~al.}(1983){Rappaport}, {Verbunt}, \& {Joss}}]{1983ApJ...275..713R}
{Rappaport}, S., {Verbunt}, F., \& {Joss}, P.~C. 1983, \apj, 275, 713

\bibitem[{{Ritter} \& {Kolb}(2003)}]{2003ritterkolb}
{Ritter}, H. \& {Kolb}, U. 2003, \aap, 404, 301

\bibitem[{{Rodriguez}(2024)}]{2024PASP..136e4201R}
{Rodriguez}, A.~C. 2024, \pasp, 136, 054201

\bibitem[{{Rodriguez} {et~al.}(2023{\natexlab{a}}){Rodriguez}, {Galiullin}, {Gilfanov}, {Kulkarni}, {Khamitov}, {Bikmaev}, {van Roestel}, {Yungelson}, {El-Badry}, {Sunayev}, {Prince}, {Buntov}, {Caiazzo}, {Drake}, {Gorbachev}, {Graham}, {Gumerov}, {Irtuganov}, {Laher}, {Masci}, {Medvedev}, {Purdum}, {Sakhibullin}, {Sklyanov}, {Smith}, {Szkody}, \& {Vanderbosch}}]{2023ApJ...954...63R}
{Rodriguez}, A.~C., {Galiullin}, I., {Gilfanov}, M., {et~al.} 2023{\natexlab{a}}, \apj, 954, 63

\bibitem[{{Rodriguez} {et~al.}(2023{\natexlab{b}}){Rodriguez}, {Kulkarni}, {Prince}, {Szkody}, {Burdge}, {Caiazzo}, {van Roestel}, {Vanderbosch}, {El-Badry}, {Bellm}, {G{\"a}nsicke}, {Graham}, {Mahabal}, {Masci}, {Mr{\'o}z}, {Riddle}, \& {Rusholme}}]{2023rodriguez_polars}
{Rodriguez}, A.~C., {Kulkarni}, S.~R., {Prince}, T.~A., {et~al.} 2023{\natexlab{b}}, \apj, 945, 141

\bibitem[{{Schreiber} {et~al.}(2021){Schreiber}, {Belloni}, {G{\"a}nsicke}, {Parsons}, \& {Zorotovic}}]{2021NatAs...5..648S}
{Schreiber}, M.~R., {Belloni}, D., {G{\"a}nsicke}, B.~T., {Parsons}, S.~G., \& {Zorotovic}, M. 2021, Nature Astronomy, 5, 648

\bibitem[{{Schreiber} {et~al.}(2024){Schreiber}, {Belloni}, \& {Schwope}}]{2024A&A...682L...7S}
{Schreiber}, M.~R., {Belloni}, D., \& {Schwope}, A.~D. 2024, \aap, 682, L7

\bibitem[{{Schreiber} {et~al.}(2016){Schreiber}, {Zorotovic}, \& {Wijnen}}]{2016MNRAS.455L..16S}
{Schreiber}, M.~R., {Zorotovic}, M., \& {Wijnen}, T.~P.~G. 2016, \mnras, 455, L16

\bibitem[{{Schwope} {et~al.}(2022{\natexlab{a}}){Schwope}, {Buckley}, {Kawka}, {K{\"o}nig}, {Lutovinov}, {Maitra}, {Mereminskiy}, {Miller-Jones}, {Pichardo Marcano}, {Rau}, {Semena}, {Townsend}, \& {Wilms}}]{2022schwope}
{Schwope}, A., {Buckley}, D. A.~H., {Kawka}, A., {et~al.} 2022{\natexlab{a}}, \aap, 661, A42

\bibitem[{{Schwope} {et~al.}(2022{\natexlab{b}}){Schwope}, {Buckley}, {Malyali}, {Potter}, {K{\"o}nig}, {Arcodia}, {Gromadzki}, \& {Rau}}]{2022schwope_b}
{Schwope}, A., {Buckley}, D. A.~H., {Malyali}, A., {et~al.} 2022{\natexlab{b}}, \aap, 661, A43

\bibitem[{{Smethurst} {et~al.}(2022){Smethurst}, {Masters}, {Simmons}, {Garland}, {G{\'e}ron}, {H{\"a}u{\ss}ler}, {Kruk}, {Lintott}, {O'Ryan}, \& {Walmsley}}]{2022MNRAS.510.4126S}
{Smethurst}, R.~J., {Masters}, K.~L., {Simmons}, B.~D., {et~al.} 2022, \mnras, 510, 4126

\bibitem[{{Souchay} {et~al.}(2022){Souchay}, {Secrest}, {Lambert}, {Zacharias}, {Taris}, {Barache}, {Arias}, \& {Makarov}}]{2022A&A...660A..16S}
{Souchay}, J., {Secrest}, N., {Lambert}, S., {et~al.} 2022, \aap, 660, A16

\bibitem[{{Sprague} {et~al.}(2022){Sprague}, {Culhane}, {Kounkel}, {Olney}, {Covey}, {Hutchinson}, {Lingg}, {Stassun}, {Rom{\'a}n-Z{\'u}{\~n}iga}, {Roman-Lopes}, {Nidever}, {Beaton}, {Borissova}, {Stutz}, {Stringfellow}, {Ram{\'\i}rez}, {Ram{\'\i}rez-Preciado}, {Hern{\'a}ndez}, {Kim}, \& {Lane}}]{apogeenet2022}
{Sprague}, D., {Culhane}, C., {Kounkel}, M., {et~al.} 2022, \aj, 163, 152

\bibitem[{{Spruit} \& {Ritter}(1983)}]{1983A&A...124..267S}
{Spruit}, H.~C. \& {Ritter}, H. 1983, \aap, 124, 267

\bibitem[{{Stone-Martinez} {et~al.}(2024){Stone-Martinez}, {Holtzman}, {Imig}, {Nitschelm}, {Stassun}, \& {Brownstein}}]{distmassapogee2024}
{Stone-Martinez}, A., {Holtzman}, J.~A., {Imig}, J., {et~al.} 2024, \aj, 167, 73

\bibitem[{{Sunyaev} {et~al.}(2021){Sunyaev}, {Arefiev}, {Babyshkin}, {Bogomolov}, {Borisov}, {Buntov}, {Brunner}, {Burenin}, {Churazov}, {Coutinho}, {Eder}, {Eismont}, {Freyberg}, {Gilfanov}, {Gureyev}, {Hasinger}, {Khabibullin}, {Kolmykov}, {Komovkin}, {Krivonos}, {Lapshov}, {Levin}, {Lomakin}, {Lutovinov}, {Medvedev}, {Merloni}, {Mernik}, {Mikhailov}, {Molodtsov}, {Mzhelsky}, {M{\"u}ller}, {Nandra}, {Nazarov}, {Pavlinsky}, {Poghodin}, {Predehl}, {Robrade}, {Sazonov}, {Scheuerle}, {Shirshakov}, {Tkachenko}, \& {Voron}}]{2021A&A...656A.132S}
{Sunyaev}, R., {Arefiev}, V., {Babyshkin}, V., {et~al.} 2021, \aap, 656, A132

\bibitem[{{Walter} \& {Bowyer}(1981)}]{1981rscvn}
{Walter}, F.~M. \& {Bowyer}, S. 1981, \apj, 245, 671

\bibitem[{{Warner}(2003)}]{2003cvs..book.....W}
{Warner}, B. 2003, {Cataclysmic Variable Stars}

\bibitem[{{Weisskopf} {et~al.}(2002){Weisskopf}, {Brinkman}, {Canizares}, {Garmire}, {Murray}, \& {Van Speybroeck}}]{2002PASP..114....1W}
{Weisskopf}, M.~C., {Brinkman}, B., {Canizares}, C., {et~al.} 2002, \pasp, 114, 1

\bibitem[{{Weisskopf} {et~al.}(2000){Weisskopf}, {Tananbaum}, {Van Speybroeck}, \& {O'Dell}}]{2000SPIE.4012....2W}
{Weisskopf}, M.~C., {Tananbaum}, H.~D., {Van Speybroeck}, L.~P., \& {O'Dell}, S.~L. 2000, in Society of Photo-Optical Instrumentation Engineers (SPIE) Conference Series, Vol. 4012, X-Ray Optics, Instruments, and Missions III, ed. J.~E. {Truemper} \& B.~{Aschenbach}, 2--16

\bibitem[{{Wenger} {et~al.}(2000){Wenger}, {Ochsenbein}, {Egret}, {Dubois}, {Bonnarel}, {Borde}, {Genova}, {Jasniewicz}, {Lalo{\"e}}, {Lesteven}, \& {Monier}}]{2000A&AS..143....9W}
{Wenger}, M., {Ochsenbein}, F., {Egret}, D., {et~al.} 2000, \aaps, 143, 9

\bibitem[{{Wevers} {et~al.}(2017){Wevers}, {Torres}, {Jonker}, {Nelemans}, {Heinke}, {Mata S{\'a}nchez}, {Johnson}, {Gazer}, {Steeghs}, {Maccarone}, {Hynes}, {Casares}, {Udalski}, {Wetuski}, {Britt}, {Kostrzewa-Rutkowska}, \& {Wyrzykowski}}]{2017MNRAS.470.4512W}
{Wevers}, T., {Torres}, M.~A.~P., {Jonker}, P.~G., {et~al.} 2017, \mnras, 470, 4512

\bibitem[{{Wilms} {et~al.}(2000){Wilms}, {Allen}, \& {McCray}}]{2000ApJ...542..914W}
{Wilms}, J., {Allen}, A., \& {McCray}, R. 2000, \apj, 542, 914

\bibitem[{{Yang} {et~al.}(2022){Yang}, {Hare}, {Kargaltsev}, {Volkov}, {Chen}, \& {Rangelov}}]{2022ApJ...941..104Y}
{Yang}, H., {Hare}, J., {Kargaltsev}, O., {et~al.} 2022, \apj, 941, 104

\end{thebibliography}

\end{document}